\documentclass[onecolumn,superscriptaddress,nofootinbib]{revtex4}

\pdfoutput=1
\usepackage{amssymb,amsmath,amsthm,graphicx}
\usepackage{graphics, color}
\usepackage{subfigure} 
\usepackage{latexsym}
\usepackage{bm}
\usepackage{epsfig}
\usepackage[pdftex]{hyperref}

\usepackage[normalem]{ulem}
\usepackage{amsmath}
\usepackage{enumerate}
\usepackage{amsfonts}
\usepackage{yfonts}

\usepackage{subfigure}
\usepackage{psfrag}

\usepackage{epsfig}
\usepackage[utf8]{inputenc}
\usepackage{float}
\usepackage{graphicx}
\usepackage{cancel}
\usepackage{mathrsfs}
\usepackage{amssymb}
\usepackage{amsfonts}
\usepackage{amsmath}
\usepackage{slashed}
\usepackage{comment}

\usepackage{graphicx}
\usepackage{bm}

\def\ie{{\it i.e.}}

\def\({\left(}
\def\){\right)}
\def\[{\left[}
\def\]{\right]}
\def\<{\langle}
\def\>{\rangle}

\newcommand{\be}{\begin{equation}}
\newcommand{\ee}{\end{equation}}
\newcommand{\bea}{\begin{eqnarray}}
\newcommand{\eea}{\end{eqnarray}}
\newcommand{\bwt}{\begin{widetext}}
\newcommand{\ewt}{\end{widetext}}

\newcommand{\bi}{\begin{itemize}}
\newcommand{\ei}{\end{itemize}}
\newcommand{\ben}{\begin{enumerate}}
\newcommand{\een}{\end{enumerate}}
\newcommand{\bca}{\begin{cases}}
\newcommand{\eca}{\end{cases}}
\newcommand{\bln}{\begin{align}}
\newcommand{\eln}{\end{align}}
\newcommand{\bst}{\begin{split}}
\newcommand{\est}{\end{split}}

\begin{document}

\title {Hairy black holes and the endpoint of AdS$_4$ charged superradiance}


\author{\'Oscar J. C. Dias}
\email{ojcd1r13@soton.ac.uk}
\affiliation{STAG research centre and Mathematical Sciences, University of Southampton, UK \vspace{1 cm}}

\author{Ramon Masachs}
\email{rmg1e15@soton.ac.uk}
\affiliation{STAG research centre and Mathematical Sciences, University of Southampton, UK \vspace{1 cm}}

\begin{abstract}
We construct  hairy black hole solutions that merge with the anti-de Sitter (AdS$_4$) Reissner-Nordstr\"om black hole at the onset of superradiance. These hairy black holes have, for a given mass and charge, higher entropy than the corresponding AdS$_4$-Reissner-Nordstr\"om black hole. Therefore, they are natural candidates for the endpoint of the charged superradiant instability. On the other hand, hairy black holes never dominate the canonical and grand-canonical ensembles. The zero-horizon radius of the hairy black holes is a soliton (i.e. a boson star under a gauge transformation). We construct our solutions perturbatively, for small mass and charge, so that the properties of  hairy black holes can be used to testify and compare with the endpoint of initial value simulations. We further discuss the near-horizon scalar condensation instability which is also present in global AdS$_4$-Reissner-Nordstr\"om black holes. We highlight the different nature of the near-horizon and superradiant instabilities and that hairy black holes ultimately exist because of the non-linear instability of AdS.
\end{abstract}

\today

\maketitle

\tableofcontents



\section{Introduction and summary of main results}\label{section:Introduction}

One of the most challenging open questions of  general relativity concerns finding the time evolution and endpoint of the {\it superradiant instability} in the Kerr-AdS black hole (or in systems that mimic this system). When a wave with frequency $\omega$ and azimuthal angular momentum $m_\phi$ scatters a  Kerr-AdS black hole with angular velocity $\Omega_H$, its amplitude can grow if $\omega<m_\phi \Omega_H$ \cite{Zeldovich:1971,Starobinsky:1973,Teukolsky:1974yv} (see \cite{Brito:2015oca,Ganchev:2016zag} for recent reviews on superradiance). This follows from a linear perturbation analysis \cite{Cardoso:2004hs,Dias:2013sdc,Cardoso:2013pza}. In global AdS, the amplified wave reflects at the AdS boundary and keeps being amplified driving the system unstable. Finding the properties of the instability evolution and its endpoint requires a non-trivial $( 3+1)$-dimensional numerical simulation not yet available. 

However, valuable insights on the possible evolution can be obtained without solving the actual initial value problem. Indeed, it was found that the zero mode (i.e. onset) of the instability signals the existence of a new branch of black hole solutions with scalar hair \cite{Reall:2002bh,Kunduri:2006qa,Dias:2011at,Stotyn:2011ns} or even gravitational hair  \cite{Dias:2015rxy}. {\it Per se} the existence of these hairy black holes is interesting for two main reasons: 1) they immediately rule out any attempts of proving no-hair theorems (under non-restrictive assumptions), and 2) they have a single Killing vector field that is also the horizon generator, thus showing that the assumptions of Hawking's rigidity theorem can be evaded \cite{Dias:2011at,Dias:2015rxy,Dias:2015nua}. Of relevance for the initial value problem, it turns out that, given a mass and angular momentum, these hairy black holes always have higher entropy than the Kerr-AdS black hole. This property and the fact that they are connected to the onset of superradiance suggest that these black holes could  be the endpoint of rotational superradiance. However, this is not the case because all of those constructed to date have $\Omega_H L>1$ (where $L$ is the AdS radius) and are thus still superradiantly unstable \cite{Dias:2011at,Dias:2015rxy,Green:2015kur}, this time to superradiant modes with higher $m_\phi$. This, together with other observations that can be found in the original articles, led the authors of \cite{Dias:2011at,Dias:2015rxy,Niehoff:2015oga} to conjecture 
that the superradiant instability might evolve into one of two possibilities: 
1) the endpoint of the superradiant instability is a singular solution reached in finite time  which implies  cosmic  censorship  violation, or 2) the time  evolution  develops  higher and higher  $m_\phi$ structure (with increasing entropy) and the system never settles down. In the latter case quantum gravity effects would become relevant at some point and cosmic censorship would also be violated, at least in spirit. In either case, we would have an explicit example of cosmic censorship violation in four dimensions and in a system that can be formed by gravitational collapse. This would show that the violation found in \cite{Lehner:2010pn,Figueras:2015hkb} can also occur in a four-dimensional geometry.   

To downsize the level of difficulty of the problem while still keeping some essential ingredients and associated questions, we can turn-off the rotation of the black hole and consider a static, but electrically charged, black hole. Indeed the global AdS Reissner-Nordstr\"om black hole still has a generalized (effective) ergoregion where negative energy states can be excited \cite{Denardo:1973pyo,Bekenstein:1973mi,Hawking:1999dp,Uchikata:2011zz,Green:2015kur} and the initial value problem is now a simpler $(1+1)$-dimensional problem. Such a black hole with chemical potential $\mu$ can then be unstable to superradiance when a scalar field with frequency $\omega$ and charge $q$ scatters it if  $\omega<\mu q$ \cite{Bekenstein:1973mi,Hawking:1999dp,Uchikata:2011zz,Green:2015kur}. Similar to the rotating geometry case, the onset of charged superradiance is associated to novel hairy  black holes with a charged scalar field floating above the horizon: electric repulsion balances the scalar condensate against gravitational collapse. In global AdS$_5$, these charged hairy black holes were constructed perturbatively and then numerically in \cite{Basu:2010uz,Bhattacharyya:2010yg,Dias:2011tj,Gentle:2011kv,Markeviciute:2016ivy}. 

For a given energy and charge, the hairy solutions of \cite{Basu:2010uz,Bhattacharyya:2010yg,Dias:2011tj,Markeviciute:2016ivy} have higher entropy than the AdS$_5$ Reissner-Nordstr\"om black hole. Therefore, unless the phase diagram of the theory contains a (unlikely) solution with even higher entropy, these hairy black holes should be the endpoint of charged superradiance. Unlike the rotating case, this is a robust statement since $q$ is fixed once we choose the theory and thus the hairy black holes are not unstable to superradiance (in the rotating case $m_\phi$ is a quantized integer that is not fixed by the theory so hairy black holes associated with a given $m_\phi$ are always unstable to higher $m_\phi$'s). This is where the charged superradiant system becomes a less interesting toy-model to discuss the endpoint of rotating superradiance.\footnote{\label{foot1}Note however that the charged hairy black holes can still be unstable to superradiance if we introduce a new scalar field in the system.} 

Nevertheless, moving beyond the initial motivation, there are good reasons to investigate the charged system. Firstly, the properties of the evolution of the unstable charged system should be known. This was addressed by a recent numerical time evolution simulation which confirmed that, in AdS$_4$, an unstable Reissner-Nordstr\"om black hole indeed evolves towards a charged hairy black hole \cite{Bosch:2016vcp}. 

A second reason is related to the {\it non-linear, weakly turbulent, instability} of global AdS. If we perturb global AdS with a scalar field or with a wave packet of gravitons, the system evolves towards the formation of a black hole, for arbitrarily small initial data, as several studies suggest \cite{Bizon:2011gg,Dias:2012tq,Buchel:2012uh,Balasubramanian:2014cja,Craps:2014vaa,Craps:2014jwa,Green:2015dsa,Bizon:2015pfa}. If the spherically symmetric initial data is made of a neutral scalar field the final black hole is an AdS-Schwarzschild black hole. However, if the scalar field is charged then the endpoint of the nonlinear instability should be a hairy black hole since it has higher entropy than the AdS-Reissner-Nordstr\"om with same charge and energy. This was confirmed to be the case by numerical simulations in global AdS$_4$ and AdS$_5$
 \cite{Buchel:2012uh,Buchel:2013uba,Arias:2016aig}.

Since non-linear simulations are computationally costly, the numerical evolution studies of \cite{Buchel:2012uh,Buchel:2013uba,Bosch:2016vcp,Arias:2016aig} were necessarily limited in the range of parameters and initial data. Thus, it would be desirable to construct the hairy black hole solutions of Einstein-Maxwell theory in global AdS$_4$ directly as solutions of the elliptic boundary value problem. This would confirm (or not) that the qualitative features of hairy black holes of AdS$_5$ extend to AdS$_4$. It would also give quantitative thermodynamic quantities that can be used to compare with the endpoint of  future numerical simulations similar to \cite{Buchel:2012uh,Buchel:2013uba,Bosch:2016vcp,Arias:2016aig}. 
This is a main aim of our manuscript. 

A third reason to be interested on the charged AdS system, even in the absence of rotation, is related to a third instability of AdS known as the {\it near-horizon scalar condensation instability} \cite{Gubser:2008px}. 
The superradiant instability is present only in global AdS black holes and requires (for charged superradiance) that both the scalar field and the geometry are charged. On the other hand, the near-horizon scalar condensation instability (and associated hairy black holes) was first found in planar AdS and, in the AdS/CFT duality context, triggered the research field of ``holographic superconductors" and AdS/condensed matter correspondence \cite{Gubser:2008px,Hartnoll:2008vx,Hartnoll:2008kx}. This instability is present in any AdS geometry that has an extremal (zero temperature) configuration even if the spacetime and the scalar field are neutral \cite{Hartnoll:2008kx,Dias:2010ma}. It occurs whenever the scalar field satisfies the asymptotic AdS$_d$  Breitenlohner-Freedman (BF) bound \cite{Breitenlohner:1982jf} of the full theory but violates the AdS$_2$ BF bound (associated with the near-horizon geometry of the extremal black hole of the theory).
In particular, the near-horizon instability is present in the global AdS-Reissner-Nordstr\"om black hole, where it co-exists with the superradiant instability \cite{Dias:2010ma,Dias:2011tj}. These two instabilities are entangled but in certain limits their distinct nature emerges: {\it large} ({\it small}) global AdS-Reissner-Nordstr\"om black holes, in AdS radius units, only have the near-horizon (superradiant) instability. In Section \ref{sec:instability}, we will discuss the sharp distinction  of these two instabilities in AdS$_4$.  

Since the superradiant instability is present in small AdS-Reissner-Nordstr\"om black holes, the associated hairy black holes can be constructed perturbatively around global AdS as a small expansion in the horizon radius and in the scalar condensate amplitude \cite{Basu:2010uz,Bhattacharyya:2010yg,Dias:2011tj}. (On the other hand,  large hairy black holes that emerge from the near-horizon instability do not have such a perturbative expansion). A few observations immediately suggest the existence of a perturbative expansion in the  superradiant case. As mentioned above, a complex scalar field with frequency $\omega$ and charge $q$ scattering a charged black hole with chemical potential $\mu$ is superradiantly amplified when \cite{Bekenstein:1973mi,Hawking:1999dp}\footnote{Note that we can make, as we will typically do in this manuscript, the gauge choice $A(r_+)=0$. In this case the criterion \eqref{superradiance} for instability reads instead  ${\rm Re}(\omega)<0$. For example, in this gauge \eqref{AdS4normal} reads $\omega \simeq \frac{3+2p}{L}-q\mu+\mathcal{O}\left( r_+\right)$.}
\begin{equation}\label{superradiance}
{\rm Re}\,\omega < q\,\mu.
\end{equation} 
For definiteness, consider a Reissner-Nordstr\"om black hole in global AdS$_4$ (RN-AdS$_4$), with cosmological radius $L$, horizon radius $r_+$ and chemical potential $\mu$. Its gauge potential is thus $A=\mu\left( 1-r_+/r\right) dt$. Alternatively, we can parametrize this solution with $r_+$ and the temperature $T$ in which case $\mu=\sqrt{2}\sqrt{1+\frac{3r_+^2}{L^2}-4\pi r_+ T}$.
These black holes are regular all the way to extremality (where their temperature vanishes), and the chemical potential is
\begin{equation}\label{extremalitycond}
\mu{\bigr |}_{ext} =\sqrt{2}\sqrt{1+\frac{3r_+^2}{L^2}}\: \lesssim  \sqrt{2}+\mathcal{O}\left( r_+^2\right),
\end{equation} 
where in the last approximation we considered the small horizon radius limit of the condition.
Consider now a scalar perturbation in this RN-AdS$_4$ background that we Fourier decompose as $\phi(t,r)=e^{-i\omega t}\phi(r)$. If we impose reflecting asymptotic boundary conditions that preserve the energy and charge of the system, RN-AdS$_4$ behaves as a confined box whereby waves can be dissipated only through the horizon. Therefore, only certain frequencies can fit inside the AdS$_4$ box: the frequency spectrum is quantized. The inverse of the imaginary part of the frequency  gives the dissipation timescale (if ${\rm Im}\,\omega<0$) or an instability growth rate (if ${\rm Im}\,\omega>0$). In the limit where $r_+\to 0$, one has ${\rm Im}\,\omega \to 0$ and the frequency spectrum reduces to the normal modes of global AdS$_4$,
\begin{equation}\label{AdS4normal}
\omega \simeq \frac{3+2p}{L}+\mathcal{O}\left( r_+\right),
\end{equation} 
where $p$ is the radial overtone of the perturbation that gives the number of radial nodes of the associated wavefunction (in this manuscript we will be interested on the scalar solutions with lowest energy which correspond to $p=0$). Altogether, conditions \eqref{superradiance}$-$\eqref{AdS4normal} imply that incident scalar waves on a small near-extremal RN-AdS$_4$ black hole drive it unstable to superradiance if  
\begin{eqnarray}\label{eSuperradianceBound}
&& \frac{3+2p}{L}+\mathcal{O}\left( r_+\right)\:  \lesssim\:  {\rm Re}\,\omega\:  \lesssim \: q\sqrt{2}+\mathcal{O}\left( r_+^2\right)  \quad \Leftrightarrow \quad e\gtrsim  e_c +\mathcal{O}\left( r_+\right)\quad \hbox{with} \quad e_c \equiv \frac{3+2p}{\sqrt{2}}, \nonumber\\
&&
\end{eqnarray} 
where we introduced the dimensionless scalar field charge $e\equiv q L$.
Thus, for $p=0$ small near-extremal black holes are unstable for $e\gtrsim  \frac{3}{\sqrt{2}}$. By continuity, they should also be unstable when we move away from extremality up to  a point where the instability switches-off (this point is the onset of superradiance if we do the reversed path, from large to low temperatures). 

In the absence of a horizon, a normal mode of global AdS$_4$, with frequency $\omega$ and amplitude $\varepsilon$, solves the Klein-Gordon equation.
Going beyond the linear order, this scalar field sources the Einstein equation at order $\mathcal{O}\left(\varepsilon^2\right)$ and higher and back-reacts in the background geometry. If we impose reflecting boundary conditions on the boundary of global AdS$_4$, it is conceivable that the solution can be back-reacted (while being smooth everywhere) to any order in a perturbative expansion in the scalar amplitude $\varepsilon$. In the back-reaction process the normal mode frequency should itself receive corrections at each order. This possibility turns out to be correct and the back-reaction of a normal mode of global AdS is known as a {\it boson star}. Indeed they were explicitly constructed in AdS$_5$ in \cite{Basu:2010uz,Bhattacharyya:2010yg,Dias:2011tj,Gentle:2011kv} and studied with much detail in \cite{Gentle:2011kv}. Similarly, these solutions exist in AdS$_4$ and will be constructed in Section \ref{sec:Soliton}. We have the freedom to do a $U(1)$ gauge transformation that eliminates the phase of the scalar field at the expense of changing the gauge potential. So the `boson star' can be equivalently written as a `soliton', i.e. a time-independent horizonless solution that is regular everywhere. In most of our manuscript we will adopt this perspective. 

When a theory has a solitonic solution and a black hole solution that is unstable, we should consider, as a rule of thumb, the possibility that there might exist a third solution that describes a hairy black hole  constructed by placing the small black hole at the core of the soliton \footnote{In the literature there were attempts, to find no-hair theorems stating that one cannot add black holes inside boson stars or solitons \cite{Kastor:1992qy,Pena:1997cy,Astefanesei:2003qy}. However, as pointed out in \cite{Dias:2011at} for asymptotically AdS and in \cite{Herdeiro:2014goa} for asymptotically flat, these results only apply to static black holes with no coupling to the Maxwell field and follow essentially because the boson star has $e^{-i\omega t}$ time dependence and $t\to \infty$ at the horizon of a BH. The scalar field thus oscillates infinitely often near the horizon and cannot be smoothly continued inside. There are however ways to evade this no-go fate  \cite{Basu:2010uz,Dias:2011tj,Dias:2011at,Herdeiro:2014goa} if a gauge field or rotation are present in the system.}. This was first observed in the Einstein$-$Yang-Mills theory: a black hole can be added to the Bartnik-Mckinnon soliton \cite{Bartnik:1988am} of the theory leading to a coloured black hole with non-abelian Yang-Mills hair \cite{Bizon:1990sr,Kuenzle:1990is,Volkov:1990sva}.
This rule of thumb also materializes in the gravitational Abelian Higgs model (i.e. Einstein-Maxwell theory with a complex scalar field) that we are considering \cite{Basu:2010uz,Bhattacharyya:2010yg,Dias:2011tj}. The theory has a small RN-AdS$_4$ black hole that we can put on  top of the soliton that we construct in Section \ref{sec:Soliton}. 

Quite remarkably, the idea encoded in the above rule of thumb  can be realized in a simple `non-interacting thermodynamic model' that captures the {\it leading} order thermodynamics of the system {\it without} solving the equations of motion of the theory\footnote{To our knowledge, the first work that tried to concretize explicitly from first principles (i.e., beyond the numerical confirmation) the idea that we can place a black hole inside a solitonic solution goes back to \cite{Ashtekar:2000nx}. Indeed, in the context of the Einstein$-$Yang-Mills theory, \cite{Ashtekar:2000nx} proposed an `interacting thermodynamic bound state model' where the energy of the final colored (hairy) black hole is the sum of the energies of the solitonic background plus the small bare black hole (i.e. the uncolored, hairless, black hole of the theory) and, in addition, a binding energy contribution that would account for the interactions between the two previous components of the system. This model was then explored to extract some bounds and monotonic behaviors of the thermodynamic quantities of colored BHs. The non-interacting model on the other hand, neglects the interaction or binding energy contribution, and allows to find analytical expressions, not only bounds, for the thermodynamic quantities.}. The model simply assumes that at leading order the energy and charge of the hairy black hole is just the sum of the energies and charges of its RN-AdS$_4$ and solitonic components. The distribution of charges among the two constituents is  determined by the  condition that the entropy must be maximized. This is equivalent to require that the chemical potential of the RN-AdS$_4$ and solitonic component match, i.e. that the system is in thermodynamic equilibrium. We apply this model in Section  \ref{sec:noninteracting} and will find that it gets the correct leading order thermodynamics of the system.

Moving beyond the leading order inspection of the system, the hairy black holes will be constructed perturbatively in a double expansion in the dimensionless horizon radius $r_+/L$ and scalar condensate amplitude $\varepsilon$ in  Section \ref{sec:Hairy BH}. The equations of motion of the gravitational Abelian Higgs theory are solved using a standard matching asymptotic expansion procedure (see e.g.  \cite{Basu:2010uz,Bhattacharyya:2010yg,Dias:2011tj,Dias:2011at,Stotyn:2011ns}) whereby we consider a near region $\frac{r_+}{L}\leq \frac{r}{L}\ll 1 $  and a far region $r\gg r_+$ which overlap in the region $\frac{r_+}{L}\ll \frac{r}{L}\ll 1$ if we consider small black holes, $\frac{r_+}{L}\ll 1$.  
The resulting thermodynamic quantities of the hairy black hole $-$ energy, charge, chemical potential, temperature, entropy, and the Helmoltz and Gibbs free energies $-$ are then computed in \eqref{ThermoHairyBH} and obey the first law of thermodynamics. Confirming the rationale beyond their construction, hairy black holes merge with the RN-AdS$_4$ family at the onset of superradiance and their zero-horizon radius limit is a soliton. The reader not interested on the construction details can immediately jump to these expressions that describe uniquely the hairy black holes and to their physical discussion in Section \ref{sec:ThermoDiscussion}.

The hairy black holes and solitons of the theory add new competitions in the phase diagram of thermal phases of the gravitational Abelian Higgs model. The discussion of these competitions depends on the thermal ensemble $-$ microcanonical, canonical or grand-canonical $-$ that we consider. 
It turns out that, as summarized in the phase diagrams of Fig. \ref{fig:MCphaseDiag}, in the microcanonical ensemble hairy black holes are always the dominant thermal phase. However, large RN-AdS$_4$ are still the preferred thermal phase both in the canonical ensemble (see phase diagram of Fig. \ref{fig:Canonical_phase_comp}) and grand-canonical ensemble (see phase diagram of Fig. \ref{fig:GCphaseDiag}). 

A final comment on the near-horizon scalar condensation instability and its associated hairy black holes is in order. In Section \ref{sec:NH} we will find the conditions under which a scalar field in an extremal RN-AdS$_4$ is unstable to the near-horizon instability. This happens when the dimensionless scalar charge $e$ is above the bound \eqref{NHunstableCriterion}. In the limit where the dimensionless horizon radius $R_+\to\infty $, we recover the known result for planar RN-AdS$_4$ black holes, $e^2\geq \frac{3}{2}+\mathcal{O}\left(R_+^{-2}\right)$ (for a massless field) \cite{Gubser:2008px,Hartnoll:2008vx,Hartnoll:2008kx}. Global RN-AdS$_4$ black holes are unstable for larger values of the scalar field charge. However, in  the opposite limit,  $R_+\to 0$,  we get $e^2\geq \frac{1}{8R_+^2}+\mathcal{O}(1)$, which indicates that the near-horizon scalar condensation instability is {\it suppressed} for {\it small} (global) RN-AdS$_4$ black holes. This is to be contrasted with the superradiant instability and highlights the different nature of the two instabilities.
In particular, this also means that hairy black holes branching-off from the RN-AdS$_4$ family do not admit a small perturbative expansion in the dimensionless horizon radius $R_+$ (and $\varepsilon$). A numerical non-linear construction, that we leave for future work, would be required.  

Summarizing our conclusions, the phase diagram of AdS$_4$-Einstein-Maxwell theory with a complex scalar field depends on the window of the dimensionless scalar field charge $e$ that we consider. There are three cases:
\begin{itemize}
\item $e^2< \frac 3 2$: in this case RN-AdS$_4$ is stable both against superradiance and the near-horizon scalar condensation instability. Thus, there are no hairy black hole solutions. We do have a 1-parameter soliton family.  Based on the AdS$_5$ results of  \cite{Basu:2010uz,Bhattacharyya:2010yg,Dias:2011tj} and specially of \cite{Gentle:2011kv}, very likely this soliton has an intermediate Chandrasekhar limit at a critical charge and mass and then an intricate discontinuous branch structure. Finding wether this is the case for large charges also in AdS$_4$  requires a numerical study beyond the perturbative analysis done here.  

\item $\frac 3 2 \leq e^2\leq \frac 9 2$: the near horizon instability takes place near extremality for large RN-AdS$_4$ black holes, i.e. only above a critical mass and charge. Consequently, the associated {\it large} hairy black hole solutions cannot be constructed perturbatively and would require a full non-linear numerical construction. Since they do not have a perturbative construction their zero-horizon radius limit is likely singular and certainly not the soliton (this is the case in AdS$_5$ \cite{Basu:2010uz,Bhattacharyya:2010yg,Dias:2011tj}). The soliton, on the other hand has similar properties to those of the previous case. 

\item $e^2\geq \frac 9 2$: Small RN-AdS$_4$ black holes are unstable to the superradiant instability and the associated hairy black holes can be constructed perturbatively for small mass and charge (see Section \ref{sec:Hairy BH}). They are a 2-parameter family of solutions that in a phase diagram span an area bounded by the onset of superradiance (where they merge with RN-AdS$_4$) and,  at their zero-horizon radius limit, the soliton (see Fig. \ref{fig:MCphaseDiag}). Eventually, for larger values of the charges (that are not captured by our perturbative analysis) the zero-horizon radius limit of the hairy black hole might no longer be the soliton (this is the case in AdS$_5$  \cite{Dias:2011tj}) but answering this requires a numerical analysis. 
These hairy black holes always have higher entropy than the RN-AdS$_4$ black hole in the window of energy and charge where the two phases co-exist. Interestingly, in the phase diagram of solutions of the theory, the hairy black holes connect the onset of the (linear) superradiant instability with the key player $-$ the soliton $-$ of the non-linear weakly turbulent instability of AdS$_4$. Indeed, at the simplest but fundamental level, this non-linear instability follows from the appearance of irremovable secular resonances when two or more solitons are taken as initial data and collide \cite{Bizon:2011gg,Dias:2012tq,Dias:2016ewl}.  
\end{itemize}

A perturbative construction of the hairy solutions for the gravitational Abelian Higgs model in AdS$_5$ was done in  \cite{Basu:2010uz,Bhattacharyya:2010yg,Dias:2011tj} and agree extremely well (for small charges) with the full nonlinear solutions constructed in \cite{Gentle:2011kv,Dias:2011tj}. Our perturbative construction finds hairy solutions for the AdS$_4$  theory. 
Qualitatively, we do not find differences between the AdS$_5$ and AdS$_4$ phase diagrams. Therefore, \cite{Gentle:2011kv,Dias:2011tj} and our results suggest that (at least for small black holes) the qualitative phase diagram of gravitational Abelian Higgs theory is independent of the dimension. In addition, our quantitative results can be used to compare against and testify the hairy solutions that are the endpoint of initial value simulations like the ones reported in \cite{Buchel:2012uh,Buchel:2013uba,Bosch:2016vcp,Arias:2016aig}. Hairy black hole solutions of gravitational Abelian Higgs model in AdS$_4$ for a scalar field mass of $m^2=-4/L^2$ were recently constructed numerically in \cite{Basu:2016mol} and discussed in the grand-canonical ensemble. Their qualitative results agree with our conclusions for the grand-canonical ensemble (namely, hairy black holes are always subdominant with respect to RN-AdS$_4$) indicating that the qualitative conclusions, at least in the grand-canonical ensemble, might be similar for different scalar masses. The reader interested on superradiance and asymptotically AdS$_5\times$S$^5$ hairy black holes that are solutions of supergravity can find an exhaustive discussion in \cite{Bhattacharyya:2010yg,{Markeviciute:2016ivy}}.

The plan of the manuscript is as follows.
In Section \ref{sec:Model} we introduce the theory, its equations of motion and the boundary conditions for the elliptic problem, and describe how the thermodynamic quantities can be computed. Section \ref{sec:instability} briefly reviews the RN-AdS$_4$ solution and discusses in detail the conditions where they are unstable to near-horizon scalar condensation and/or superradiant instabilities. The soliton (boson star) of the theory is constructed perturbatively in Section \ref{sec:Soliton}. The leading order thermodynamics of hairy black holes is discussed in Section \ref{sec:noninteracting}  using the non-interacting thermodynamic model. In Section \ref{sec:Hairy BH}, hairy black hole solutions are constructed perturbatively using a matched asymptotic expansion analysis. The thermodynamical and physical properties of the hairy solutions are analysed in detail in the three thermal ensembles in Section \ref{sec:ThermoDiscussion}. 
For the benefit of a reader who wants to reproduce our results, the technical appendices \ref{App:Superradiance}, \ref{App:Soliton}, and \ref{App:hairyBH} give the expressions for the field coefficients that appear in the perturbative expansions of the superradiant instability of Section \ref{sec:superradiance}, of the soliton construction of Section \ref{sec:Soliton} and of the hairy black hole of Section \ref{sec:Hairy BH}.

\section{Model \label{sec:Model}}

\subsection{Field ansatz, equations of motion and boundary conditions}

We consider 4-dimensional Einstein-Maxwell gravity with negative cosmological constant $\Lambda=-\frac{3}{L^2}$, minimally coupled to a charged complex scalar field $\phi$ with charge $q$. The associated action is 
\begin{align}\label{action}
S=\frac{1}{16 \pi G_4}\int \mathrm d^4 x \sqrt{-g}\left[R-2\Lambda - \frac 1 2 F_{\mu\nu} F^{\mu\nu} - 2D_{\mu}\phi (D^{\mu}\phi)^\dagger +2V(|\phi|) \right],
\end{align}
where $R$ is the Ricci scalar, $F=dA$ with $A$ being a gauge potential and $D_{\mu}\phi=\nabla_{\mu} \phi -iqA_{\mu}\phi$ is the associated gauge covariant derivative. We consider a potential $V(|\phi|)= m^2 \phi\phi^*$, with $m$ being the mass of the scalar field. For definiteness, onwards we set $m=0$ but our nonlinear construction can be extended to a massive scalar field or to a more generic potential. We fix Newton's constant as $G_4=1$. 

We are interested on solitonic and black hole solutions of \eqref{action} that are static, spherically symmetric and asymptotically global AdS$_4$. We can use reparametrization of the time $t$ and radial coordinates $r$, $r\to \tilde{r}(r)$ and $t\to \tilde{t}=t+ H(t,r)$ to fix the gauge to be such that the radius of a round S$^2$ is $r$ and there is no cross term $dt dr$ (this is often called the radial or Schwarzschild gauge)\footnote{In this gauge, the horizon radius $r_+$, if present, is also a gauge invariant quantity since it is proportional to the entropy $r_+=\sqrt{S/4\pi}$.}. A field ansatz that accommodates the desired symmetries is 
\begin{align}\label{ansatz}
\mathrm{d}s^2=-f(r)\mathrm{d}t^2+g(r)\mathrm{d}r^2+r^2\mathrm{d}\Omega_{(2)}^2, \qquad A_{\mu}\mathrm{d}x^{\mu}=A(r)\mathrm{d}t, \qquad \phi=\phi^\dagger=\phi(r).
\end{align}
where $\mathrm{d}\Omega_{(2)}^2$ describes the line element of a unit radius $S^2$ (parametrized by $\{\theta,\psi\}$, say).

Once a static solution is found we can always use a $U(1)$ gauge transformation, $\varphi \to \varphi +q \,\chi\,,A_t \to A_t +\nabla_t \chi$  to rewrite a scalar field  $\phi=|\phi|e^{i\varphi}$ in a gauge where the scalar field oscillates with a frequency $\omega$, $\phi=|\phi|e^{-i\omega t}$. However, since the energy-momentum tensor of the scalar field only depends on $\phi \phi^\dagger$ and $\partial \phi (\partial \phi)^\dagger$, the gravitational and Maxwell fields are always invariant under the action of the Killing vector field $\partial_t$. In this gauge, the equations of motion require that $A{\bigl |}_{r_+}=\omega/q$, if the solution has a horizon at $r=r_+$. In particular, we shall adopt this gauge in sections \ref{sec:NH} and \ref{sec:normalMode}.

At this point we have fixed all the diffeomorphic and $U(1)$ gauge freedom to simplify our field ansatz.  However, our system still has two scaling symmetries that we can use for further simplifications \footnote{For planar solutions, i.e. solutions that asymptote to {\it local} AdS  there is a third scaling symmetry that allows to set the Poincar\`e horizon at $r=r_+\equiv 1$ (see e.g. \cite{Hartnoll:2008kx}). This is the reason why we can have small and large black holes in global AdS but not in planar AdS. Actually, in planar AdS a black hole is always `large' in the sense that it does not have the superradiant instability (see further discussions of Section \ref{sec:instability}).}.
The first scaling symmetry is  
\begin{align}\label{scaling1}
\{t,r,\theta,\psi\}\to \{\lambda_1 t,\lambda_1 r,\theta,\psi\},\quad \{ f,g,A,\phi \}\to \{ f,g,A,\phi \},\quad \{q,L,r_+\}\to \left\{\frac{q}{\lambda_1},\lambda_1 L,\lambda r_+ \right\},
\end{align}
 which leaves the equations of motion invariant and rescales the line element and the gauge field 1-form as $\mathrm{d}s^2\to \lambda_1^2 \mathrm{d}s^2$ and $A \mathrm{d}t \to \lambda_1 \,A \mathrm{d}t$. We use this scaling symmetry to work with dimensionless coordinates and measure our thermodynamic quantities in units of $L$, the natural scale of AdS (this amounts to set $L\equiv 1$),
\begin{align}\label{adimTR}
T=\frac t L\,,\qquad R=\frac r L\,;\qquad e=q L.
\end{align}
The second scaling symmetry is  
\begin{align}\label{scaling2}
\{t,r,\theta,\psi\}\to \{\lambda_2 t, r,\theta,\psi\},\quad \{ f,g,A,\phi \}\to \{\lambda_2^{-2} f,g,\lambda_2^{-1}A,\phi \},\quad  \{q,L,r_+\}\to \{q,L,r_+\}.
\end{align}
A Taylor expansion of the equations of motion at the asymptotic boundary yields $f\simeq c_0(R^2+1)+\cdots$ and $g^{-1}\simeq R^2+1+\cdots$. We use the scaling symmetry \eqref{scaling2} 
to require that $g$ approaches  $1/f$ as $r\to \infty$, i.e. to fix $c_0\equiv 1$. This choice can also be motivated by the following observation (see e.g. \cite{Hartnoll:2008kx}). In the context of the AdS$_4$/CFT$_3$ duality, AdS$_4$ black holes are dual to thermal states on the boundary CFT.  The choice $g=\frac{1}{f}$ as $r\to \infty$ fixes the normalization of the time coordinate with respect to the gravitational redshift normalization of the radial coordinate to be such that the Hawking temperature of the black hole in the bulk matches the temperature of the dual CFT on the boundary.
 
Variation of \eqref{action} yields the equations of motion for the fields $f,g,A$ and $\phi$:

\begin{subequations}\label{eom}
\begin{align}
&\phi''(R)+\frac{\phi'(R)}{R} \left(1-\frac{R^2 A'(R)^2}{2 f(R)}+g(R)(3 R^2+1)\right)+\frac{e^2 A(R)^2 g(R) \phi(R)}{f(R)}=0,\label{eomphi}\\
&A''(R)+\frac{A'(R)}{R} \left(2-\frac{e^2 R^2 A(R)^2 \phi(R)^2 g(R)}{f(R)}-R^2 \phi'(R)^2\right)-2 e^2 \phi(R)^2 g(R) A(R) =0,\label{eomA}\\
&f'(R)+\frac{f(R)}{R} \left(1-g(R)\left(1+3 R^2\right)-R^2 \phi'(R)^2\right)+\frac{R}{2} \left(A'(R)^2-2 e^2 A(R)^2 \phi(R)^2 g(R)\right)=0,\label{eomf}\\
&g'(R)-\frac{g(R)}{R} \left(1+\frac{R^2 A'(R)^2}{2 f(R)}+R^2 \phi'(R)^2\right)+\frac{g(R)^2}{R} \left(1+3 R^2-\frac{e^2 R^2 A(R)^2 \phi(R)^2}{f(R)}\right)=0.\label{eomg}
\end{align}
\end{subequations}

We can use \eqref{eomf} to get an algebraic relation for $g$ in terms of the other fields and their first order derivatives:
\begin{equation}\label{algebraig_g}
g(R)=-\frac{2 f(R) \left(R^2 \phi'(R)^2-1\right)-R \left(R A'(R)^2+2 f'(R)\right)}{2 \left(e^2 R^2 A(R)^2 \phi(R)^2+ f(R) \left(1+3 R^2\right)\right)}.
\end{equation}
We insert this algebraic relation into \eqref{eomphi}, \eqref{eomA} and \eqref{eomg} to obtain a coupled system of three second order ODE's. Onwards, these are the three equations of motion that we will solve  to find the three fields $\{f,A,\phi \}$ (and thus also $g$). 

The most notable solution of \eqref{action} is global AdS$_4$ spacetime: $f(R)=g^{-1}(R)=1+R^2$ $A(R)=\phi(R)=0$.  We are interested in solutions that asymptote to this background.

We can rewrite our ansatz and equations of motion in Fefferman-Graham coordinates (defined such that $g_{zz}=1/z^2$ and $g_{zb}=0$, where $z$ is the radial distance with boundary at $z=0$, and $x^b=\{t,\theta,\psi\}$ are the boundary coordinates)  \cite{FG:1985,Graham:1999jg,Anderson:2004yi}.
Any asymptotically AdS$_4$ spacetime has the following Taylor expansion of the metric around the holographic boundary $z=0$ \cite{deHaro:2000xn,Henningson:1998gx}:
\begin{eqnarray}\label{introFG}
&&ds^2 = \frac{1}{z^2}\left[\mathrm{d}z^2+g_{ab}(z,x) \mathrm{d}x^{a}\mathrm{d}x^{b}\right]\,, \nonumber\\
&& g_{ab}{\bigl |}_{z\to 0}=g_{ab}^{(0)}(x)+\cdots+z^3 g_{ab}^{(3)}(x)+\cdots\,, \qquad \hbox{with} \quad \langle T_{ab}(x)\rangle\equiv \frac{3}{16\pi G_4}\,g_{ab}^{(3)}(x)
\end{eqnarray}
and where $g^{(0)}(x)$ and $g^{(3)}(x)$ are the two integration ``constants" of the expansion\footnote{Our solutions are static and spherically symmetric so there is no $x$ dependence.}; the first dots include only even powers of $z$ (smaller than $3$) and depend only on $g^{(0)}$ (thus being the same for {\it any} solution that asymptotes to global AdS$_4$) while the second dots depend on the two independent terms $g^{(0)}, g^{(3)} $. Within the AdS/CFT duality we are (typically) interested in Dirichlet boundary conditions (BCs) that do not deform the conformal metric $g^{(0)}$: the dual holographic CFT is formulated on this fixed background. This is the static Einstein Universe $\mathbb{R}_t \times S^2$, $ds^2_\partial=-dT^2+d\Omega_{(2)}^2$. 
Given this Dirichlet BC our task is then to solve the equations of motion in the bulk, subject to regular BCs at the horizon or radial origin, to read $g^{(3)}$. Using the holographic renormalization formalism, this gives us the expectation value of the holographic stress tensor $\langle T_{ab}(x)\rangle$ which specifies and describes the boundary CFT \cite{Henningson:1998gx,Balasubramanian:1999re,deHaro:2000xn}. This Dirichlet BC also ensures that there is no dissipation of energy at the asymptotic boundary (see Appendix A of \cite{Cardoso:2013pza}).
For this reason these Dirichlet BCs can be denoted as `reflecting boundary conditions' \footnote{We can however have other reflecting BCs that preserve the charges but not the static Einstein Universe conformal boundary.}.

We still need to discuss the asymptotic BCs for the Maxwell and massless scalar fields. A Taylor expansion of the equations of motion at the conformal boundary yields,
\begin{eqnarray}
\label{asympAphi}
 && A_t(R) \simeq  \mu+ \frac{\rho}{R} +\cdots \nonumber \\
&& \phi(R)\simeq \frac{\sigma}{R^{\Delta_-}}
+\cdots+ \frac{\varepsilon}{R^{\Delta_+}} +\cdots\,, \quad
 \hbox{with} \quad \Delta_\pm=\frac{3}{2}\pm \frac{3}{2}\;,
\end{eqnarray}
where $\mu$ and $\rho$ are the chemical potential and charge density, respectively, and $\varepsilon$ and $\sigma$ are the two arbitrary integration constants describing the decay of the massless scalar field. In $d=4$ the unitarity bound for the scalar field mass is  $m_{\rm unit}^2L^2=-5/4$. We have chosen to work with a massless scalar field. Hence it is above the unitarity bound. It follows that only the mode $\varepsilon/r^{\Delta_+}$ with faster fall-off is normalisable (since it has finite canonical energy) and $\sigma/r^{\Delta_-}$ is a non-normalisable mode \cite{Breitenlohner:1982jf,Mezincescu:1984ev}. We thus choose the BC $\sigma=0$.  According to the AdS/CFT dictionary, the boundary CFT is then not sourced and the scalar field amplitude $\varepsilon$ is  proportional to the expectation value $\langle {\cal O}\rangle$ of the boundary operator $\mathcal{O}$ that has dimension $\Delta_+=3$ \cite{Klebanov:1999tb}.
 
To summarize, at the asymptotic boundary ($R\to\infty$) we will impose Dirichlet BCs such that our solutions at large $R$ behave as
\begin{eqnarray}\label{BCboundary}
&& f(R)=1+R^2+\frac{m_0}{R} +\mathcal{O}(1/R^2), \nonumber\\
&& A(R)=\mu+ \frac{\rho}{R}+ \mathcal{O}(1/R), \nonumber\\
&& \phi(R)=\frac {\varepsilon} {R^3}+\mathcal{O}(1/R^5)
\end{eqnarray}
with the dots representing terms that are a function of ${m_0,\mu,\rho,\varepsilon}$.

Consider now the inner boundary of our solution. This is a {\it fictitious} boundary since it is a coordinate singularity at the edge of our integration domain, but is at a finite proper distance from other points \cite{Dias:2015nua}. In the present study, this can be the origin of the radial coordinate, $R=0$, if we look into a soliton (boson star). Or, it can be  a horizon at $R=R_+$ if the solution is a black hole, a condition that is imposed by demanding that $f(R_+)=0$. In both cases to find the physical BCs we can Euclideanise the metric with a Wick rotation and require regularity of the metric and matter fields in Cartesian coordinates, as explained in detail in the review \cite{Dias:2015nua} \footnote{For the gauge field, this refers to the gauge invariant field strength tensor $F=\mathrm{d}A$, rather than the gauge potential $A$}. 

For a soliton, a Taylor expansion of the equations of motion at the origin yields,
\begin{eqnarray}\label{TaylorOrigin}
f(R)&=& f_0+ \mathcal{O}\left(R^2 \right), \nonumber \\
A(R)&=& a_0+ \mathcal{O}\left(R^2 \right), \nonumber \\
\phi(R)&=& \phi_0+ \mathcal{O}\left(R^2 \right),
\end{eqnarray}
where $\{ f_0,a_0,\phi_0 \}$ are arbitrary integration constants and all higher order terms in \eqref{TaylorOrigin} are even powers of $R$ with coefficients fixed in terms of $\{ f_0,a_0,\phi_0 \}$. 
Here we have already imposed smoothness of the fields at the origin by imposing a {\it defining} BC (in the nomenclature of \cite{Dias:2015nua}). Namely, we set to zero three integration constants (one for each field $f$, $g$, and $\phi$) which are associated to terms that diverge as $R\to 0$. Following the nomenclature of  \cite{Dias:2015nua}, if we wish, we can now use this defining BC together with the equations of motion to impose a {\it derived} BC to find the soliton. For example, it follows from \eqref{TaylorOrigin} that a good derived BC is a Neumann BC for the three fields at $R=0$ since the derivative of all of them vanishes.

On the other hand, if the solution is a black hole a Taylor expansion around its horizon $-$ defined as the locus  $f(R_+)=0$ $-$ yields
\begin{eqnarray}\label{TaylorH}
f(R)&=& f_0(R-R_+)+ \mathcal{O}\left((R-R_+)^2 \right), \nonumber \\
A(R)&=& a_0(R-R_+)+ \mathcal{O}\left((R-R_+)^2 \right), \nonumber \\
\phi(R)&=& \phi_0+ \mathcal{O}\left((R-R_+)^2 \right),
\end{eqnarray}
after imposing defining BCs that set three integration constants to zero (one for each function) to have smoothness as $R\to R_+$. We are left with the remaining three arbitrary integration constants $\{ f_0,a_0,\phi_0 \}$, and all higher order terms in \eqref{TaylorH} are fixed as a function of $\{ f_0,a_0,\phi_0 \}$.  

At this stage we can ask how many parameters we need to describe the soliton and hairy black hole that we want to construct. The theory \eqref{action} is fixed once we choose the mass and charge of the scalar field. At the UV asymptotic boundary we have a total of five free parameters $\{c_0,m_0,\mu,\rho,\varepsilon\}$ to which we need to add the cosmological radius $L$ and the horizon radius $R_+$ (if present). As explained previously, we can use the two scaling symmetries \eqref{scaling1} and  \eqref{scaling2} to fix $L$ and $c_0$. We are therefore left with four UV free parameters $\{m_0,\mu,\rho,\varepsilon\}$ and the horizon radius $R_+$. On the other hand, in the IR we have a total of three free parameters $\{ f_0,a_0,\phi_0 \}$ (both in the soliton and black hole case).
  Therefore, any black hole of the theory is described by $4_{UV}+1_H -3_{IR}=2$ parameters. The AdS$_4$ $-$Reissner-N\"ordstrom black hole is indeed a 2-parameter family of black holes. More importantly, we also find that the hairy black holes we want to construct are described by two parameters. These are the mass and the electric charge of the black hole. Alternatively (but equivalently) we can take these to be the horizon radius $R_+$ and the amplitude of the scalar condensate $\varepsilon$. On the other hand, any soliton of the theory is described by $4_{UV}-3_{IR}=1$ parameter. This can be the mass of the soliton (which fixes its charge) or, alternatively, the amplitude of the scalar condensate $\varepsilon$.

\subsection{Conserved and thermodynamic quantities}\label{sec:conservedThermo} 

Once we have found our fields $\{f,g,A,\phi \}$ we can read the gauge invariant thermodynamic quantities that describe hairy black holes or solitons. 

To find the energy we can use either holographic renormalization  \cite{Henningson:1998gx,Balasubramanian:1999re,deHaro:2000xn} or the Ashtekar-Das formalism \cite{Ashtekar:1999jx}, \cite{Das:2000cu}. We describe briefly the latter formalism.
Energy is a conserved asymptotic quantity associated to the Killing vector field $\xi^a \partial_a =\partial_t$. Consider a conformal transformation of the metric \eqref{ansatz} defined as : $\hat g_{ab}=\Omega g_{ab}$. The appropriate conformal factor to use here is $\Omega=\frac{1}{R}$. Then consider the Weyl tensor of the conformal metric: $\hat C_{abcd}$. Consider also the conformal hypersurface defined by $\Sigma|_{\Omega=0}$ and the normal vector to it: $n_{a}=\partial_{a}\Omega$. We define the tensors $\hat K_{abcd}=\Omega^{3-d} \hat C_{abcd}|_{\Omega=0}$ and $\hat\varepsilon_{ac}=\hat K_{abcd} n^b n^d$. Consider, on the conformal hypersurface $\Sigma|_{\Omega=0}$, the timelike surfaces of constant $t$, $\tilde\Sigma|_{\Omega=0, t=const}$ with normal vector $t_{a}=\partial_{a}t$. This surface has dimension 2 and induced metric $h_{ab}$. Then the conserved energy is defined as:
\begin{equation}
M=-\frac{1}{8 \pi}\int_{\tilde\Sigma}\hat\varepsilon_{ac} \;\xi^a t^{b} \mathrm d \tilde\Sigma
\end{equation}
which, in terms of the expansion parameters defined in \eqref{BCboundary} yields,
\begin{equation}
\frac{M}{L}=-\frac{m_0}{2}.
\end{equation}

To define the conserved electric charge, we use Gauss' law. Consider a 2-sphere $S^2$ at constant time $t$ and $R$ in the limit $R\to \infty$. The electrical charge is
\begin{equation}
Q=\lim_{R\to \infty}\frac{1}{16 \pi}\int_{S^2} \star F,
\end{equation} 
where $\star F$ is the Hodge dual of the field strength. As a function of the asymptotic expansion parameters defined in \eqref{BCboundary} this gives,
\begin{equation}
\frac{Q}{L}=-\frac{\rho}{2}.
\end{equation}

The value of the temperature for a soliton is undefined, and its entropy is zero as it does not have a horizon. Our black holes have a Killing horizon at $R=R_+$ generated by the Killing vector field $K=\partial_t$. Their temperature is thus given by its surface gravity over $2\pi$ which yields,
\begin{equation}
T_H L=\frac 1 {4 \pi}\frac{f'(R)}{\sqrt{f(R) g(R)}}\Big|_{R_+}.
\end{equation}
The entropy of the black hole is a quarter of horizon area. In terms of the expansion parameters introduced in \eqref{TaylorH}, the temperature and entropy are then
\begin{align}
T_H \,L&  =\frac{f_0 \left(3 R_+^2+1\right)}{2 \sqrt{2} \pi  \sqrt{R_+\left(3 R_+^2+1\right) \left(a_0^2 R_+ +2 f_0 \right)}},\nonumber \\
\frac{S}{L^2}&=\pi R_+^2.
\end{align}
The chemical potential $\mu$ is defined in \eqref{BCboundary}. It is the difference between the value of the electromagnetic field at infinity and at the horizon\footnote{Recall that we work in the gauge where $A{\bigl |}_{r_+}=0$. With a gauge transformation, we can instead work in the gauge  $A{\bigl |}_{\infty}=0$ and the scalar field acquires a harmonic time dependence.}. 

To discuss the  competition between thermal phases in the canonical and in the grand-canonical ensembles we have to compute the associated thermodynamic potentials, namely the Helmholtz free energy $F=E-TS$ and the Gibbs free energy, $G=M-TS-\mu Q$, respectively.

Solutions of \eqref{eom} must satisfy the first law of thermodynamics:
\begin{align}
&\mathrm d M=T_H \mathrm d S+\mu\mathrm d Q,\quad \text{for black holes},\label{firstlawBH}\\
&\mathrm d M=\mu\mathrm d Q,\qquad \qquad \text{\ \ for solitons}.\label{firstlawsoliton}
\end{align}

\section{AdS$_4$$-$Reissner-N\"ordstrom black hole and its instabilities}\label{sec:instability}

\subsection{AdS$_4$$-$Reissner-N\"ordstrom black holes}\label{section:RN}
When the scalar field vanishes, global AdS$_4$ $-$Reissner-N\"ordstrom (RN-AdS$_4$) black hole is a known solution of \eqref{eom}. This is a 2-parameter family of solutions, that we can take to be the horizon radius $R_+$ and the chemical potential $\mu$. It is described by ansatz \eqref{ansatz} with 
\begin{eqnarray}\label{RNsolution}
&& f(R)=\frac{1}{g(R)}=1 + R^2 - \frac{R_+}{R} \left[1 + R_+^2 + \frac{\mu^2}{2} \left(1 -  \frac{R_+}{R}\right)\right], \nonumber \\
&& A(R)=\mu\left[1-\frac{R_+}{R}\right],\nonumber \\
&& \phi(R)=0.
\end{eqnarray}
Alternatively, we can parametrize RN-AdS$_4$ with $R_+$ and the temperature $T$ in which case the chemical potential is given by
\begin{equation}\label{muRpT}
\mu=\sqrt{2}\sqrt{1+3 R_+^2-4\pi R_+ T}.
\end{equation}
It follows that regular RN-AdS$_4$ black holes exist if their chemical potential  satisfies \begin{equation}\label{conditionhorizon}
\mu\leq \sqrt{2}\sqrt{1+3 R_+^2}.
\end{equation}
When $\mu=0$ one recovers the  AdS$_4$-Schwarzschild solution while the upper bound of \eqref{conditionhorizon} occurs for the extremal configuration ($T=0$) where the entropy is finite but the temperature vanishes. 

The mass, charge, entropy and temperature expressed as a function of $R_+$ and $\mu$  are
\begin{eqnarray}\label{RNthermo}
&&M_{RN}/ L=\frac{R_+}{2}\left(1+\frac{\mu^2}{2}+R_+^2\right),
\qquad  Q_{RN}/L=\frac{1}{2}\,\mu \,R_+\,,\nonumber \\
&& S_{RN}/L^2=\pi R_+^2,
\qquad T_{RN} L=\frac{1}{8 \pi R_+}\left(2-\mu^2+6R_+^2\right).
\end{eqnarray}

In Section \ref{sec:noninteracting}, we will find the leading order thermodynamics of hairy black holes assuming that they can be described as a non-interacting mixture of a soliton and a small RN-AdS$_4$ black hole. To complete that analysis, we will only consider the leading  order (in the $R_+$ expansion) thermodynamics of the RN-AdS$_4$ black hole. For future reference this is:
\begin{eqnarray}\label{RNthermoLeading}
&&M_{RN}/ L \simeq\frac{R_+}{2}\left(1+\frac{\mu^2}{2}\right)+\mathcal O(R_+^3),
\qquad  Q_{RN}/L=\frac{1}{2}\,\mu \,R_+\,,\nonumber \\
&& S_{RN}/L^2=\pi R_+^2,
\qquad T_{RN} L\simeq \frac{1}{8 \pi R_+}\left(2-\mu^2\right)+\mathcal O(R_+).
\end{eqnarray}


When perturbed by an infinitesimal scalar field perturbation, global AdS$_4$-RN black holes can develop two types of linear instabilities that have different physical nature. One is the near-horizon scalar condensation instability and the other is the superradiant instability. The former is already present in planar RN-AdS$_4$ black holes (where it was first found) while the later is only present in global AdS RN black holes. The zero mode of these instabilities is closely connected to the existence of hairy black hole solutions. Therefore, in the next two subsections we discuss the main properties of these two instabilities with some detail.  

\subsection{Near horizon scalar condensation instability of large black holes \label{sec:NH}}

Consider a scalar field with mass $m$ and charge $q$ in an asymptotically AdS$_4$ background. A Taylor expansion around the asymptotic boundary yields the two independent solutions\footnote{In this subsection it is illuminating to restore the factors of $L$ and coordinates $\{t,r\}$ $-$ see \eqref{adimTR} $-$ and consider a massive scalar filed.}
\begin{equation}
\label{BFdecays}
 \phi(R)\simeq \frac{\sigma}{R^{\Delta_-}}
+\cdots+ \frac{\varepsilon}{R^{\Delta_+}} +\cdots\,, \quad
 \hbox{with} \quad \Delta_\pm=\frac{3}{2}\pm \sqrt{\frac{9}{4}+m^2L^2}\;,
\end{equation}
which reduces to \eqref{asympAphi} when $m=0$. Breitenl\"ohner and Freedman (BF) found that a scalar field in AdS$_4$ is normalizable, i.e. it has finite energy, as long has its mass obeys the AdS$_4$ BF bound, 
 \begin{equation}
\label{BFbound}
m^2 \ge m^2_{\mathrm{BF}}\equiv -\frac{9}{4}\,\frac{1}{L^2}.
 \end{equation}
When this is the case we can have asymptotically AdS$_4$ solutions that are  stable {\it in the UV} region. 

But this analysis is blind to the behaviour of the scalar field in the interior of the spacetime. In particular, the properties of the scalar field in the IR region of a particular spacetime might drive the system unstable. This is indeed the case: as first found in the context of holographic superconductors \cite{Gubser:2008px,Hartnoll:2008vx,Hartnoll:2008kx}, planar AdS black holes can be unstable to a near horizon scalar condensation mechanism \cite{Gubser:2008px}. The zero mode of this instability then signals a second order phase transition to planar hairy black holes that are dual to holographic superconductors \cite{Hartnoll:2008vx,Hartnoll:2008kx}. It was later understood that the near horizon scalar condensation instability is quite universal and present for any asymptotically AdS background that has a black hole with an extremal (zero temperature) configuration \cite{Hartnoll:2008kx,Dias:2010ma} (this includes even some systems with neutral black holes and uncharged scalar fields). The criterion for this instability is the following. Start with a scalar field whose mass obeys the UV BF bound \eqref{BFbound}.  When it violates the AdS$_2$ BF bound,  
 \begin{equation}
\label{BFboundAdS2}
m^2 \ge m^2_{\mathrm{AdS_2\, BF}}\equiv -\frac{1}{4}\,\frac{1}{L_{AdS_2}^2}\,,
 \end{equation}
the system, although asymptotically stable, is unstable in its near-horizon region. We might then expect the system to evolve into a hairy black hole with the same UV asymptotics but different IR behaviour.

We can work out in more detail the consequences of this instability for our current system, following the original analysis in \cite{Dias:2010ma} and \cite{Dias:2011tj}. Global AdS$_4$-RN black holes described by \eqref{ansatz} and \eqref{RNsolution} have an extremal configuration when the bound \eqref{conditionhorizon} is saturated, $\mu= \sqrt{2}\sqrt{1+3\frac{R_+^2}{L^2}}$.
Take the near-horizon limit of this extremal RN-AdS$_4$ black hole, i.e. introduce the new coordinates $\{\tau,\tilde{\rho}\}$,
\begin{equation}\label{NHlim}
 t=L_{AdS_2}^2 \,\frac{\tau}{\lambda},\qquad r=r_+ +\lambda \tilde{\rho},
\end{equation}
and let $\lambda\to 0$. Setting,
\begin{equation}\label{NHlim2}
L_{AdS_2}\equiv \frac{L \, r_+}{\sqrt{L^2+6r_+^2}}
\end{equation}
this procedure yields the solution of \eqref{eom}, 
\begin{eqnarray}\label{NHansatz}
&& \mathrm{d}s^2=L_{AdS_2}^2\left( -\tilde{\rho}^2 \mathrm{d}\tau^2+\frac{\mathrm{d}\tilde{\rho}^2}{\tilde{\rho}^2} \right)+r_+^2\mathrm{d}\Omega_{(2)}^2, \nonumber \\
&& A_{\mu}\mathrm{d}x^{\mu}=\alpha\, \tilde{\rho} \,\mathrm{d}t 
 \qquad \hbox{with}\:\: \alpha \equiv\frac{L_{AdS_2}}{r_+}\sqrt{r_+^2+L_{AdS_2}^2}.
\end{eqnarray}
So, the near-horizon limit of extremal RN-AdS$_4$ is the direct product of AdS$_2$ with a sphere $S^2$. Applying this near-horizon limit to the Klein-Gordon equation that describes a perturbation $\delta\phi(t,r)=e^{-i\omega t}R(r)$ of a scalar field (with mass $m$ and charge $q$)  about the extremal RN-AdS$_4$ black hole yields 
\begin{equation}\label{NHlimKG}
\partial_{\tilde{\rho}}\left(\tilde{\rho}^2\partial_{\tilde{\rho}} R\right)+\left(\frac{\left(\omega+q \,\alpha \,\tilde{\rho} \right)^2}{\tilde{\rho}^2}-m^2 L_{AdS_2}^2\right)R=0,
\end{equation}
which, not surprisingly, is the Klein-Gordon equation for a scalar field around AdS$_2$ with an electromagnetic potential $A_\tau=\alpha\, \tilde{\rho}$. A Taylor expansion of \eqref{NHlimKG} yields
\begin{equation}
\label{BFdecaysAdS2}
R{\bigl |}_{\tilde{\rho}\to\infty}\simeq a \,\tilde{\rho}^{\,-\widetilde{\Delta}_-}
+\cdots+ +b \,\tilde{\rho}^{\,-\widetilde{\Delta}_+} +\cdots\,, \quad
 \hbox{with} \quad \widetilde{\Delta}_\pm=\frac{1}{2}\pm \frac{1}{2}\sqrt{1+m_{eff}^2L_{AdS_2}^2}\;,
\end{equation}
which dictates the effective mass of the scalar field from the perspective of a near-horizon observer,
\begin{eqnarray}
\label{nearmass}
m_{eff}^2 L_{AdS_2}^2&\equiv& m^2 L_{AdS_2}^2 -q^2 \,\alpha^2 \nonumber\\
 &=&  m^2 L_{AdS_2}^2-2 e^2 R_+^2 \frac{1+3 R_+^2}{\left(1+6 R_+^2\right)^2},
\end{eqnarray}
where in the second line we re-introduced the dimensionless quantities $R_+=r_+/L$ and $e=q L$.

A scalar field with mass \eqref{nearmass} in AdS$_2$ is unstable if it violates the  AdS$_2$ BF bound \eqref{BFboundAdS2}. It follows that extremal RN-AdS$_4$ black holes are unstable whenever the charge of the scalar field obeys
\begin{equation}\label{NHunstableCriterion}
e^2\geq \left( 1+4 m^2 L^2 \frac{R_+^2}{1+6R_+^2} \right) \frac{\left(1+6 R_+^2\right)^2}{8 R_+^2 \left(1+3 R_+^2\right)}\,.
\end{equation}
Taking the large $R_+$ expansion, this yields
\begin{equation}\label{NHunstableCriterion2}
e^2\geq \left(\frac{3}{2}+m^2L^2 \right)- \frac{m^2L^2}{6}\frac{1}{R_+^2}+\left(1+\frac{4}{3}\,m^2L^2\right) \frac{1}{24\,R_+^4} +\mathcal{O}(1/R_+^6).
\end{equation}
In the strict limit $R_+\to\infty $, we recover the known result for planar RN-AdS$_4$ black holes, $e^2\geq \left(\frac{3}{2}+m^2L^2 \right)$ \cite{Gubser:2008px,Hartnoll:2008vx,Hartnoll:2008kx}. Global RN-AdS$_4$ black holes are unstable for larger values of the scalar field charge.

In the opposite $R_+\to 0$ limit, an expansion of \eqref{NHunstableCriterion} yields
\begin{equation}\label{NHunstableCriterion3}
e^2\geq \frac{1}{8R_+^2}+\mathcal{O}(1),
\end{equation}
which indicates that the near-horizon scalar condensation instability is suppressed for small (global) RN-AdS$_4$ black holes. This is to be contrasted with the superradiant instability (discussed in the next subsection) which is present for small black holes but not large black holes. This property highlights the different nature of the two instabilities.

The near-horizon instability criterion that we have been discussing applies strictly to extremal black holes. Continuity indicates that this instability should extend to near-extremal black holes. Thus large, near extremal RN-AdS$_4$ black holes are unstable to condensation of the scalar field when condition \eqref{NHunstableCriterion} is satisfied. 

We have chosen to work with a massless ($m=0$) scalar field, so the near horizon instability is present in our system for large black holes with $e>\sqrt{\frac {3}{2}}$.

\subsection{Normal modes of AdS$_4$\label{sec:normalMode}}

Consider a massive charged scalar field perturbation $\phi(t,r)$ in global AdS$_4$ with a constant gauge field $A=\mu dt$, which is governed by the Klein-Gordon equation. The background is time independent so we can do a Fourier decomposition of the perturbation in time such that a particular mode is described by $\phi(T,R)=e^{-i \omega T}\psi(R)$ ($\omega$ is the frequency of the mode).
If the scalar field mass $m$ is above the BF bound \eqref{BFbound}, we can impose the boundary condition $\sigma=0$ in \eqref{BFdecays} and this yields the normalizable solution
\begin{equation}\label{normaleigenvector}
\psi_n(R)= \frac{\varepsilon}{(R^2+1)^{\frac{1}{2}\Delta_+}} \, _2F_1\left[\frac{1}{2} \left(\Delta_+ - e\mu -\omega_n \right),\frac{1}{2} \left(\Delta_+ + e\mu +\omega_n\right),\Delta_+-\frac{1}{2};\frac{1}{1+R^2}\right],
\end{equation}
where $\Delta_+$ is defined in  \eqref{BFdecays}, $\varepsilon$ is an arbitrary amplitude and $_2F_1$ is the hypergeometric function. To have a regular solution at the origin, the frequency spectrum must then be quantized as\footnote{This follows from the Gamma function property $\Gamma[-n]=\infty$ for $n=0,1,2,\cdots$} 
\begin{equation}\label{normalw}
\omega_n=\Delta_+ +2 n -e\mu,
\end{equation}
where the non-negative integer $n$ is the radial overtone that gives the number of radial zeros of $\psi(R)$. Equations \eqref{normaleigenvector} and \eqref{normalw} give, respectively, the eigenvectors and eigenvalues of the normal modes of a scalar field. 
Note however that we can use the $U(1)$ gauge transformation of the system to choose the gauge potential to be such that, e.g.  $\mu= \left(\Delta_+ +2 n \right)/e$ and the normal mode frequency (i.e. for a given $n$) vanishes, $\omega_n=0$.

In our current study, we are interested in the case $m=0$, which implies that $\Delta_+=3$, and onwards we choose to work with the solution that has the lowest frequency (and thus energy), namely we set $n=0$. Then, \eqref{normalw} reads
\begin{equation}\label{normalw0}
\omega_0=3 -e\mu\,.
\end{equation}

\subsection{Superradiant instability of small black holes \label{sec:superradiance}}

In subsection \ref{sec:NH} we have seen that large RN-AdS$_4$ black holes are unstable to the near-horizon scalar condensation instability. In the present subsection we show that, in the opposite limit, small RN-AdS$_4$ black holes are unstable to the superradiant instability. For that we solve the Klein-Gordon equation for a massless scalar field perturbation with charge $e\geq\frac{3}{\sqrt{2}}$ (see discussion associated with \eqref{eSuperradianceBound}),
\begin{equation}\label{A_Fourier}
\phi(T,R)=e^{-i \omega T}\phi(R)
\end{equation}
around the RN-AdS$_4$ in a perturbative expansion in the horizon radius $R_+$. We will find that the imaginary part of the perturbation frequency $\omega$ is positive, and thus the system is unstable, when $\mu\geq \frac 3 e$. 

To impose smoothness of the perturbations at the horizon boundary it is convenient to work in ingoing Eddington-Finkelstein coordinate
\begin{equation}\label{A_Edd-Fink}
v=T+\int \frac{1}{f(R)}\mathrm d R.
\end{equation}

The Klein-Gordon equation cannot be solved  analytically at each order. To circumvent this limitation, we consider small black holes, $R_+ \ll 1$, and use the matched asymptotic expansion method to solve the resulting Klein-Gordon equation at each order in $R_+$. Namely, we consider two different regions: the far region, $R\gg R_+$, and the near region, $R\ll 1$ which have an overlapping region $R_+\ll R\ll 1$ since we are assuming  $R_+ \ll 1$. In the near and far regions, certain contributions in the equation of motion are sub-dominant and can be discarded. We are left with ODEs, with a source term that depends on the previous order solutions and their derivatives, that can now be solved analytically. At each order, the near and far region solutions are then  matched in the overlapping region.

Both in the far (superscript $^{out}$) and in the near (superscript $^{in}$) regions, we write the perturbative expansion in $R_+$ of $\phi(R)$ and $\omega$ as:
\begin{align} \label{superradiantExpansion}
& \phi^{out}(R)=\sum_{k\geq 0} R_+^k\,\phi_k^{out}(R), \qquad \phi^{in}(R)=\sum_{k\geq 0} R_+^k\,\phi_k^{in}(R), \nonumber \\
& \omega=\sum_{k\geq 0} R_+^k\,\omega_k \quad \text{with }\ \omega_0=3-\mu e,\end{align}
and solve the Klein-Gordon equation order by order. The choice of $\omega_0$ is justified because the perturbation mode of RN-AdS$_4$ with lowest frequency reduces to the lowest normal mode of AdS \eqref{normalw0} in the limit $R_+\to 0$. At any order $k\geq 1$ we have five parameters to determine: two integration constants from $\phi_k^{out}(R)$, say $\alpha_k$ and $\tilde{\alpha}_k$, two integration constants from $\phi_k^{in}(R)$, say $\beta_k$ and $\tilde{\beta}_k$, and the frequency coefficient $\omega_k$. For our purposes, it is enough to present the results up to order $k=2$: this is the order at which the frequency acquires an imaginary contribution.

\noindent {\tt Far region analysis:}

In the far region, $R\gg R_+$, the Klein-Gordon equation for  $\phi^{out}(R)$ around the RN-AdS$_4$ black hole \eqref{RNsolution} effectively describes linearized perturbations around global AdS. Indeed at each order in the $R_+$ expansion it reads,
\begin{equation}\label{KGfar}
\bar{D}^{\mu}\bar{D}_\mu \phi^{out}_k = S^{out}_k\left(\phi^{out}_{j<k}, \partial \phi^{out}_{j<k}\right),
\end{equation}
where $\bar{D}_{\mu}=\bar{\nabla}_mu -i e A_{\mu}$ is the gauge covariant derivative of  the global AdS background $\bar{g}$ with a gauge potential $A=\mu dt$, and the RHS is a source term that depends on the lower order fields $\phi^{out}_{j<k}$ and their derivatives. 
As justified when discussing \eqref{asympAphi}, at any expansion order, in the far region we impose the Dirichelet boundary condition:
\begin{equation}\label{A_bc}
\phi^{out}{\bigl |}_{R\to\infty}=\frac{\varepsilon}{R^3}+\mathcal{O}(1/R^5).
\end{equation}

At leading order, $k=0$ the far region solution is
\begin{equation} \label{pert:out0}
\phi^{out}_0(R)=\frac{3 \alpha_0 R+\tilde{\alpha}_0 \left(R^4+6 R^2-3\right)}{3 R \left(R^2+1\right)^{3/2}}\,e^{-i (\mu e-3) \arctan R}.
\end{equation}
The boundary condition \eqref{A_bc} requires we set $\tilde{\alpha}_0=0$. Moreover, without loss of generality we can set $\alpha_0\equiv 1$, and \eqref{A_bc} and \eqref{pert:out0} then fix the amplitude of the scalar field to be $\varepsilon=-i e^{-\frac{1}{2} i \pi  e \mu}$.  

For higher orders, $k\geq 1$, the boundary condition \eqref{A_bc} always fixes the two integration constants $\alpha_k$ and $\tilde{\alpha}_k$ but leaves the frequency coefficient $\omega_k$ undetermined. It is fixed by matching the far region with the near region. The final far region solutions $\phi^{out}_k$ for $k=0,1,2$, after imposing the boundary condition \eqref{A_bc} and fixing the frequency coefficients, are listed in \eqref{superradiantExp:far} of Appendix \ref{App:Superradiance}. 

\noindent {\tt Near region analysis:}

In the near region $R\ll 1$, it is convenient to work with the rescaled variable $y=\frac {R}{R_+}$. The reason for this rescaling will become fully clear in the discussion associated to \eqref{nearcoordinates} and \eqref{RNnear} that we avoid repeating here. In short, in the near region the perturbation problem simplifies since, for $R_+\ll 1$, the  electric field is {\it weak} and the perturbation system reduces to a small perturbation around the neutral solution.
After this rescaling $y=\frac {R}{R_+}$, at each order, the Klein-Gordon equation for  $\phi^{in}(y)$ around the AdS$_4$ black hole $-$ see \eqref{RNnear} $-$ reads
\begin{equation}\label{KGnear}
\bar{D}^{\mu}\bar{D}_\mu \phi^{in}_k = S^{in}_k\left(\phi^{in}_{j<k}, \partial \phi^{in}_{j<k}\right),
\end{equation}
where $\bar D _{\mu}$ is the leading order part of the gauge covariant derivative of the rescaled coordinate in the RN-AdS$_4$ black hole\footnote{In \eqref{KGnear}, the gauge covariant derivatives $\bar D _{\mu}$ are covariant derivatives $\bar{\nabla}_{\mu}$, since the electric field makes no contribution at this leading order.},  and the RHS is a source term that depends on the lower order fields $\phi^{in}_{j<k}$ and their derivatives. 
At each order, we solve \eqref{KGnear} subject to the boundary condition that the solution is regular at the horizon in Eddington-Finkelstein coordinates. 

At leading order, $k=0$, the near region solution is
\begin{equation}
\phi^{in}_0(y)=\beta_0-\frac{\tilde{\beta}_0}{\mu^2-2}\, \left[\log (y-1)-\log \left(2 y-\mu^2\right)\right],
\end{equation}
and smoothness requires that we set $\tilde{\beta}_0=0$.

At any order $k\geq 1$, the boundary condition at the horizon always fixes one of the integration constants, say $\tilde{\beta}_k$, but the other is left undetermined until we do the matching of the near and far regions. 
The final near region solutions $\phi^{in}_k$ for $k=0,1,2$, after imposing the boundary condition and fixing the integration constants, are listed in \eqref{superradiantExp:near} of Appendix \ref{App:Superradiance}. 
 
\noindent {\tt Matching the near and far region solutions:}

At this stage, at each order $k\geq 1$ we have two undetermined parameters, namely the near region integration constant $\beta_k$ and the frequency coefficient $\omega_k$. They are fixed by requiring that the large radius expansion of the near solution $\phi^{in}$ matches the small radius expansion of the far solution $\phi^{out}$. In these expansions we keep only terms that will not receive contributions from higher orders. 

Let us illustrate this matching at the leading order, $k=0$.
The small radius limit of $\phi^{out}_0$ and the large radius limit of $\phi^{in}_0$ yield
\begin{align}
&\phi^{out}_0(R)\simeq 1+\mathcal O(R),\nonumber \\
&\phi^{in}_0(R)=\beta_0.
\end{align}
Matching requires that we set  $\beta_0=1$.

As another example consider the matching at order $k=1$. The small radius expansion of $\phi^{out}_1$ and the large radius expansion of $\phi^{in}_1$ are:
\begin{align}
\phi^{out}_1(R)&\simeq 1+i R (3-e \mu ) + R_+{\biggl [} \frac{1}{R} \frac{16 e \mu -3 \left(4 \mu ^2+\pi  \omega_1 +8\right)}{16} \nonumber \\
& +\frac{1}{48}{\biggl (}  -24 i (e \mu -3) \left(2 e \mu -\mu ^2-2\right)+3 \pi  {\bigl [}\mu  (3 i e \omega_1 -8 e+12 \mu )-17 i \omega_1 +24 {\bigr ]} \nonumber \\
&-24 i \left(\mu ^2+2\right) (e \mu -3) \log (R)-16 \omega_1 {\biggr )} {\biggr ]}  +\mathcal O(R_+^2,R R_+), \\
\phi^{in}_1(R)&\simeq1+i R (3-e \mu ) + R_+ \left[\beta_1 +\frac{1}{2} i (3- e\mu) \left(\left(\mu^2+2\right) \log \left(\frac{2 R}{R_+}\right)-2\right)\right] +\mathcal O(R_+^2).\nonumber
\end{align}
At order $\mathcal{O}(R_+^0)$ the two solutions agree as a result of the previous order matching. Moving to next-to-leading order, the divergent $R_+/R$ term in the far region has no counterpart in the near region expansion so we fix $\omega_1$ to eliminate it. Matching the remaining terms, namely the contributions proportional to $R_+ R^0$ and $R_+ \log R$ then fixes uniquely the integration constant $\beta_1$.  
A similar matching procedure can be done for higher orders. 

By the end of the day, we find that the frequency coefficients in \eqref{superradiantExpansion} up to order $k=2$ are:
\begin{align}\label{superradiantFreq}
\omega_0=&3-\mu e\,, \nonumber \\
\omega_1=& -\frac{4 \left(3 \mu^2-4 \mu e+6\right)}{3 \pi }, \nonumber \\
\omega_2=&\, i\, (\mu e-3)\frac{16}{3 \pi }+\frac{1}{96} {\biggl [}\mu {\biggl (}-264 \left(\mu^2+2\right) {e}+9 \mu \left(3 \mu^2+52\right)+224 \mu {e}^2{\biggl )}+108 {\biggl ]} \nonumber\\
&-\frac{4}{27 \pi ^2} \left(45 \mu^2-52 \mu {e}+90\right)\left(3 \mu^2-4 \mu {e}+6\right).
\end{align}

The property in \eqref{superradiantFreq} to be highlighted is that at order $k=2$ the frequency acquires an imaginary part proportional to $\mu e-3$. This is negative whenever $\mu\leq\frac{3}{e}$, which signals exponential damping, but it becomes positive for $\mu>\frac{3}{e}$. The latter case describes the superradiant instability of RN-AdS$_4$ black holes. The reader interested on the properties of  RN-AdS$_4$ superradiant modes  beyond the perturbative analysis done here can find them in \cite{Uchikata:2011zz,Li:2016kws}.

The above analysis shows that  global RN-AdS$_4$ black holes are unstable to superradiance in the small horizon radius regime, $R_+\ll 1$. On the other hand, large radius global RN-AdS$_4$ black holes are expected to be stable to superradiance. Indeed, in the most extreme case of a large radius black hole whereby $R_+\to\infty$, the studies of \cite{Gubser:2008px,Hartnoll:2008vx} indicate that planar RN-AdS$_4$ black holes are unstable only to the near-horizon instability described in Section \ref{sec:NH}. Starting from this planar limit, it seems natural to expect  that, as $R_+$ decreases, the superradiant instability will kick in at (and below) a critical horizon radius (in addition, the near-horizon instability becomes supressed in the limit $R_+\to 0$ as seen in \eqref{NHunstableCriterion3}). It would be interesting to do a numerical linear perturbation analysis that spans the full parameter space of RN-AdS$_4$ black holes to find the instability properties of RN-AdS$_4$. However, it should be noticed that such an analysis will not be able to disentangle the nature of the two instabilities except in the two limiting cases $R_+\to 0$ and  $R_+\to \infty$ that are covered by the analytical analysis done in this manuscript. 

\section{Small solitons (boson stars)\label{sec:Soliton}}

In subsection \ref{sec:normalMode} we revisited the normal mode frequency spectrum of global AdS$_4$. These frequencies are quantized according to 
\eqref{normalw}. A far-reaching property of this spectrum is that the imaginary part of the frequency vanishes: the associated eingenmodes are not dissipative. Moving beyond linear order in perturbation theory in the amplitude of the scalar field, we might then consider the back-reaction of these scalar normal modes on the gravitational and electromagnetic field. Since the leading order is not radiative we might suspect that the normal mode can be back-reacted to all orders and yield a boson star, i.e.  a horizonless solution with static  electromagnetic and gravitational fields and a complex scalar field with time dependence $e^{-i \omega t}$. We should further expect that the frequency $\omega$ is given at leading order by  \eqref{normalw} but gets corrected as we climb the expansion ladder. We can further explore the U(1) gauge transformation freedom to rewrite this solution in a gauge where the frequency vanishes and the scalar field is real. This static solution is equivalent to the boson star but usually called a soliton.  
As explained in section \ref{sec:Model}, this soliton is a 1-parameter family of solutions that we can take to be the asymptotic amplitude $\varepsilon$ of the scalar field as described in \eqref{BCboundary}.

In this section, we consider the massless scalar field case $m=0$, and will confirm these ideas. Namely, we construct the ground state soliton $-$ i.e. the soliton with lowest energy that at leading order is described by the lowest normal mode \eqref{normalw0} $-$ up to fourth $\mathcal{O}(\varepsilon^4)$. A $U(1)$ gauge transformation allows to set the frequency of the solution to zero; then at leading order we have $A=\frac{3}{e}\,dT$.
To get this soliton, we solve perturbatively the three second order ODEs for $\{f,A,\phi\}$, that we obtain after replacing the algebraic equation \eqref{algebraig_g} for $g$ into \eqref{eom}, subject to the boundary conditions \eqref{BCboundary} and \eqref{TaylorOrigin} at the asymptotic boundary and at the origin, respectively. The pair of integration constants associated to each ODE is fixed by the boundary conditions. 
We build up the perturbative solution in a power series expansion of the fields in $\varepsilon$ around global AdS$_4$,
\begin{align}
\begin{split}\label{solitonexp}
&f(R)=1+R^2 +\sum_{n\geq 1}\varepsilon^{2 n}f_{2 n}(R),\\
&A(R)=\frac 3 e +\sum_{n\geq 1}\varepsilon^{2 n}A_{2 n}(R),\\
&\phi(R)=\frac{\varepsilon}{(1+R^2)^{3/2}} +\sum_{n\geq 1}\varepsilon^{2 n+1}\phi_{2 n+1}(R).
\end{split}
\end{align}
When $\varepsilon=0$ we recover global AdS$_4$ as expected. The $\mathcal{O}(\varepsilon)$ scalar field and its derivatives source the first non-trivial contribution to the electromagnetic and gravitational fields at order  $\mathcal{O}(\varepsilon^2)$.
The structure of the equations of motion is then such that the odd (even) powers of $\varepsilon$ never contribute to the electro-gravitational (scalar) fields.

The equations of motion can be solved analytically order by order for arbitrary order but the expressions for the solutions quickly become quite long. In Appendix \ref{App:Soliton} we display the final (i.e. after imposing the boundary conditions and thus fixing all integration constants) coefficient functions $f_{2 n}(R), A_{2 n}(R)$ and $\phi_{2 n+1}(R)$ up to order $n=2$. This order is enough to extract the relevant physical conclusions of our study.

Using these field expansions we can now compute the gauge invariant thermodynamic quantities discussed in Section \ref{sec:conservedThermo}. 
Namely, the mass $M_{sol}$, charge $Q_{sol}$, chemical potential $\mu_{sol}$, Helmoltz free energy $F_{sol}$ and Gibbs free energy $G_{sol}$ of the ground state soliton up to order $\mathcal O(\varepsilon^6)$ are: 
\begin{align}\label{ThermoSoliton}
M_{sol}/L =& \frac{9 \pi}{32}\, \varepsilon^2-\frac{3 \pi  \left(-301 e^2+60 \pi ^2 \left(e^2-9\right)+2898\right)}{10240}\,\varepsilon^4+\mathcal O(\varepsilon^6), \nonumber\\
Q_{sol}/L =& \frac{3}{32} \pi  e \,\varepsilon ^2-\frac{\pi  e \left(-196 e^2+60 \pi ^2 \left(e^2-9\right)+2583\right)}{10240}\,\varepsilon^4+\mathcal O(\varepsilon^6), \nonumber\\
\mu_{sol} =& \frac{3}{e}+\frac{21 \left(e^2-3\right)}{32 e}\,\varepsilon^2+\frac{\varepsilon ^4}{204800\, e}\big[-339814 e^4+3159903 e^2\\
 &  +300 \pi ^2 (92 e^4-834 e^2 +2079)-7714818\big]+\mathcal O(\varepsilon^6), \nonumber\\
F_{sol} =& M_{sol}, \nonumber\\
G_{sol}/L =& -\frac{63 \pi  \left(e^2-3\right)}{2048}\,\varepsilon ^4+\mathcal O(\varepsilon ^6).\nonumber
\end{align}
As a non-trivial check of our computation, we confirm that these quantities satisfy the first law of thermodynamics for solitons \eqref{firstlawsoliton} up to the required order $\mathcal {O}(\varepsilon ^6)$.

We postpone the physical discussion of our soliton results \eqref{ThermoSoliton}
to Section \ref{sec:ThermoDiscussion}. 

\section{Non-interacting thermodynamic model for hairy black holes}\label{sec:noninteracting}

In the last Section we have constructed the ground state AdS$_4$ soliton of the Einstein-Scalar-Maxwell theory \eqref{action}. On the other hand, this theory also has the RN-AdS$_4$ black hole as a solution. In Section \eqref{sec:superradiance} we have seen that the latter is  unstable to superradiance when scattered by a scalar field. It is natural to expect that this superradiant instability drives the system to a hairy black hole that has a charged scalar condensate floating above the horizon, with the electromagnetic repulsion balancing the gravitational collapse of the scalar condensate into the horizon. For this to be true we need that: 1) a hairy black solution of \eqref{action} does indeed exist, and 2) that for a given energy and charge, the entropy of the hairy black hole is higher than the entropy of the unstable  RN-AdS$_4$ black hole so that the system can evolve from the latter to the former solution while preserving the second law of thermodynamics. 

In this section we confirm the above expectations, i.e. we establish the existence of the hairy black hole solutions and their {\it leading order} thermodynamic properties using a very simple non-interacting thermodynamic model introduced in \cite{Basu:2010uz,Bhattacharyya:2010yg,Dias:2011tj}. In the next section we will confirm that this thermodynamic model yields the correct physical properties of the sytem by solving directly the equations of motion  \eqref{eom}-\eqref{algebraig_g}  as a boundary value problem. Our analysis thus complements a recent numerical time evolution simulation which, for some solution parameters and initial values, showed that RN-AdS$_4$ are superradiantly unstable and decay into hairy black holes \cite{Bosch:2016vcp}.

The leading order thermodynamics of the hairy black hole can be constructed using the so-called non-interacting thermodynamic model of \cite{Basu:2010uz,Bhattacharyya:2010yg} (see also \cite{Dias:2011at,Dias:2011tj,Cardoso:2013pza,Dias:2015rxy} ). One interprets the hairy BH as a {\it non}-interacting mix of a soliton and a RN-AdS$_4$ BH. One first observes that at leading order a charged boson star (soliton) is just a normal mode of the ambient background (whose frequency $\omega$ is then corrected as we walk along the perturbation expansion ladder). This is a 1-parameter family of solutions with $E=\frac{\omega}{q}\,Q+{\cal O}(Q^2)$, where $q$ is the charge of the scalar condensate. This soliton further obeys the first law, $dE=\frac{\omega}{q}\,dQ$.  We can now place a small RN-AdS$_4$ BH on top of this soliton to get a 2-parameter family of hairy BHs. At leading order, we will assume a {\it non}-interacting mixture whereby we take the mass $M$ and the charge $Q$ of the resulting hairy BH to be just the sum of the masses and charges of the BH ($M_{RN}$ and $Q_{RN}$)  and soliton ($M_{sol}$ and $Q_{sol}$) components:\footnote{Beyond the leading order this is property certainly does not hold due to the non-linear character of Einstein's equation. The validity of the non-intarecting assumption at leading order will be confirmed in the next section.}
\begin{align} \label{masscharge}
M=M_{sol}+M_{RN} \quad \text{and} \quad Q=Q_{sol}+Q_{RN}. 
\end{align}
The soliton carries no entropy, so the entropy $S$ of the hairy BH simply reads
\begin{align} \label{entropy}
S=S_{RN}(M_{RN},Q_{RN})+S_{sol}(M_{sol},Q_{sol})=S_{RN}(M-M_{sol},Q-Q_{sol}).
\end{align}

The hairy BH can partition its charge $Q$ and mass $M$ between the RN-AdS$_4$ BH and soliton components. On physical grounds one expects this distribution to be such that, for fixed mass $M$ and charge $Q$, the entropy $S$ is maximised, $dS=dS_{RN}=0$, while respecting the first law of thermodynamics, 
\begin{equation}\label{maxentropy}
 \frac{dM_{RN}}{dQ_{RN}}=-\frac{ \partial_{Q_{RN}}S_{RN}}{ \partial_{M_{RN}}S_{RN}}.
\end{equation}
For fixed total mass and charge, $dM=0=dQ$, and thus $dM_{RN}=-dM_{sol}$ and  $dQ_{RN}=-dQ_{sol}$. It follows from the first law for the soliton that the LHS in \eqref{maxentropy} yields the chemical potential of the soliton, $\frac{dM_{RN}}{dQ_{RN}}=\mu_{sol}$. On the other hand, the first law for the RN-AdS$_4$ BH implies that $\partial_{M_{RN}}S_{RN}=1/T_H$ and $\partial_{Q_{RN}}S_{RN}=-\mu_{RN}/T_H$ where $T_{RN}=T_H$ is the temperature of the RN-AdS$_4$ BH (and thus of the hairy BH) and $\mu_{RN}$ is its chemical potential. Therefore the RHS of \eqref{maxentropy} is simply $-\frac{ \partial_{Q_{RN}}S_{RN}}{ \partial_{M_{RN}}S_{RN}}=\mu_{RN}$. The simple non-interacting thermodynamic model and associated maximization of entropy therefore requires
\begin{equation}\label{chemicaleq}
\mu_{RN}=\mu_{sol}\equiv \mu \quad \hbox{and} \quad T_{RN}\equiv T_H.
\end{equation}
That is, the hairy BH inherits the chemical potential of its solitonic component and the system is in chemical and thermal equilibrium.

The mass and charge of the soliton are related by $M_{sol}=\mu Q_{sol}$, while the leading order thermodynamics of the RN-AdS$_4$ black hole is given by \eqref{RNthermoLeading}. Using these properties together with the non-interacting relation \eqref{masscharge} and equilibrium conditions \eqref{chemicaleq}, we can express the leading order thermodynamic quantities of the hairy black hole (and of its two components) in terms of its mass $M$, charge $Q$ and chemical potential $\mu$:
\begin{align}\label{noninteracting}
&R_+ L=\frac{4 }{2-\mu ^2}(M-\mu  Q)+\mathcal O\left(M^2,Q^2,M Q\right),\nonumber\\
&T =\frac{\left(2-\mu ^2\right)^2}{32 \pi  (M-\mu  Q)}+\mathcal O\left(M^2,Q^2,M Q\right),\nonumber\\
&S=\frac{16 \pi }{\left(2-\mu ^2\right)^2} (M-\mu  Q)^2+\mathcal O\left(M^3,Q^3,M^2 Q,M Q^2\right);
\end{align}
\begin{align}
&\hspace{1cm} M_{RN}=\frac{\left(2+\mu ^2\right)}{2-\mu ^2}(M-\mu  Q)+\mathcal O\left(M^2,Q^2,M Q\right),\nonumber\\
&\hspace{1cm} M_{sol}=\frac{\mu  \left(2+\mu ^2\right) Q-2 \mu ^2 M}{2-\mu ^2}+\mathcal O\left(M^2,Q^2,M Q\right),\nonumber\\
&\hspace{1cm} Q_{RN}=\frac{2 \mu }{2-\mu ^2} (M-\mu  Q)+\mathcal O\left(M^2,Q^2,M Q\right),\nonumber\\
&\hspace{1cm} Q_{sol}=\frac{\left(2+\mu ^2\right) Q-2 \mu  M}{2-\mu ^2}+\mathcal O\left(M^2,Q^2,M Q\right).\nonumber
\end{align}

The domain of existence of the hairy black hole can be  inferred from this analysis. In one extremum, the soliton component is absent and all the mass and charge of the hairy BH is carried by the RN-AdS$_4$ component. This describes the hairy black hole that merges with  the RN-AdS$_4$ black hole at the zero-mode of its superradiant instability. On the opposite extremum configuration, the  RN-AdS$_4$ component is absent and the soliton component carries all the mass and charge of the solution. This is the zero-radius or zero-entropy limit of the hairy black hole. It follows that the hairy black hole mass must be within these two boundaries:
\begin{align}\label{bdriesThermoModel}
\frac{\left(\mu ^2+2\right) }{2 \mu }Q+\mathcal O (Q^2)\geq M\geq \mu  Q+\mathcal O (Q^2).
\end{align}
Equating the two extrema configurations, yields an interval of existence for the chemical potential:
\begin{align}\label{condThermoModel}
\mu\leq\frac{\left(\mu ^2+2\right)}{2 \mu }\quad \Rightarrow  \quad \mu\leq\sqrt{2} \quad \implies e\geq \frac{3}{\sqrt{2}}\equiv e_c,
\end{align}
where in the last relation we used that for the ground state solution the leading order potential is $\mu=\omega/e=3/e +\mathcal O(R_+)$.
Note that condition \eqref{condThermoModel} is the same we found in the superradiant analysis leading to \eqref{eSuperradianceBound}.

The above leading order thermodynamic analysis must be considered with a few grains of salt. Indeed, first note that a theory can have hairy black holes that do not have a zero-radius limit, \ie$\text{ }$a solitonic limit. Second, there is no reason why the non-interacting mixture assumption \eqref{masscharge} should hold, even at leading order.

In the next section we will construct the hairy black hole. In Section \ref{section:Thermodynamics} we compute the thermodynamic quantities of the hairy black hole and we check that our intuition was correct, and that, to leading order, the hairy black hole thermodynamic quantities are indeed given by \eqref{noninteracting}.

\section{ Small hairy black holes}\label{sec:Hairy BH}

In this section we construct perturbatively the hairy black holes whose leading order thermodynamics was discussed in the previous section.  For that we  solve the elliptic equations of motion \eqref{eom}-\eqref{algebraig_g},  subject to the boundary conditions \eqref{BCboundary} and \eqref{TaylorH}, to find a perturbative analytical expression for the hairy black hole.

\subsection{Setting up the perturbation problem}

As explained in Section \ref{sec:Model}, the hairy black holes of our theory are a two-parameter family of solutions that we can take to be the asymptotic scalar amplitude $\varepsilon$ and the horizon radius $R_+$. Thus, their perturbative construction requires that we do a double expansion of the fields in powers of $\varepsilon$ and $R_+$. In order to be able to solve analytically the equations of motion \eqref{eom}-\eqref{algebraig_g} we must resort to a matched asymptotic expansion, similar to those done in a similar context in 
\cite{Basu:2010uz,Bhattacharyya:2010yg,Dias:2011at,Dias:2011tj,Stotyn:2011ns}. Much like in Section \ref{sec:superradiance}, we divide the outer domain of communications of our black hole into two regions; a near-region where $r_+\leq r\ll L$ and a far-region where $r\gg r_+$. Restricting the analysis to small black holes that have $r_+/L \ll 1$, the two regions have an overlapping zone, $r_+ \ll r\ll L$. In this overlapping region, we can match/relate the set of independent parameters that are generated by solving the perturbative equations of motion in each of the two regions.

The chemical potential of the solution should itself have a double expansion in powers of $\varepsilon$ and $R_+$, 
\begin{equation}
\mu=\sum_{n\geq 0}\varepsilon^{2 n}\sum_{k\geq 0}R^{k}\mu_{2 n,k}.
\end{equation}
Indeed, recall that the soliton is the back-reaction of a normal mode of AdS to higher orders. At leading oder the chemical potential of the  soliton is related, via a gauge transformation, to an AdS normal mode frequency. We saw that this is  corrected at higher orders. Thus, we must permit similar corrections when the horizon is present. We shall construct the hairy black hole family whose zero-radius limit is the {\it ground state} soliton of Section \ref{sec:Soliton} (so with lowest energy for a given charge).\footnote{A similar construction could be done for the excited hairy black holes.}

In the far region, $R\gg R_+$, the hairy black hole can be seen as a small perturbation in $R_+$ and $\varepsilon$ around global AdS. We use the superscript $^{out}$ when referring to far region fields which have the double expansion:
\begin{align}
\begin{split}\label{RNexp}
&f^{out}(R)=\sum_{n\geq 0}\varepsilon^{2 n}\sum_{k\geq 0}R_+^{k}\,f_{2 n,k}^{out}(R),\qquad A^{out}(R)=\sum_{n\geq 0}\varepsilon^{2 n}\sum_{k\geq 0}R_+^{k}\,A_{2 n,k}^{out}(R),\\
&\phi^{out}(R)=\sum_{n\geq 0}\varepsilon^{2 n+1}\sum_{k\geq 0}R_+^{k}\,\phi_{2 n+1,k}^{out}(R).
\end{split}
\end{align}
This expansion already anticipates that odd (even) powers of the scalar condensate do not correct the fields $f,A$ ($\phi$).

At each order $\{n,k\}$ the perturbed equations of motion can be solved analytically (with the help of {\it Mathematica}) to find the far fields up to a total of 6 integration constants (recall that the equations of motion are a system of 3 second order ODEs). In addition, the system also depends on the chemical potential corrections $\mu_{2 n,k}$.
The requirement that the far-region fields obey the asymptotic boundary conditions \eqref{BCboundary} typically fixes 4 of the integration constants (namely, those associated to $\phi_{2 n+1,k}^{out}$, one associated to $f_{2 n,k}^{out}$ and another to $A_{2 n,k}^{out}$). We are left with two integration constants and $\mu_{2 n,k}$ that will be determined by the requirement that the far fields match the near-region fields in the overlapping region.
To prepare the system for the matching we need to take the small radius $R$ expansion of $f^{(out)},A^{(out)},\phi^{out}$ (with the expansion coefficients available at the order under consideration). We find that these are singular as $R\to 0$, diverging with a power of $\frac{R_+}{R}$. This indicates that the far-region analysis  breaks down at $R\sim R_+$. This justifies why the far-region analysis is valid only for $R\gg R_+$.  Also, it follows that in the far-region we can safely do a Taylor expansion in $R \ll 1$ and $\varepsilon \ll 1$ since the large hierarchy of scales between the solution parameters and the distance guarantees that they do not compete.

Consider now the near-region, $R_+\leq R\ll 1$. This time the Taylor expansions in $R \ll 1$ and $\varepsilon \ll 1$ should proceed with some caution since the small parameters can now be of similar order as the radius $R_+$. This is closely connected with the fact that the far-region solution breaks down when  $R/R_+\sim \mathcal{O}(1)$. This suggests that to proceed with the near-region analysis we should define a new radial, $y$, and time, $\tau$, coordinates as  
\begin{equation}\label{nearcoordinates}
 y=\frac{R}{R_+}, \qquad \tau= \frac{T}{R_+}.
\end{equation}
The near region now corresponds to $1 \leq y \ll R_+^{-1}$. If we further require that $R_+\ll 1$ one sees that the near region corresponds to $y\geq 1 \gg R_+$ (and $y\gg \varepsilon$). In particular, we can now safely do Taylor expansions in $R \ll 1$ and $\varepsilon \ll 1$ since the radial coordinate $y$ and the black hole parameters have a large hierarchy of scales \footnote{At the heart of the matching expansion procedure, note that a factor of $R_+$ (one of the expansion parameters) is absorbed in the new coordinates.}. To have further physical insight it is also instructive to rewrite the RN-AdS$_4$ solution \eqref{RNsolution} in the new coordinate system \eqref{nearcoordinates} 
\begin{eqnarray}\label{RNnear}
&&\mathrm d s^2=R_+^2\left(-f(y)\mathrm d \tau^2+\frac{\mathrm d y^2}{f(y)}+y^2\mathrm d \Omega_{2}^2\right),\\
&& \hspace{1cm}f(y)=1-\frac{1}{y}-\frac{1}{y}\left(\frac{1}{2} \mu^2 \left(1-\frac{1}{y}\right)+R_+^2 \left(1-y^3\right)\right),\nonumber \\
&& \hspace{1cm}A_{\tau}(y)=R_+\;A_T(y)=R_+\;\mu\left(1-\frac{1}{y}\right),\nonumber \\
&& \hspace{1cm}\phi(y)=0.\nonumber
\end{eqnarray}
The explicit factor of $R_+\ll 1$ in $A_{\tau}(y)$ indicates that in the near region the electric field is weak and the system can be seen as a small perturbation around the neutral solution. The same should hold when we add a small scalar condensate to the system. The near fields of the hairy black hole thus have the double expansion,
\begin{align}
\begin{split}\label{RNexpNear}
&f^{in}(y)=\sum_{n\geq 0}\varepsilon^{2 n}\sum_{k\geq 0}R_+^{k}\,f_{2 n,k}^{in}(y),\qquad A^{in}(y)=\sum_{n\geq 0}\varepsilon^{2 n}\sum_{k\geq 0}R_+^{k}\,A_{2 n,k}^{in}(y),\\
&\phi^{in}(y)=\sum_{n\geq 0}\varepsilon^{2 n+1}\sum_{k\geq 0}R_+^{k}\,\phi_{2 n+1,k}^{in}(y).
\end{split}
\end{align}

At each order $\{n,k\}$, we can  solve the perturbed equations of motion analytically to find the near fields up to a total of 6 near region integration constants. Imposing the horizon boundary conditions \eqref{TaylorH} typically fixes 3 of the integration constants (namely, one associated to $\phi_{2 n+1,k}^{in}$, one to $f_{2 n,k}^{in}$ and another to $A_{2 n,k}^{in}$). The 3 leftover near field integration constants are left undetermined until we match the far and near region fields in the overlapping region. For the matching, we first need to restore the original coordinates $\{T,R\}$ and take the large radius limit of  $f^{in}(R),\, A^{in}(R),\, \phi^{in}(R)$ (with the expansion coefficients available at the given order). We find that these diverge as a power of $R$, which shows that the near region analysis  breaks down at $R\sim 1$. This explains why the near region analysis is valid only for $R\ll 1$.

In the following subsection, we give some {\it low order} detailed examples of the matching asymptotic expansion implementation. We are necessarily constrained on this exposition to lowest orders since, as the expansion order grows, the analytical expressions for the far and near fields become quite large.   For the benefit of the reader that wants to reproduce our results in full, we give the far and near field  functions  in Appendices \ref{AppendixBH:Far} and \ref{AppendixBH:Near}, respectively. These are the final fields, i.e. after fixing all the integration constants of the system using the boundary conditions \eqref{BCboundary} and \eqref{TaylorH} and the matching of far and near fields. 

We leave to subsection \eqref{section:Thermodynamics} in the main text, the physically relevant outcome of our matching asymptotic construction, namely the thermodynamic quantities that describe uniquely the hairy black holes.

\subsection{Examples illustrating the matching asymptotic expansion}

\subsubsection{Matching asymptotic expansion at $\mathcal O\left(\varepsilon^1,R_+^k\right)$}

We insert the double expansion \eqref{RNexp} (far region) or \eqref{RNexpNear} (near region) into the equations of motion \eqref{eom}-\eqref{algebraig_g} and solve the resulting perturbed equations order by order.

At lowest order, $\mathcal{O}(\varepsilon^0)$ in the scalar amplitude expansion, the scalar field is naturally absent and the far field coefficients $\{f^{out}_{0,k}(R),A^{out}_{0,k}(R)\}$ can be read directly from an $R_+\ll 1$ expansion of the  RN-AdS$_4$ solution \eqref{RNsolution}. A similar Taylor expansion of \eqref{RNnear} yields the near field coefficients $\{f^{in}_{0,k}(y),A^{in}_{0,k}(y)\}$. The $\mathcal{O}(\varepsilon^1)$  correction turns-on the scalar field $\phi$ without back-reacting yet in the gravitational background: it describes a small perturbation of the scalar field around the RN-AdS$_4$ black hole. That is, the non-trivial equations of motion \eqref{eom}-\eqref{algebraig_g}  reduce to the Klein-Gordon equation without a source\footnote{At higher orders in $n$, the equation of motion for $\phi$ still has the form of a Klein-Gordon equation but with an inhomogeneous term sourced by the lower order fields and their derivatives.}. 

Next, we illustrate how the matching asymptotic expansion determines $\phi_{1,k}$. A similar procedure yields the fields $\phi_{2n+1,k}$ at the odd orders $\mathcal O(\varepsilon^{2 n+1})$. 

$\noindent \bullet$ {\tt Far region, $R\gg R_+$:}

The definition of $\varepsilon$ introduced in the boundary condition \eqref{BCboundary} determines completely the solution at order $\mathcal O(R_+^0)$:\footnote{\label{foot:epsilon}To be precise, $\varepsilon$ is an integration constant that parametrizes our solution and, without loss of generality, we choose not to allow corrections to its definition \eqref{BCboundary} at any order.}
\begin{equation}
\phi_{1,0}^{out}(R)=\frac{1}{(1+R^2)^{3/2}}\,,
\end{equation}
and the equation of motion is obeyed only for the choice $\mu_{0,0}=3/e$ (which corresponds, via a gauge transformation, to the lowest normal mode frequency of AdS$_4$).
We now expand the equation to $\mathcal O(R_+^1)$. We get a Klein-Gordon equation with an inhomogeneous term sourced by a derivative of $\phi_{1,0}^{out}(R)$ which can be solved analytically. The two integration constants are directly determined by the asymptotic boundary condition \eqref{BCboundary} (see also footnote \ref{foot:epsilon}). This yields
\begin{align}
\phi_{1,1}^{out}(R)=&
\frac{1}{48 e^2 R \left(R^2+1\right)^{5/2}}\Big(36 \left(2 e^2-18 \left(R^3+R\right) \arctan R+9 \left(R \left(\pi  R^2-2 R+\pi \right)-1\right)\right)\nonumber\\
&+e^3 \left(R^2+1\right) \mu_{0,1} (-2 R \left(3 R^2+17\right)+3 \pi  \left(R^4+6 R^2-3\right)-6 (R^4+6 R^2\nonumber\\
&-3) \arctan R)\Big).
\end{align}
The value of the chemical potential correction $\mu_{0,1}$ will be determined only in  the matching of the far region with the near region.

As explained above, we skip here the details of the computation of the field coefficient $\phi_{1,2}^{out}(R)$. However, it is perhaps useful to highlight that it is typically a good idea to complete the matching procedure, that fixes constants not determined by the boundary conditions, before proceeding to the next order.  
This keeps the size of the expressions that carry to next order smaller.

$\noindent \bullet$ {\tt Near region, $R_+\leq R\ll 1$.}

The homogeneous Klein Gordon equation in the near region at $\mathcal O (R_+^0)$ has the solution
\begin{align}
\phi^{in}_{1,0}(y)=C_1+\frac{C_2 \left(-\log \left(3 e^2 y^2-32\right)+\log \left(y^2-1\right)-i \pi \right)}{6 e^2-64}.
\end{align}
We need to impose the boundary conditions \eqref{TaylorH} at the horizon  $y=1$. Namely regularity of the scalar field at the horizon requires that we eliminate the divergent logarithmic term by choosing $C_2=0$. Hence, the regular near region solution is 
\begin{align}\label{phi10note}
\phi^{in}_{1,0}(y)=C_1.
\end{align}

At order $\mathcal O (R_+^1)$, $\phi^{in}_{1,1}(y)$ still solves the homogeneous Klein-Gordon equation\footnote{A would be source term is proportional to the derivatives of the previous order field $\phi^{in}_{1,0}$, but the latter is just a constant, see \eqref{phi10note}.}. Therefore, its regular solution is 
\begin{align}
\phi^{in}_{1,1}(y)=\widehat{C}_1.
\end{align}

$\noindent \bullet$ {\tt Matching in the overlapping region $R_+ \ll R\ll 1$:}

In order to do the matching we take the small $R$ limit of the far region solution and the large $R$ limit of the near solution. Notice that at this order only terms up to $R^0 R_+^1$ and $R^1 R_+^0$ are correctly accounted for (terms higher than this will receive corrections from the next order). 
The small $R$ expansion of the far region is:
\begin{align}
\phi_{1,0}^{out}(R)+R_+\phi_{1,1}^{out}(R)\Big|_{R\to 0}= &1+R_+\left(-\frac{\left(\frac{3 \pi  e}{16}  \mu_{0,1}+\frac{27}{4 e^2}-\frac{3}{2}\right)}{R}+\left(\frac{27 \pi }{4 e^2}-\frac{e}{3} \mu_{0,1}\right)+\mathcal O(R)\right)\nonumber\\
&+\mathcal O(R_+^2).
\end{align} 
On the other hand, the large $R$ expansion of the near region solution is:
\begin{align}
\phi_{1,0}^{in}\left(\frac{R}{R_+}\right)+R_+\phi_{1,1}^{in}\left(\frac{R}{R_+}\right)\Big|_{R\to\infty}=C_1+R_+ \widehat{C}_1.
\end{align}
The divergent term in the far region expansion has no counterpart in the near region and needs to be eliminated by a judicious choice of $\mu_{0,1}$. Then, the matching of the two leftover terms fixes the integration constants $C_1$ and  $\widehat{C}_1$ that were not fixed by the boundary conditions. Altogether we find,
\begin{equation}
\mu_{0,1}= \frac{4 \left(2 e^2-9\right)}{\pi  e^3}\,,\qquad C_1=1,\qquad
\widehat{C}_1=\frac{-32 e^2+81 \pi ^2+144}{12 \pi  e^2}\,.
\end{equation}

With this we have completed the solution up to $\mathcal O(\varepsilon^1,R_+^1)$. We can now proceed to order $\mathcal O(\varepsilon^1,R_+^2)$ following a similar procedure. The final far and near field coefficients $\{f^{out}_{0,2},A^{out}_{0,2},\phi^{out}_{1,2}\}$ and $\{f^{in}_{0,2},A^{in}_{0,2},\phi^{in}_{1,2}\}$, after imposing the boundary conditions and the matching are listed in Appendices \ref{AppendixBH:Far} and \ref{AppendixBH:Near}, respectively. 

\subsubsection{Matching asymptotic expansion at $\mathcal O\left(\varepsilon^2,R_+^k\right)$}

At  order $\mathcal O\left(\varepsilon^2\right)$, the order $\mathcal O\left(\varepsilon^1\right)$ scalar field back-reacts on the metric and gauge potential and the hairy black hole starts differentiating from the RN-AdS$_4$ solution. On the other hand the Klein Gordon equation is trivially satisfied.
Next, we illustrate how the matching asymptotic expansion determines $f_{2,k}$ and $A_{2,k}$. The procedure applies similarly for all the even orders $\left(\varepsilon^{2n}\right)$. 

$\noindent \bullet$ {\tt Far region, $R\gg R_+$:}

Starting at order $\mathcal O\left(\varepsilon^2,R_+^0\right)$ the far field equation of motion can be solved analytically to get  $f_{2,0}^{out}$ and  $A_{2,0}^{out}$. Each one of these fields has two integration constants. Imposing the asymptotic Dirichlet boundary condition \eqref{BCboundary} we fix one of the constants in each pair. We are left with the integration constants $C_f$ and $C_A$ that will be fixed in the matching step. At this stage the fields read:
\begin{eqnarray}\label{hairy2far}
f_{2,0}^{out}(R)&=&\frac{C_f}{R}-\frac{3 \left(R \left(3 R^2+5\right)+3 \left(R^2+1\right)^2 \arctan R\right)}{8 R \left(R^2+1\right)^2},\\
A_{2,0}^{out}(R)&=&\mu_{2,0}-\frac{C_A}{R}-\frac{e \left(3 R^2+5\right)}{8 \left(R^2+1\right)^2}-\frac{3 e \arctan R}{8 R}\,.
\end{eqnarray}

$\noindent \bullet$ {\tt Near region, $R_+\leq R\ll 1$.}

Also at order $\mathcal O\left(\varepsilon^2,R_+^0\right)$, we solve for the near fields  $f_{2,0}^{in}$ and  $A_{2,0}^{in}$ up to a pair of integration constants. Requiring that these fields vanish linearly at the horizon $y=1$, as required by the horizon boundary condition \eqref{TaylorH}, we fix two of the constants. 
The near region fields are then still a function of two arbitrary constants
 $K_f$ and $K_A$:
\begin{eqnarray}\label{hairy2near}
f_{2,0}^{in}(y)&=&\frac{K_f\, e y-3 K_A}{e y^2}\,(y-1),\\
A_{2,0}^{in}(y)&=&\frac{K_A}{y}\,(y-1).
\end{eqnarray}

$\noindent \bullet$ {\tt Matching in the overlapping region $R_+ \ll R\ll 1$:}

To match the far and near fields, we take the small $R$ expansion of the far fields \eqref{hairy2far}:
\begin{align}
&f^{out}_{2,0}(R)\Big|_{R\to 0}=\frac{C_f}{R}-3+\mathcal O\left(R^1,R_+^1\right),\\
&A^{out}_{2,0}(R)\Big|_{R\to 0}=(\mu_{2,0}-e)-\frac{C_A}{R}+\mathcal O\left(R^2\right).
\end{align}
and the large $R$ expansion of the near fields \eqref{hairy2near}:
\begin{align}
&f^{in}_{2,0}\left(\frac{R}{R_+}\right)\Big|_{R\to\infty}=K_f+\mathcal O(R_+^1),\\
&A^{in}_{2,0}\left(\frac{R}{R_+}\right)\Big|_{R\to\infty}=K_A+\mathcal O(R_+^1).
\end{align}
Note that we keep only terms that will not be corrected by  higher order contributions. 

Matching the two expansions fixes the four integration constants, that were not determined by the boundary conditions, as 
\begin{equation}
C_f=0,\qquad C_A=0,  \qquad K_f= -3, \qquad K_A= \mu_{2,0}-e.
\end{equation}

At this stage the chemical potential correction $\mu_{2,0}$ is still left undetermined. It is fixed at order $\mathcal O\left(\varepsilon^3,R_+^0\right)$, when the scalar field $\phi_{3,0}$ is found.

\subsection{Thermodynamic quantities}\label{section:Thermodynamics}

The  final results for the field coefficients introduced in \eqref{RNexp} and \eqref{RNexpNear} are given in Appendices \ref{AppendixBH:Far} (for the far region) and \ref{AppendixBH:Near} (for the near field). These are the final fields, i.e. after fixing all the integration constants of the system using the boundary conditions \eqref{BCboundary} and \eqref{TaylorH} and the matching of the far and near fields. 

In the presentation of our results,  we take  $\epsilon\ll 1$ and $R_+\ll 1$ and we assume that $\mathcal{O}(\epsilon^2)\sim\mathcal{O}(R_+)$. The latter assumption implies that terms with the same $(n+k)$ contribute equally to the perturbative expansion, i.e.  $\mathcal{O}\left(\epsilon^{0},R_+^2\right)\sim \mathcal{O}\left(\epsilon^{2},R_+\right)\sim \mathcal{O}\left(\epsilon^{4},R_+^0\right)$. We did the consistent perturbative expansion analysis up to $(n+k)=2$, which is sufficient for our purposes.  Extending the perturbative analysis to higher orders is  possible, but increasingly cumbersome.

The fields presented in Appendix \ref{App:hairyBH} allow to compute the relevant physical properties of hairy black holes. Namely, as described in \ref{sec:conservedThermo}, we can compute the thermodynamic quantities of the hairy solutions. The dimensionless mass $M$, electric charge $Q$, chemical potential $\mu$, entropy $S$, temperature $T$ are: 
\begin{align}
M/L=&\Big[\Big(\frac{9}{4 e^2}+\frac{1}{2}\Big) R_++\frac{6 (2 e^2-9)}{\pi  e^4}\,R_+^2+\mathcal O(R_+^3)\Big]+\varepsilon ^2 \Big[\frac{9 \pi }{32}+\frac{9 (27 \pi ^2-2 (e^2+9))}{128 e^2}\,R_+\nonumber\\
&+\mathcal O(R_+^2)\Big]+\varepsilon^4\Big[-\frac{3 \pi  (-301 e^2+60 \pi ^2 (e^2-9)+2898)}{10240}+\mathcal O(R_+^1)\Big]+\mathcal O(\varepsilon^6) ,\nonumber\\
Q/ L=&\Big[\frac{3}{2 e}\, R_++\frac{2 (2 e^2-9)}{\pi  e^3}\,R_+^2+\mathcal O(R_+^3)\Big]+\varepsilon ^2 \Big[\frac{3 \pi  e}{32}+\frac{(-38 e^2+81 \pi ^2+90)}{128 e}\,R_+\nonumber\\
&+\mathcal O(R_+^2)\Big]+ \varepsilon^4\Big[-\frac{\pi  e (-196 e^2+60 \pi ^2 (e^2-9)+2583)}{10240}+\mathcal O(R_+^1) \Big]+\mathcal O(\varepsilon^6),\nonumber\\
\mu=&\Big[\frac{3}{e}+\frac{4 (2 e^2-9)}{\pi  e^3}\,R_++\frac{(2 e^2-9) [-256 e^2+9 \pi ^2 (10 e^2-9)+2688]}{32 \pi ^2 e^5}\,R_+^2\nonumber\\
&+\mathcal O(R_+^3)\Big]+\varepsilon^2 \Big[\frac{21 (e^2-3)}{32 e}+\frac{1}{9600 \pi  e^3}\big[75 \pi ^2 (236 e^4-1809 e^2+4212)\nonumber\\
&-8 (25330 e^4-264861 e^2+685503) R_+\big] +\mathcal O(R_+^2)\Big]+\varepsilon^4\Big[\frac{1}{204800 e}\big[-339814 e^4 \nonumber\\
&+3159903 e^2+300 \pi ^2 (92 e^4-834 e^2+2079)-7714818\big]+\mathcal O(R_+^1)\Big]+\mathcal O(\varepsilon^6) ,\nonumber\\
S/L^2=&\pi R_+^2 ,\label{ThermoHairyBH}\\
T L=&\frac{1}{4\pi R_+}\Bigg\{\Big[\frac{2 e^2-9}{2 e^2}-\frac{12 \left(2 e^2-9\right)}{\pi  e^4}\,R_+ +\frac{1}{32 \pi ^2 e^6} \big(96 \pi ^2 e^6-540 \pi ^2 e^4+512 e^4\nonumber\\
&+2916 \pi ^2 e^2-13824 e^2-2187 \pi ^2+51840\big) R_+^2+\mathcal O(R_+^3)\Big] +\varepsilon^2\Big[-\frac{3 \left(5 e^2+9\right)}{32 e^2}\nonumber\\
&-\frac{3 R_+}{6400 \pi  e^4} (7300 \pi ^2 e^4-89760 e^4-87750 \pi ^2 e^2+982992 e^2+289575 \pi ^2-2640816)\nonumber\\
&+\mathcal O(R_+^2)\Big]+\varepsilon^4\Big[\frac{1}{204800 e^2} (-66600 \pi ^2 e^4+716782 e^4+653400 \pi ^2 e^2-6564933 e^2\nonumber\\
&-1761750 \pi ^2+16179426)  +\mathcal O(R_+^1)\Big]+\mathcal O(\varepsilon^6)\Bigg\},\nonumber
\end{align}
and the Helmoltz free energy $F$ and Gibbs free energy $G$ are: 
\begin{align}
F/ L=&\Big[\left(\frac{27}{8 e^2}+\frac{1}{4}\right) R_+ +\left(\frac{18}{\pi  e^2}-\frac{81}{\pi  e^4}\right) R_+^2+\mathcal O(R_+^3)\Big]+\varepsilon ^2 \Big[\frac{9 \pi }{32}+\Big(\frac{243 \pi ^2}{128 e^2}-\frac{135}{128 e^2}\nonumber\\
&-\frac{3}{128} \Big) R_+ +\mathcal O(R_+^2)\Big]+\varepsilon ^4\Big[-\frac{9 \pi ^3 e^2}{512} +\frac{903 \pi  e^2}{10240}+\frac{81 \pi ^3}{512}-\frac{4347 \pi }{5120}+\mathcal O(R_+^1)\Big] \nonumber\\
&+\mathcal O(\varepsilon^6),\nonumber\\
G /L=&\Big[\left(\frac{1}{4}-\frac{9}{8 e^2}\right) R_++\left(\frac{27}{\pi  e^4}-\frac{6}{\pi  e^2}\right) R_+^2+\mathcal O(R_+^3)\Big]+\varepsilon ^2\Big[\left(\frac{405}{128 e^2}-\frac{111}{128}\right) R_+\nonumber\\
&+\mathcal O(R_+^2)\Big] +\varepsilon ^4\Big[\frac{189 \pi }{2048}-\frac{63 \pi  e^2}{2048}+\mathcal O(R_+^1)\Big]+\mathcal O(\varepsilon^6). \nonumber
\end{align}

In the next section we discuss physical checks to these quantities and withdraw physical conclusions.


\section{Discussion of physical properties}\label{sec:ThermoDiscussion}

\subsection{Checking the hairy solutions and their interpretation}

The thermodynamic quantities \eqref{ThermoHairyBH} for the hairy black hole must pass a few  checks that already give valuable information about their physical interpretation. 

Firstly, when we set the horizon radius to zero, $R_+ \to 0$, we do get the thermodynamical quantities \eqref{ThermoSoliton} of the soliton. Recall that the latter was obtained without any matching asymptotic expansion. This also confirms that the soliton is the zero horizon radius limit of the hairy black hole. We have assumed this was the case in our construction and the fact that the computation can be done is consistent with it.

Secondly, when we set the scalar condensate amplitude to zero, $\varepsilon \to 0$, we get the thermodynamical quantities \eqref{RNthermo} of the RN-AdS$_4$ black hole, with a very specific chemical potential namely,
\begin{equation}\label{RNatOnset}
\mu=\frac{3}{e}+\frac{4 (2 e^2-9)}{\pi  e^3}\,R_++\frac{(2 e^2-9) [-256 e^2+9 \pi ^2 (10 e^2-9)+2688]}{32 \pi ^2 e^5}\,R_+^2+\mathcal O(R_+^3).
\end{equation}
That is, although the RN-AdS$_4$ black hole is a 2-parameter family of solutions, the $\varepsilon \to 0$ limit of the hairy black hole gives a particular 1-parameter sub-family of RN-AdS$_4$ parametrized by $R_+$ and chemical potential fixed by \eqref{RNatOnset}. This is precisely the RN-AdS$_4$ 1-parameter family at the onset of the superradiant instability studied in Section \ref{sec:superradiance}. To see this is indeed the case start by recalling that in Section \ref{sec:superradiance} we studied scalar field perturbations with Fourier time dependence $\phi(T,R)\sim e^{-i \omega T}\phi(R)$ and found that the frequency is quantized as described in \eqref{superradiantExpansion}  and \eqref{superradiantFreq}. In particular, the frequency has a positive imaginary part for $\omega> e \mu$ signalling the superradiant instability. The onset of the superradiant instability occurs when the imaginary part of the frequency vanishes and we can work in the $U(1)$ gauge where the real part of the frequency vanishes. If we impose these superradiant onset conditions on \eqref{superradiantFreq} and solve with respect to the chemical potential in a series expansion in $R_+$ up to order $\mathcal{O}\left(R_+^2\right)$ we precisely get \eqref{RNatOnset}. This is an important check to our computations and, together with the information from the last paragraph, confirms that hairy black holes are indeed a 2-parameter family of solutions that spans an area in a phase diagram that has the onset curve of the RN-AdS$_4$ superradiance and the soliton curve as boundaries. At leading order, these are the boundaries \eqref{bdriesThermoModel} predicted by the simple thermodynamical model 
of Section \ref{sec:noninteracting} and that will be very clear in the left panel of Fig. \ref{fig:MCphaseDiag} (red and black curves).

The first law of thermodynamics provides the third check. Indeed, we can explicitly verify that the thermodynamic quantities \eqref{ThermoHairyBH} do obey the first law \eqref{firstlawBH} up to order $(n+k)=2$. We emphasize this is a fundamental and non-trivial check of our results.

To summarize, we have constructed analytically the hairy black hole within perturbation theory up to order $(n+k)=2$ (and we could extend this construction to higher order). 
This was done solving directly the perturbed equations of motion \eqref{eom}-\eqref{algebraig_g}  of the theory as a boundary value problem. 

We can confirm that the  simple non-interacting thermodynamic model of  Section \ref{sec:noninteracting} $-$ which does {\it not} use the equations of motion $-$ yields the correct leading order thermodynamics, i.e. the leading terms of \eqref{ThermoHairyBH}.
First note that at leading order it follows from \eqref{ThermoHairyBH} that $\mu=\frac 3 e$.
Assuming, as justified above, that $\mathcal{O}(\epsilon^2)\sim\mathcal{O}(R_+)$, the leading order contributions of the expansion \eqref{ThermoHairyBH} allows to express analytically $R_+$ and $\varepsilon$ in terms of $M$ and $Q$,
\begin{equation}
R_+=\frac{4 e (e M-3 Q)}{2 e^2-9} +\mathcal O\left(M^2,Q^2,M Q\right), \quad \varepsilon =\sqrt{\frac{32 \left(\left(2 e^2+9\right) Q-6 e M\right)}{3 \pi  e \left(2 e^2-9\right)}}+\mathcal O\left(M^2,Q^2,M Q\right),
\end{equation}
which we insert in the expressions for the other thermodynamic quantities to find that at leading order in $M$ and $Q$ one has:
\begin{align}
\begin{split}
& \mu=\frac 3 e+\mathcal O\left(M, Q\right), \\
&S=\frac{16 \pi  e^2 (M e-3 Q)^2}{\left(9-2 e^2\right)^2}+\mathcal O\left(M^2,Q^2,M Q\right),\\
&T=\frac{\left(9-2 e^2\right)^2}{32 \pi  e^3 (M e-3 Q)}+\mathcal O\left(M^2,Q^2,M Q\right).
\end{split}
\end{align}
These quantities do match the result of the non-interacting model \eqref{noninteracting}. This confirms that the non-interacting thermodynamic model, in spite of its crude simplicity, is quite robust and does indeed capture the fundamental leading order properties of the hairy black hole system at very low cost. This explicit confirmation adds to those done in similar hairy black hole systems in \cite{Basu:2010uz,Bhattacharyya:2010yg,Dias:2011at,Dias:2011tj,Cardoso:2013pza,Dias:2015rxy}.

A more detailed discussion of the regime of validity of our perturbation theory is also in order. By construction, it should be valid only for $\varepsilon\ll 1$ and $R_+\ll 1$.  On the other hand, the scalar charge of the hair  must obey $e\gtrsim  \frac{3}{\sqrt{2}}  \sim 2.12$ to have superradiant hairy black holes: see discussion that leads to \eqref{eSuperradianceBound}. However, since the black holes are a small expansion around AdS$_4$, $e$ should not be too large to avoid large back-reactions. This suggests that it is appropriate to require at most $\epsilon \lesssim 0.1$, $R_+ \lesssim 0.1$ and $ \frac{3}{\sqrt{2}}  \lesssim e\lesssim 3$, say. Inserting these bounds into the thermodynamic formulas \eqref{RNthermo}, \eqref{ThermoSoliton}, and \eqref{ThermoHairyBH} we get approximate upper bounds for the physical charges. For example, for $e=2.5$ we should look into masses and charges that, in AdS units, are themselves below $10^{-1}$.  However the reader interested on making a direct comparison between our perturbative results and exact numerical constructions that emerge from a nonlinear code or from the endpoint of the superradiant time evolution might require more precise statements. We will have the opportunity to give  precise criteria in the following discussion: see footnotes \ref{foot:ValidityMC} and \ref{foot:ValidityC}.

We have found that the gravitational Higgs model with action \eqref{action} admits several solutions. Namely, global AdS$_4$, the RN-AdS$_4$ black hole, the ground state soliton, the ground state hairy black hole and an infinite tower of excited solitons and hairy black holes\footnote{Recall that the ground state solutions have their perturbative root in the lowest normal mode frequency of global AdS$_4$, and the excited states emerge from the remaining infinite tower of normal mode frequencies.}. For a given electric charge, the latter excited solutions always have larger energy than the ground state partners so we do not discuss them further.  

Naturally, these thermal phases of the theory compete with each other. This competition can be framed in one of the three possible ensembles: the microcanonical, canonical or grand-canonical ensembles. 
We discuss the phase diagram in these three ensembles in the next three subsections. Recall that the thermodynamic quantities of the RN-AdS$_4$ black hole, (ground state) soliton and hairy black hole are revisited in \eqref{RNthermo}, \eqref{ThermoSoliton}, and \eqref{ThermoHairyBH}, respectively \cite{Chamblin:1999tk,Chamblin:1999hg}. 

Our results are better illustrated with some plots. Recall that the gravitational Higgs theory with action \eqref{action}  is only fully defined once the particular value for the scalar field charge $q=e/L$ is specified. For definiteness we will choose the particular value of $e=2.5$ in our plots. 
 
\subsection{Phase diagram in the microcanonical ensemble}

In the microcanonical ensemble the energy $M$ and electric charge $Q$ of the system are fixed and the the entropy $S$ is the relevant thermodynamic potential to discuss the competition between the several thermal phases: the phase with higher entropy is the favoured one (global AdS$_4$ and the soliton have vanishing entropy). 

\begin{figure}[t]
\centerline{
\includegraphics[width=.48\textwidth]{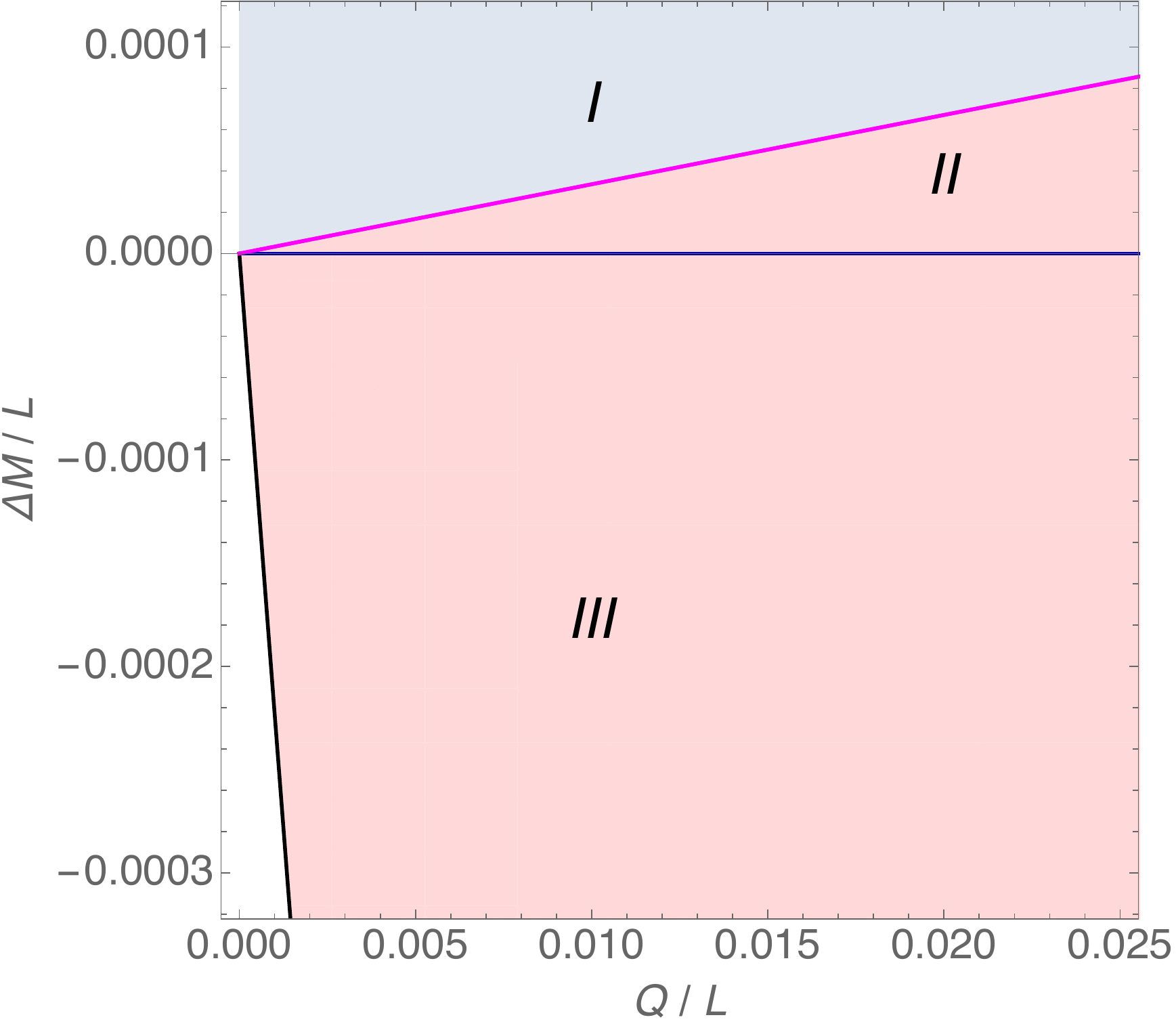}
\hspace{0.3cm}
\includegraphics[width=.48\textwidth]{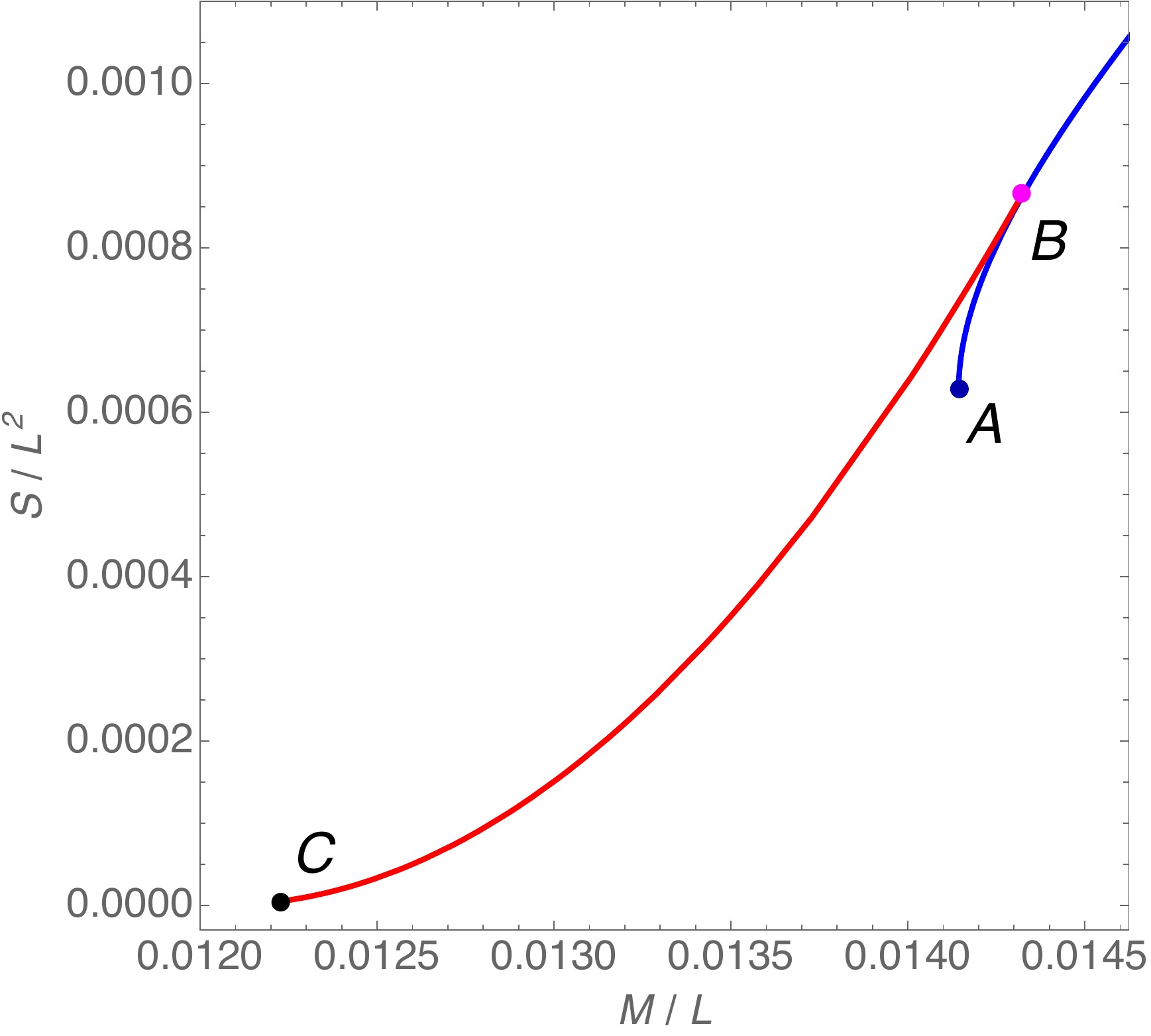}
}
\caption{{\bf Left Panel:} Microcanonical ensemble phase diagram $\Delta M=M-M_{ext}$ vs $Q$  for $e=2.5$. RN-AdS$_4$ black hole exist in regions $I$ and $II$, with the magenta line with positive slope describing the onset curve of superradiant instability. The black line with negative slope describes the soliton. Hairy black holes exist in between the two above lines, i.e. in regions $II$ and $III$ (red shaded).  {\bf Right Panel:} Entropy as a function of the mass at constant value of the charge $Q/L=0.01$ and $e=2.5$. The unbounded blue line starting at $A$ (extremality) is the RN-AdS$_4$ black hole and the red line $BC$ is the hairy back hole branch that ends on the soliton $C$. The merger point $B$ is the superradiant zero-mode. }
\label{fig:MCphaseDiag}
\end{figure}

In Fig. \ref{fig:MCphaseDiag} we display the phase diagram $M$ vs $Q$ of the microcanonical ensemble for $e=2.5$ (our perturbative expressions are valid for small $M,Q$). To make the diagram more clear, in the vertical axis we actually plot the difference between the mass of the solution and the mass of the extremal RN-AdS$_4$ with the same electric charge, $\Delta M\equiv M-M_{ext}$.  So RN-AdS$_4$  black holes exist in regions $I$ and $II$ where $\Delta M \geq 0$. The  black line with negative slope describes the soliton. Hairy black holes exist in regions $II$ and $III$. Namely they fill the area limited by the soliton curve (where $R_+\to 0$) all the way up to the magenta line with positive slope (with $\varepsilon = 0$). The latter also describes RN-AdS$_4$ black holes at the onset of superradiance. So RN-AdS$_4$ black holes below (above) the magenta line are unstable (stable) to superradiance. 

In region $II$ there is no uniqueness since RN-AdS$_4$ and hairy black holes coexist with the same $M$ and $Q$.
To find the preferred phase in the microcanonical ensemble we must compare their entropy $S$. We find that for a given electric charge $Q$ the hairy black hole always has higher entropy and is thus the favoured solution, whenever they coexist. This is illustrated with a particular example in  the right panel of Fig. \ref{fig:MCphaseDiag} where we fix the electric charge to be $Q/L=0.01$ and plot $S$ vs $M$ (again for $e=2.5$). The blue line describes the RN-AdS$_4$ black hole and extends from arbitrarily large $M$ and $S$ all the way down to the extremal configuration $A$. It is stable along this path until it reaches point $B$ that signals the onset of the superradiant instability (so $BA$ describes unstable RN-AdS$_4$). At this point $B$ there is a {\it second order phase transition}\footnote{\label{foot:ValidityMC} This provides a further check of our computations. Indeed, note that the first law requires that at a second order phase transition the slope $dS/dM$ is the same for the two branches since $T$ is the same at the bifurcation point (and $dQ=0$ in the right panel of Fig. \ref{fig:MCphaseDiag}). This is clearly the case in our plot. However, if we start departing from the regime of validity of our perturbation analysis we increasingly find that the merger is not perfect and the slopes of the two branches no longer match. This is the best criterion to identify the regime of validity of \eqref{ThermoHairyBH}.} to the hairy black hole branch that extends all the way down to the zero-horizon radius ($S=0$) where it meets the soliton (point $C$). 
This dominance of the hairy black hole in the microcanonical ensemble extends to all values of the electric and scalar charges (again, in the regime where our perturbative results hold).   

Our findings also  permit robust conclusions about the endpoint of the superradiant instability of the RN-AdS$_4$ black holes. Typically, numerical simulations are done at fixed energy and charge. Consider starting with a RN-AdS$_4$ that is within the curve $AB$ and thus unstable to superradiance. By the second law of thermodynamics, the entropy of the system can only increase. Assuming that there is no other black hole solution in the spectrum of the theory besides the ones discussed so far, the system must evolve towards the hairy black hole that has the same $M$ and $Q$ but higher entropy. The latter can be pinpointed in \eqref{ThermoHairyBH} or in Fig. \ref{fig:MCphaseDiag}. By construction, this hairy black hole is stable to superradiance and we have no arguments suggesting that it is unstable to any other mechanism (see however footnote \ref{foot1}).
Of course our findings say nothing about how the system actually evolves in time but the endpoint of the numerical simulation that gives this information can be tested against \eqref{ThermoHairyBH}.\footnote{Also recall that for $\sqrt{3/2}<e<3/\sqrt{2}$ there are hairy black holes that emerge from the near-horizon scalar condensation instability that are not captured by our ``superradiant" solutions \eqref{ThermoHairyBH}. But they can be found as the endpoint of a time evolution simulation or by solving numerically the elliptic system of equations  \eqref{eom}-\eqref{algebraig_g}.}

\subsection{Phase diagram in the  canonical ensemble}\label{subsec:canonical}

In the canonical ensemble the temperature and electric charge of the system are held fixed. The relevant  thermodynamic potential in this ensemble is the Helmholtz free energy $F=E-TS$ and the preferred phase is the one with lowest free energy.

\begin{figure}[t]
\centerline{
\includegraphics[width=.48\textwidth]{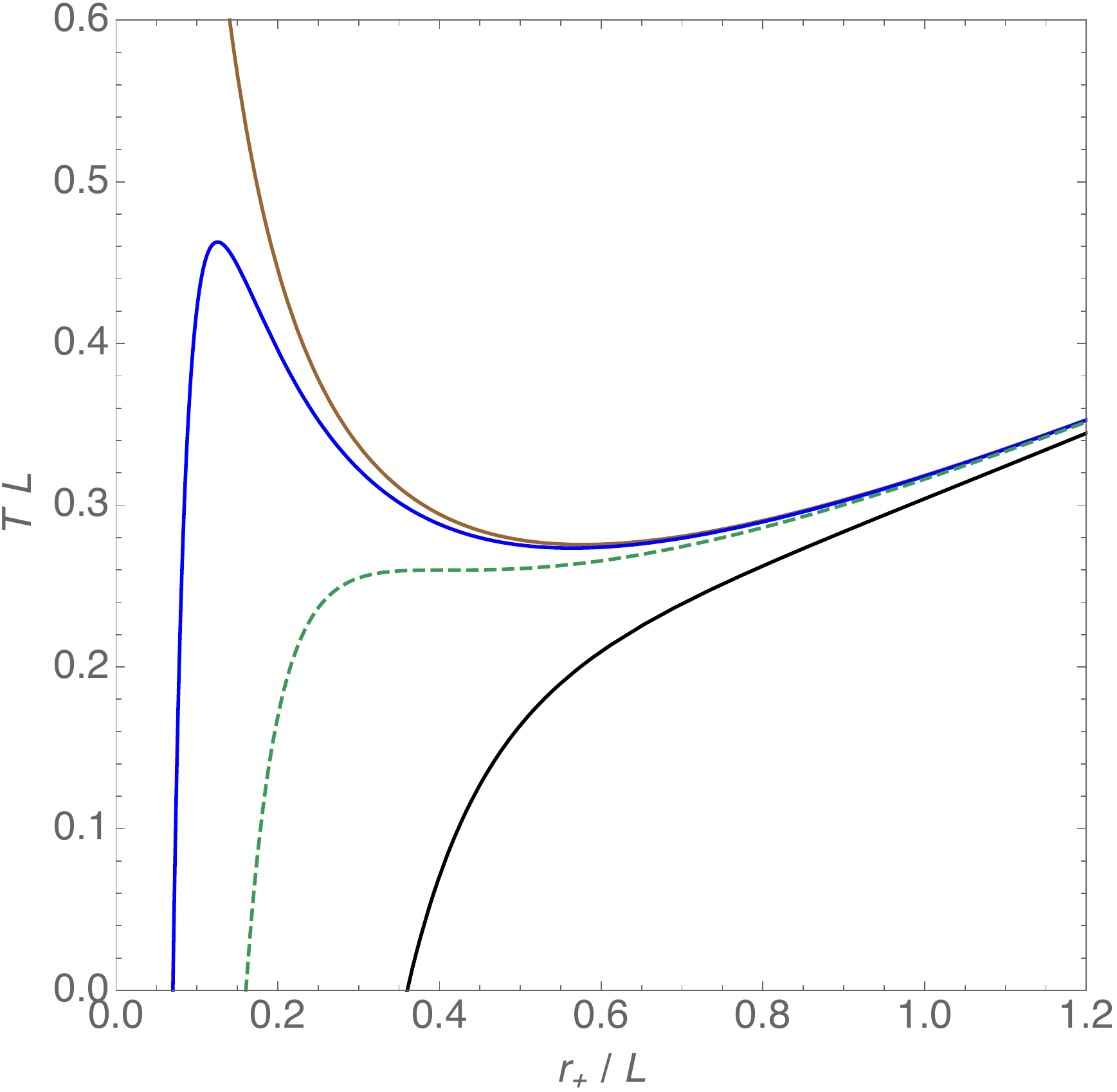}
\hspace{0.3cm}
\includegraphics[width=.48\textwidth]{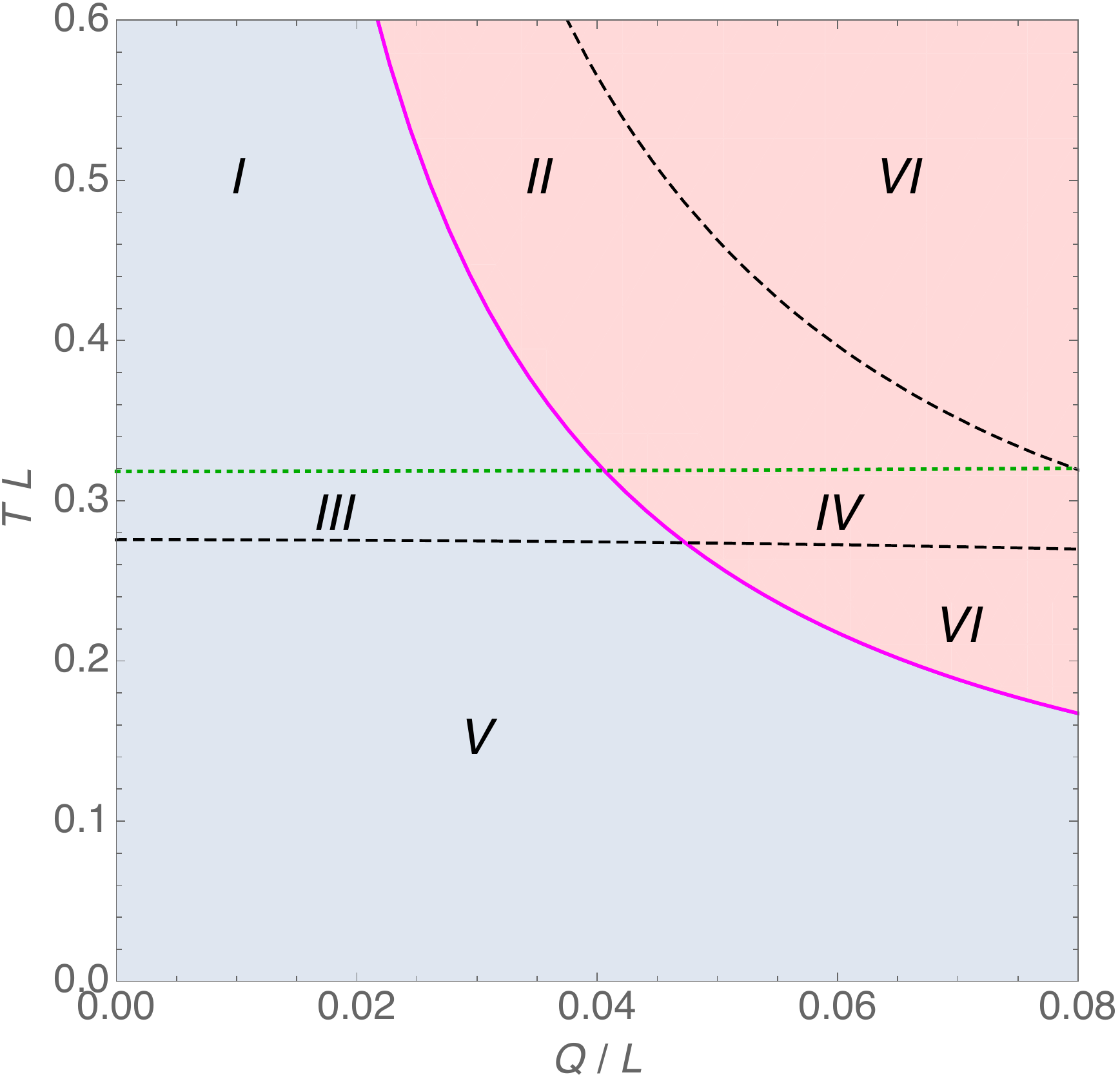}
}
\caption{{\bf Left Panel:} Temperature $T L$ of RN-AdS$_4$ black hole as a function of the horizon radius $r_+/L$ for four different charges. From top to bottom, the curves are for $Q/L=0$ (Schwarzschild case for reference), $Q/L=0.05$, $Q=Q_{crit}$, and $Q/L=0.3$, respectively. 
{\bf Right panel}: Canonical ensemble phase diagram $T$ vs $Q$ (for $e=2.5$). RN-AdS$_4$ black holes as well as `thermal' solitons exist in all regions.  Regions I$-$IV have three (namely, `small', `intermediate' and `large') branches of RN-AdS$_4$ black holes. The dashed upper line is the curve described by the `right' cusp of the middle panel of Fig. \ref{fig:Canonical_phase_comp}, while the dashed lower line is the curve followed by the `top' cusp of the same figure (the upper and lower dashed curves merge and terminate at a value of $Q$ not shown). On the other hand regions $V$ and $VI$ have a single RN-AdS$_4$ branch.  Hairy black holes exist in regions $II$, $IV$ and $VI$. The magenta line (above which hairy black holes exist) describes the onset of superradiance ($\varepsilon=0$). The dotted green line signals the Hawking-Page transition (its $T L$ grows as $Q/L$ increases). Below this line, thermal AdS$_4$ is the dominant phase while above, the preferred thermal phase is the large branch of RN-AdS$_4$ black holes.}
\label{fig:CphaseDiag}
\end{figure}

Before discussing the hairy phase and thermal competition it is convenient to review the thermal properties of the RN-AdS$_4$ solution. These were discussed in detail in the seminal works of \cite{Chamblin:1999tk} and \cite{Chamblin:1999hg}. The left panel of Fig. \ref{fig:CphaseDiag}  displays the temperature of RN-AdS$_4$ as a function of the horizon radius for four different values of the charge. This figure reveals two key properties (the upper, brown, curve describes the Schwarzschild-AdS$_4$ solution and is present only for reference). First, notice that there is a critical charge , $Q_{crit}/L=\frac{1}{6 \sqrt{2}}$, associated to the dashed green curve. For charges above this (lower black curve) there is only one RN-AdS$_4$ black hole for a given $\{Q,T\}$ pair. However, for $Q>Q_{crit}$ (second, blue, curve from the top) we see that there is a window of $T$ where $\{Q,T\}$ do not uniquely describe the RN-AdS$_4$ solution since we can have three different RN-AdS$_4$ branches that  can be denoted as the `small', `intermediate' and `large' black holes depending on their horizon radius (which is proportional to the square root of their entropy).\footnote{For reference, when $Q=0$ (top brown curve in left panel of Fig.~\ref{fig:CphaseDiag}) we have not three but two black hole branches because $T\to\infty $ as $R_+\to 0$. These are commonly designated by the `small' and `large' Schwarzschild-AdS$_4$ black holes. For $Q\leq Q_{crit}$, the first branch, that corresponds to smaller values of the horizon radius, goes from extremality to a maximum found at $R_+=\frac{\sqrt{1-\sqrt{1-72 Q^2}}}{\sqrt{6}}$. The second branch goes from this maximum to the minimum at $R_+=\frac{\sqrt{1+\sqrt{1-72 Q^2}}}{\sqrt{6}}$. And the third branch from this minimum to larger horizon radius. The intervals for the temperatures of the three RN-AdS$_4$ branches are:
\begin{eqnarray}
0\leq T_1\leq \frac{\sqrt{\frac{3}{2}} \left(1-24 Q^2-\sqrt{1-72 Q^2}\right)}{\pi  \left(1-\sqrt{1-72 Q^2}\right)^{3/2}}\leq T_2 \leq \frac{\sqrt{\frac{3}{2}} \left(-24 Q^2+\sqrt{1-72 Q^2}+1\right)}{\pi  \left(\sqrt{1-72 Q^2}+1\right)^{3/2}} \leq T_3< \infty .\nonumber
\end{eqnarray}
The critical charge $Q_{crit}$ is the case where the maximum and the minimum of $T(R_+)$ coincide (inflection point).
}

A second property observed in the left panel of Fig. \ref{fig:CphaseDiag}  is that $T\to 0$ as we decrease $R_+$ towards the extremal configuration (which is absent when $Q=0$). Therefore, small horizon radius corresponds to small temperatures. However, $T=0$ is reached at a finite minimal radius that increases with $Q$. We will come back to this observation later when discussing the validity regime of our perturbation theory.    

We can now introduce these RN-AdS$_4$ solutions in a phase diagram for the canonical ensemble and populate it with the novel hairy solutions. 
The canonical phase diagram consists of plotting the free energy $F$ as a function of $T$ for a fixed $Q$.\footnote{Ideally we would show the 3-dimensional plot $F(T,Q)$ but it is not very clear.  The plots $F(T)$ for fixed $Q$ (or $F(Q)$ for fixed $T$) illustrate better the relevant conclusions.}

\begin{figure}[t]
\centerline{
\includegraphics[width=.31\textwidth]{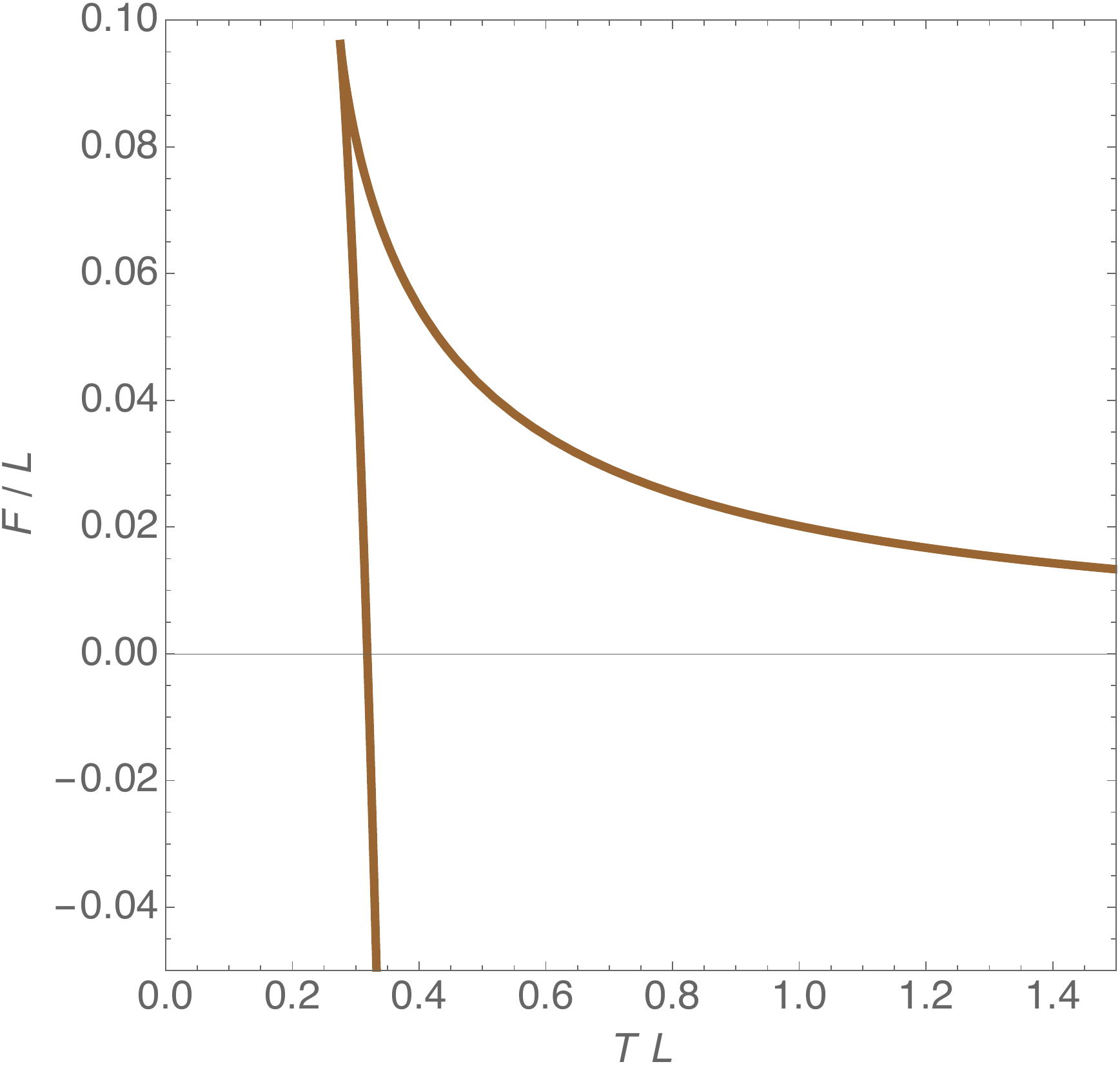}
\hspace{0.3cm}
\includegraphics[width=.31\textwidth]{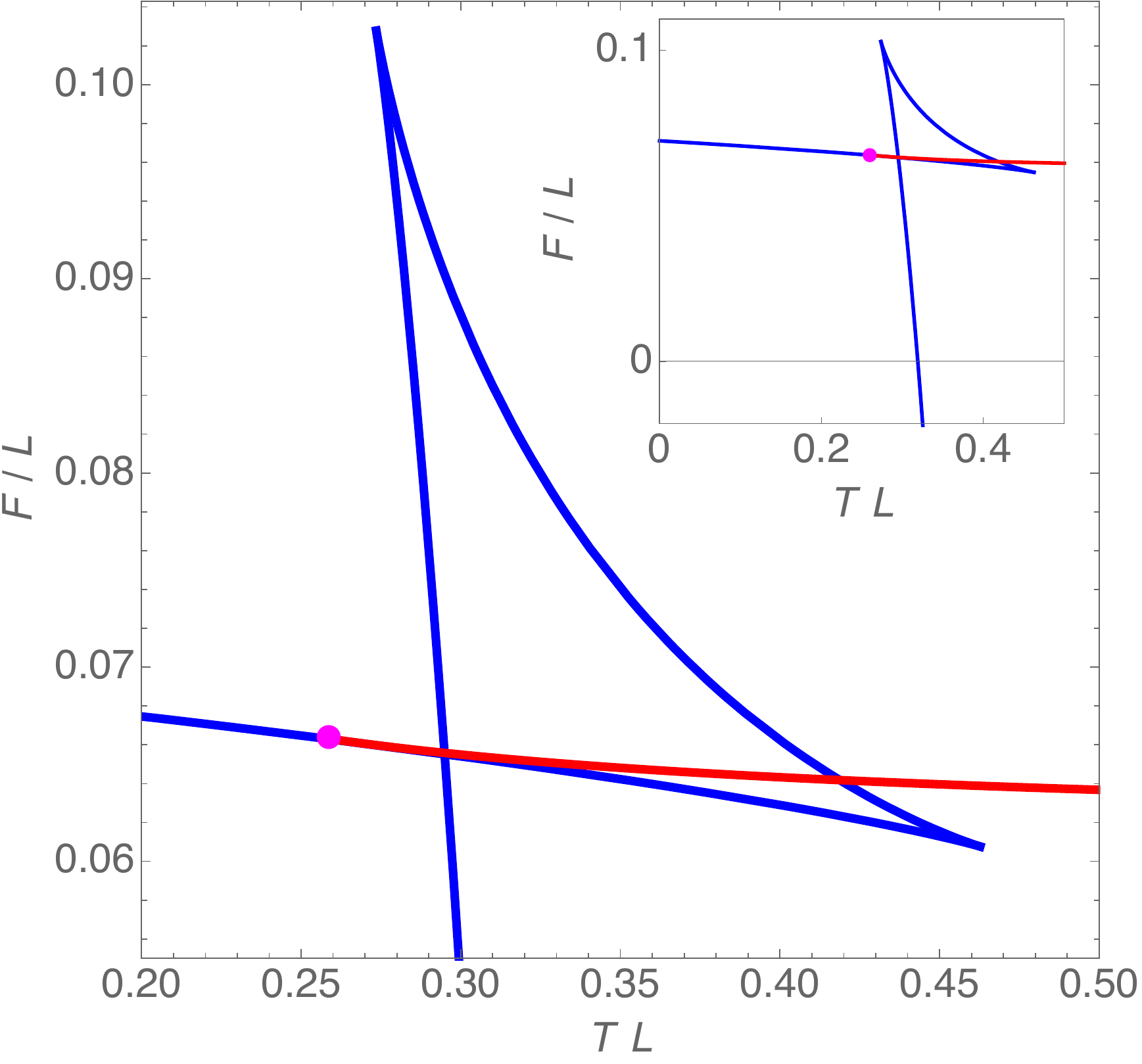}
\hspace{0.3cm}
\includegraphics[width=.31\textwidth]{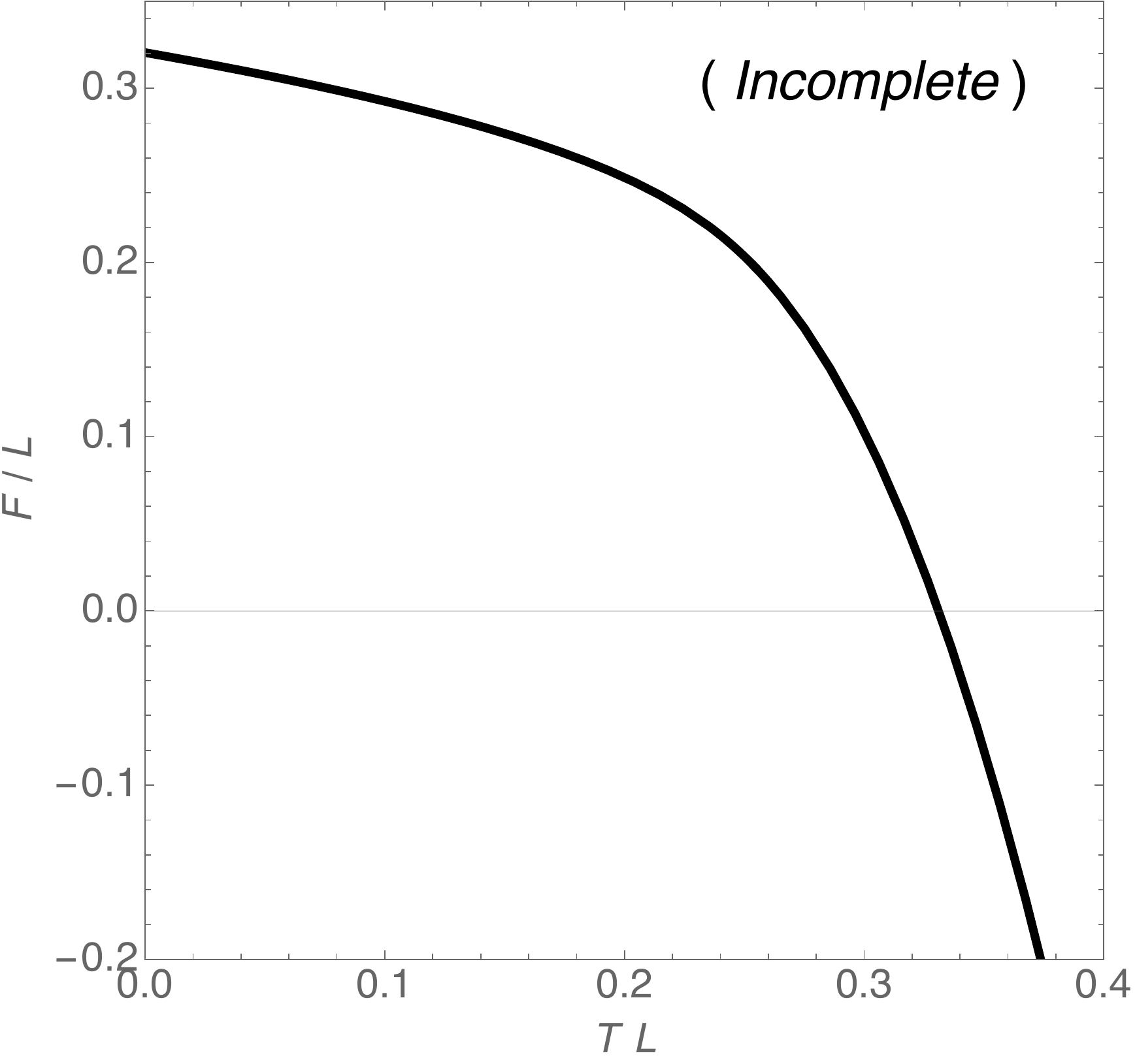}
}
\caption{Canonical ensemble phase diagram $F$ vs $T$ at fixed $Q$ and $e=2.5$. {\bf Left Panel:} $Q=0$ case with the large and small Schwarzschild black holes, but no hairy phases. {\bf Middle panel:} $Q/L=0.05<Q_{crit}/L$ case with the blue curve representing the RN-AdS$_4$ family with its 3 branches and the simple red curve describing the hairy black hole. {\bf Right Panel:} $Q/L=Q_{crit}/L+ 0.1$. This case falls outside the regime of validity of our perturbation theory, hence we opt to display only the (single) RN-AdS$_4$ branch.}
\label{fig:Canonical_phase_comp}
\end{figure}

In Fig. \ref{fig:Canonical_phase_comp} we show this phase diagram $F(T)$ for the three relevant cases: $Q=0$, $0<Q<Q_{crit}$ and $Q>Q_{crit}$ (left, middle and right panels). For the phases without hair we keep the same colour code as in the left panel of Fig. \ref{fig:CphaseDiag}. Moreover we necessarily do the plots for specific values of $Q$ and $e$ but they are qualitatively similar for {\it all} other values where the perturbation results \eqref{ThermoHairyBH} are valid. The {\it left panel} with $Q=0$ describes the Schwarzschild (Schw-AdS$_4$) black hole and is given here again for a familiar reference \cite{Hawking:1982dh}. We have the `large' branch with negative specific heat (left) and the `small' branch with positive specific heat (right) .  The free energies of large Schw-AdS$_4$ are always lower than that of the corresponding small Schw-AdS$_4$ with the same $T$. Therefore, large Schw-AdS$_4$ are favoured over small Schw-AdS$_4$ black holes. In the phase diagram of the left panel of Fig. \ref{fig:CphaseDiag}  the large and small Schw-AdS$_4$ branches meet at the regular cusp (where the specific heat vanishes). This is not the whole story since there is a third phase $-$ thermal AdS$_4$ $-$ which is just the Euclidean solution of global AdS$_4$ with an arbitrary period (and thus $T$) chosen for the Euclidean time circle. For a temperature below the Hawking-Page $T_{HP}$ (defined as $F(T_{HP})=0$),  thermal AdS$_4$ has lower free energy than both large and small  Schw-AdS$_4$ black holes.  However, at temperatures above $T_{HP}$, large Schw-AdS$_4$ black holes are the preferred phase.  This is the familiar Hawking-Page (HP) first-order phase transition \cite{Hawking:1982dh}.  In the AdS/CFT context,  this is interpreted as a confinement/deconfinement transition in the dual conformal field theory \cite{Witten:1998qj}.

Consider now the case $0<Q<Q_{crit}$, whose phase diagram (for $Q/L=0.05$) is plotted in the {\it middle panel} in Fig. \ref{fig:Canonical_phase_comp}. As explained above we have now  three RN-AdS$_4$  branches. The large branch is the natural extension to $Q\neq 0$ of the large Schw-AdS$_4$ branch. However, the small branch now extends all the way down to the extremal, $T=0$, solution (for $Q=0$ it extends instead to $T\to \infty$). These two branches are  connected by a third, `intermediate' branch. When they coexist small RN-AdS$_4$ black holes are the dominant branch below a critical temperature (determined by the intersection of the small and large branches in the middle panel). Above this critical temperature, large RN-AdS$_4$ black holes have lower free energy than the small and intermediate branches and thus dominate the canonical ensemble. Above $T_{HP}$ (defined such that $F(T_{HP})=0$) large RN-AdS$_4$ are the preferred phase, but below it there is a first order Hawking-Page phase transition and thermal AdS$_4$ is the favoured phase for all $T<T_{HP}$. In particular, as $T$ decreases to $T=0$ and small RN-AdS$_4$ becomes the only branch, thermal AdS$_4$ is still the preferred phase. When $Q\neq 0$ the theory also has hairy black hole solutions. In the perturbative regime where our results are valid\footnote{\label{foot:ValidityC} Much like in the microcanonical ensemble (see footnote \ref{foot:ValidityMC}) the first law of thermodynamics (with $dQ=0$) requires that the slope $dF/dT$ is the same for the two branches at the bifurcation point (since they have the same $S$). This is definitely the case in our plot. However, if we start departing from the regime of validity of our perturbation analysis we increasingly find that the merger is not perfect and the slopes of the two branches no longer match. This is the best criteria to identify the regime of validity of \eqref{ThermoHairyBH} for the canonical ensemble.}, the hairy family bifurcates from the branch of small RN-AdS$_4$ at the onset of superradiance and extends to larger values of the temperature all the way up to the soliton limit ($R_+\to 0$) that is reached at a finite value of $T$.\footnote{We can identify the Euclidean time circle of the soliton with any period so that it can have an arbitrary temperature. However, the free energy of this `thermal' soliton coincides with its mass, hence it is always positive and the soliton is always dominated by thermal AdS$_4$.} However, the free energy of the hairy black hole (and soliton) is always larger that the free energies of small and/or large RN-AdS$_4$ and of thermal AdS$_4$. Therefore, hairy black holes and the soliton are never the preferred phase in the canonical ensemble. This is thus in sharp contrast with the situation in the microcanonical ensemble.

Finally we have to discuss the case $Q>Q_{crit}$ ({\it right panel} in Fig. \ref{fig:Canonical_phase_comp}). Recall from the discussion associated with the left panel of Fig. \ref{fig:CphaseDiag}   that in this case there is a single branch of RN-AdS$_4$ black holes. Its free energy is plotted in the {\it right panel} in Fig. \ref{fig:Canonical_phase_comp}. As before, the Hawking-Page phase transition is present at $T=T_{HP}$ with thermal AdS$_4$ (RN-AdS$_4$) being the preferred phase at $T<T_{HP}$ ($T>T_{HP}$). For  $Q>Q_{crit}$ there are also hairy black hole solutions. We choose {\it not} to show them in the right plot of Fig. \ref{fig:Canonical_phase_comp} because we do not expect our perturbative results \eqref{ThermoHairyBH} to be valid for such large charge.
To understand the reason we go back to the left panel of  Fig. \ref{fig:CphaseDiag}. For $Q>Q_{crit}$, RN-AdS$_4$ has horizon radius $R_+\gtrsim 0.2$. But our perturbative results for the hairy black holes that merge with the RN-AdS$_4$ solution are valid for $R_+\ll 1$ so we should not expect our perturbative hairy results to be accurate for this case, at least around the merger point.  

Given that the canonical ensemble keeps the temperature $T$ and electric charge $Q$ fixed, it is also relevant to summarize the conclusions above for the thermal phases of the theory in a phase diagram $T$ vs $Q$. This effectively describes a projection of the 3-dimensional plot $F(T,Q)$.
This is done in the {\it right panel} of Fig. \ref{fig:CphaseDiag}, and the reader can find its discussion in the caption of the figure. 

So far we have discussed the preferred {\it global} thermodynamic phases of the canonical phase diagram. Given that in this ensemble we keep an intensive thermodynamic quantity $-$ the temperature $-$ fixed we can also discuss the {\it local} thermodynamic stability of the solutions. Local thermodynamic stability in the canonical ensemble requires that the specific heat at constant charge 
\begin{equation}
C_Q=T \left( \frac{\partial S}{\partial T} \right)_Q = -T \left( \frac{\partial^2 F}{\partial T^2} \right)_Q
\end{equation}
is non-negative.
To get the second relation we used the first law, $dF=-SdT+\mu dQ$. We find that small hairy black holes have $C_Q<0$. Thus, they are locally thermodynamically unstable. This is illustrated in Fig. \ref{fig:Canonical_phase_comp} (for $Q=0.05$ and $e=2.5$): the sign of $C_Q=-T\left(\partial_T^2 F \right)_Q$ is negative \footnote{\label{LocalThermoFoot}The local thermodynamic stability of the RN-AdS$_4$ branches can also be inferred from the concavity/convexity of the free energy. This stability was already discussed in detail in \cite{Chamblin:1999tk} and \cite{Chamblin:1999hg} and we do not repeat it here. Recent discussions of local thermodynamic instability can be found in \cite{Monteiro:2008wr,Monteiro:2009tc,Hollands:2012sf} and references therein.}. Essentially, a thermal fluctuation that increases the temperature of the small hairy black hole leads to a decrease of its entropy (horizon size).

\subsection{Phase diagram in the  grand-canonical ensemble}

In the grand-canonical ensemble the system is kept at fixed temperature $T$ and fixed chemical potential $\mu$. The dominant thermal phase is the one that minimizes the Gibbs free energy, $G=M-TS-\mu Q$.

\begin{figure}[t]
\centerline{
\includegraphics[width=.47\textwidth]{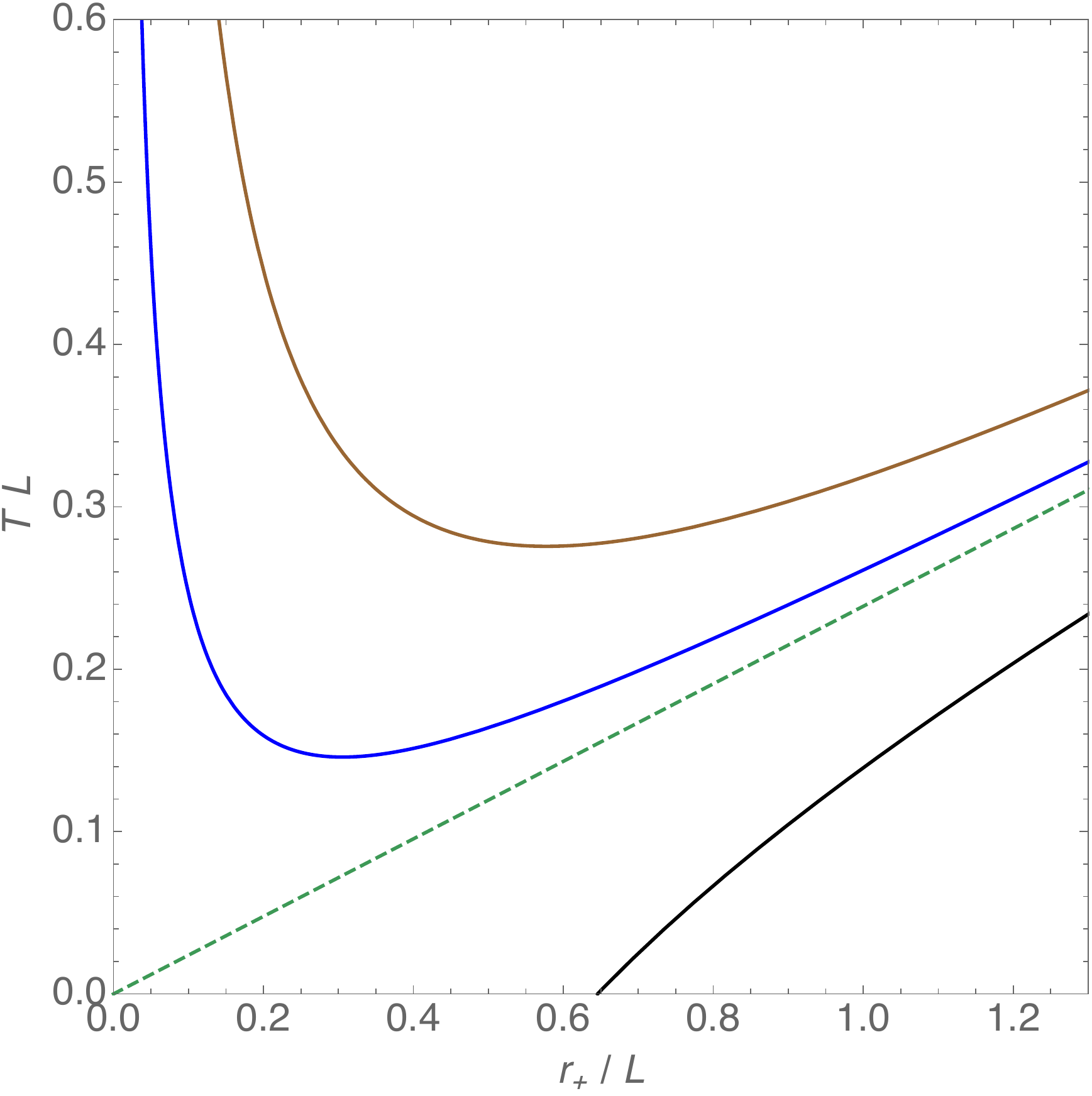}
\hspace{0.6cm}
\includegraphics[width=.48\textwidth]{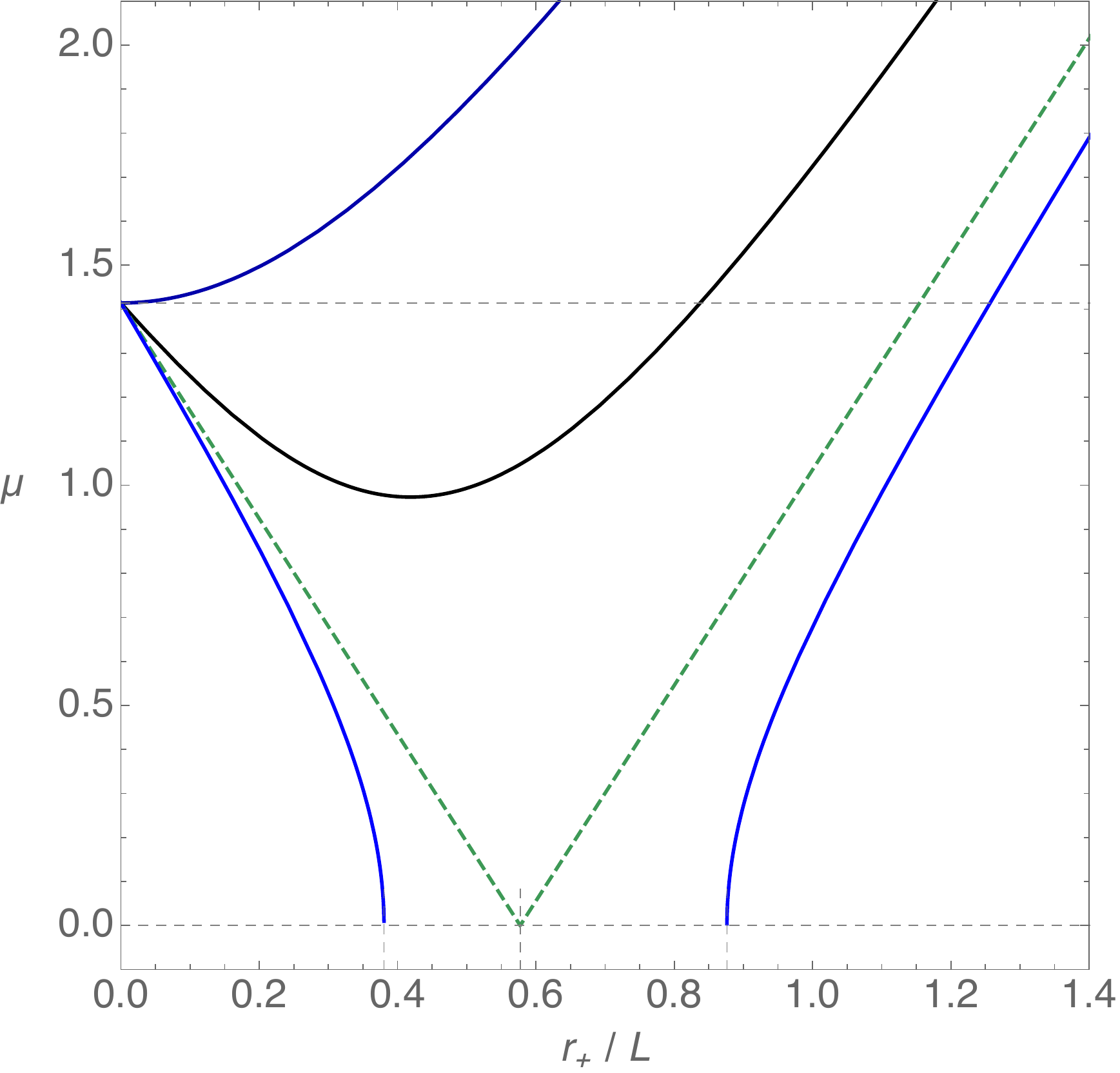}
}
\caption{{\bf Left Panel:}  Temperature $T L$ as a function of the horizon radius $r_+/L$ for values of the chemical potential (from top to bottom): $\mu=0$ (brown, Schw-AdS$_4$ case), $\mu=\frac{3}{2.5}<\mu_{c}$ (blue), $\mu=\mu_{c}=\sqrt 2$ (dashed green), $\mu=\frac{3}{2}\mu_{c}$ (black).  {\bf Right Panel:}  RN-AdS$_4$ chemical potential $\mu$ vs horizon radius $r_+/L$ for four fixed values of $T$ (curves from top to bottom): $T=0$ (dark blue), $T=0.2< T_{c}$ (black), $T_{c}=\frac{\sqrt{3}}{2\pi}$ (dashed green) and $T=0.3>T_{c}$ (blue). }
\label{fig:Grand_radius_T}
\end{figure}

The discussion of the hairy solutions in this ensemble requires that we revisit first the properties of the RN-AdS$_4$ black hole \cite{Chamblin:1999tk,Chamblin:1999hg}. Taking the RN-AdS$_4$ temperature \eqref{RNthermo} and solving it with respect to $R_+$ we find 
\begin{align}
R_+{\bigl |}_{\pm}=\frac{1}{6} \left(4 \pi T L\pm \sqrt{2} \sqrt{3 \mu^2+8 \pi ^2 (T L)^2-6}\right),
\end{align}
which indicates that for $\mu<\sqrt{2}$ there are two branches of RN-AdS$_4$ solutions: the `small' and `large' black holes described by $R_+{\bigl |}_{-}$ and $R_+{\bigl |}_{+}$, respectively. For $\mu>\sqrt{2}$ only the large branch is present. This is best illustrated in the plots and caption of Figs. \ref{fig:Grand_radius_T} where we display $T(R_+)$ for several fixed $\mu$'s ({\it left panel}) and $\mu(R_+)$ for several fixed $T$'s ({\it right panel}). 
For later use, notice that small horizon radius corresponds to large temperatures.

\begin{figure}[t]
\centerline{
\includegraphics[width=.3\textwidth]{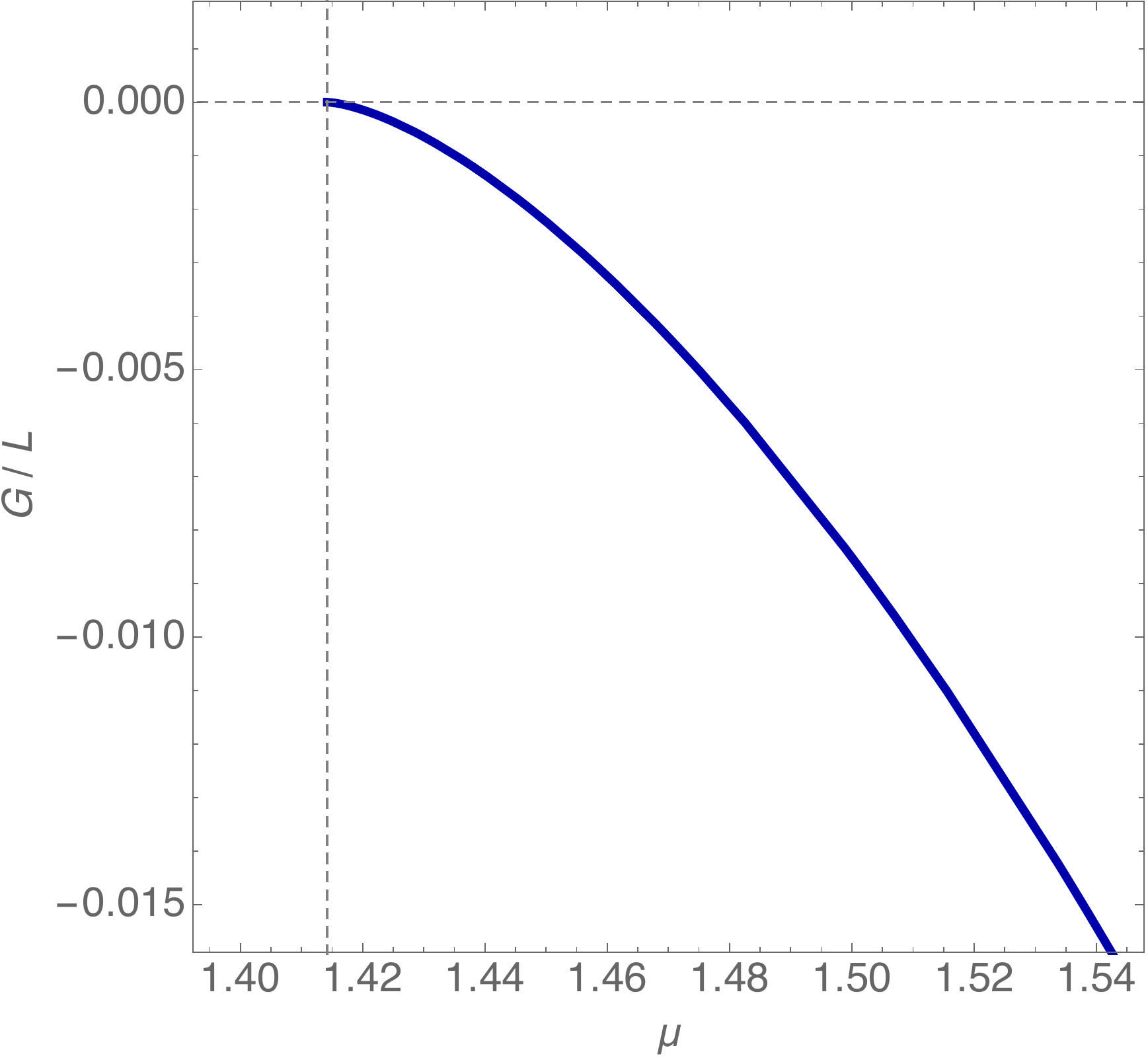}
\hspace{0.3cm}
\includegraphics[width=.3\textwidth]{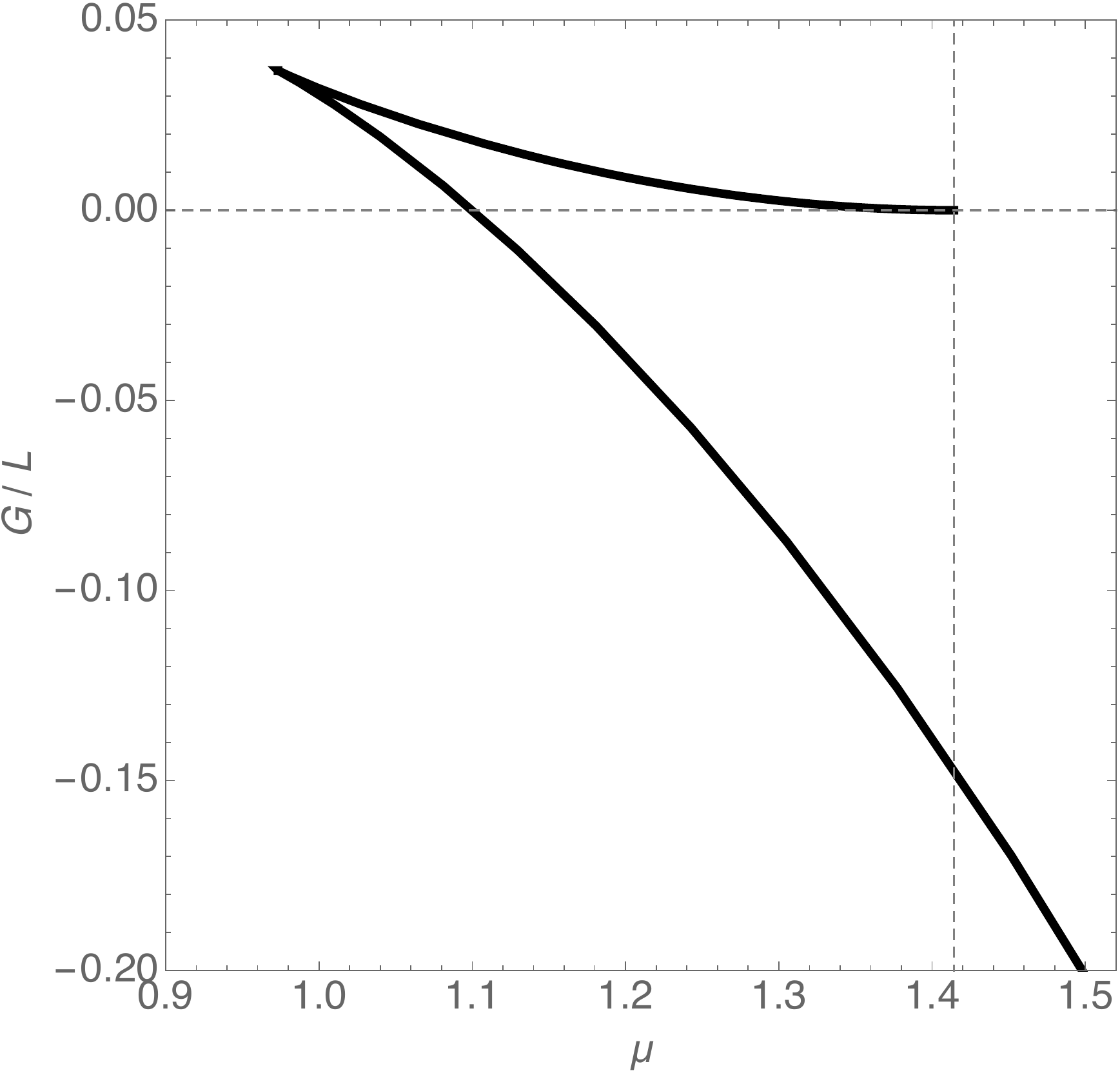}
\hspace{0.3cm}
\includegraphics[width=.3\textwidth]{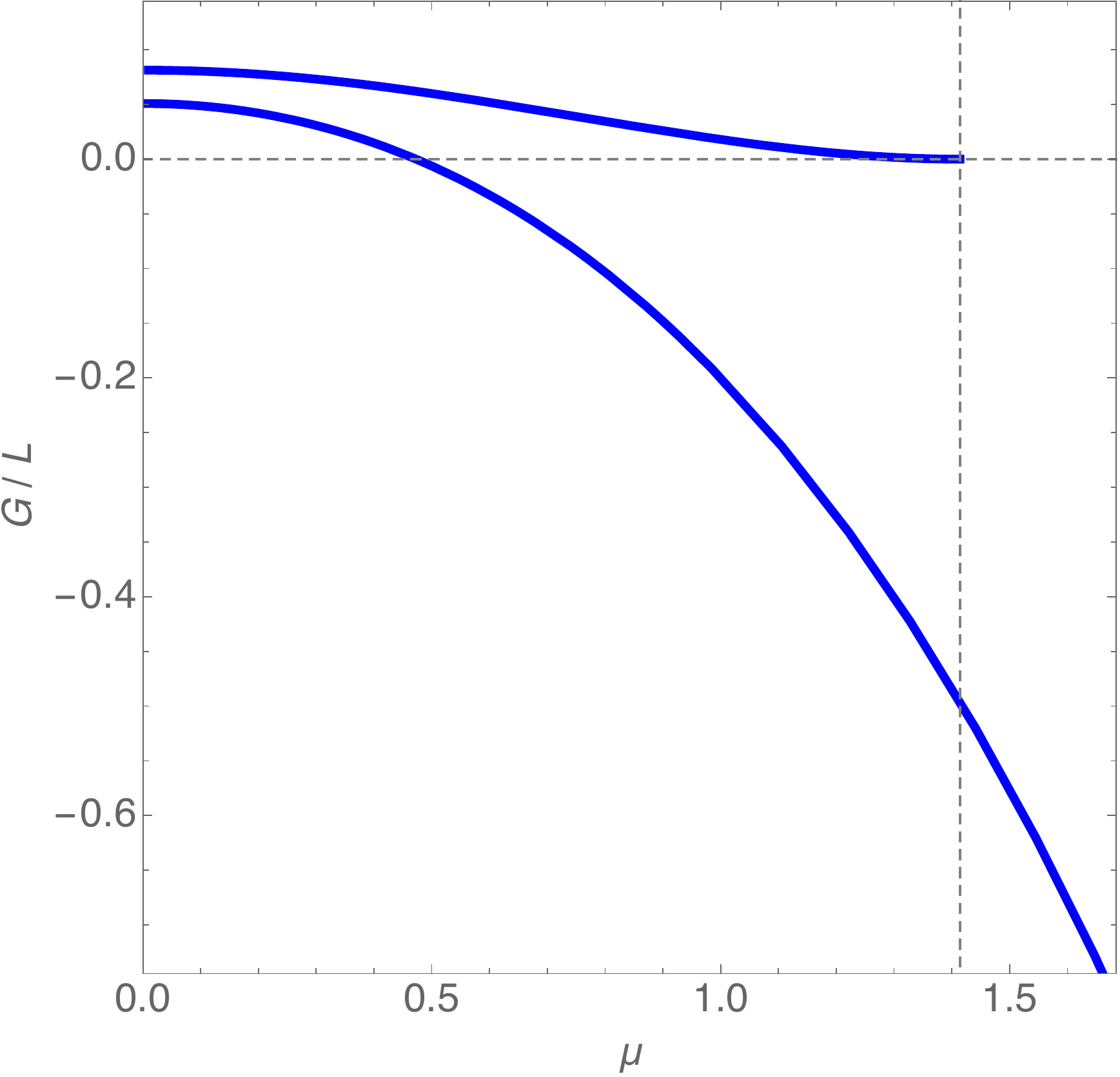}
}
\caption{Gibbs free energy $G$ vs chemical potential $\mu$ of the RN-AdS$_4$ black hole for: $T=0$ ({\bf left panel}), $T=0.2< T_{c}$ ({\bf middle panel}) and $T=0.3>T_{c}$ ({\bf right panel}). For $T_{c}=\frac{\sqrt{3}}{2\pi}\sim 0.28 $ the cusp seen in the middle panel plot hits $\mu=0$. The upper (lower) curve describes small (large) RN-AdS$_4$ black holes. The vertical dashed line is at $\mu=\sqrt{2}$ (see also Fig. \ref{fig:Grand_radius_T}). For  $0<T<T_{c_2}=\frac{1}{\pi}\sim 0.32$, the Hawking-Page phase transition occurs when large black holes attain $G(\mu_{HP})=0$ (see middle and right panels). However, for  $T>T_{c_2}=\frac{1}{\pi}$ (not shown in this figure; case represented in the left panel of Fig. \ref{fig:GCphaseDiag}), large black holes always have $G\leq 0$ and thus there is no Hawking-Page phase transition.}
\label{fig:Grand_RN_energy}
\end{figure}

\begin{figure}[t]
\centerline{
\includegraphics[width=.48\textwidth]{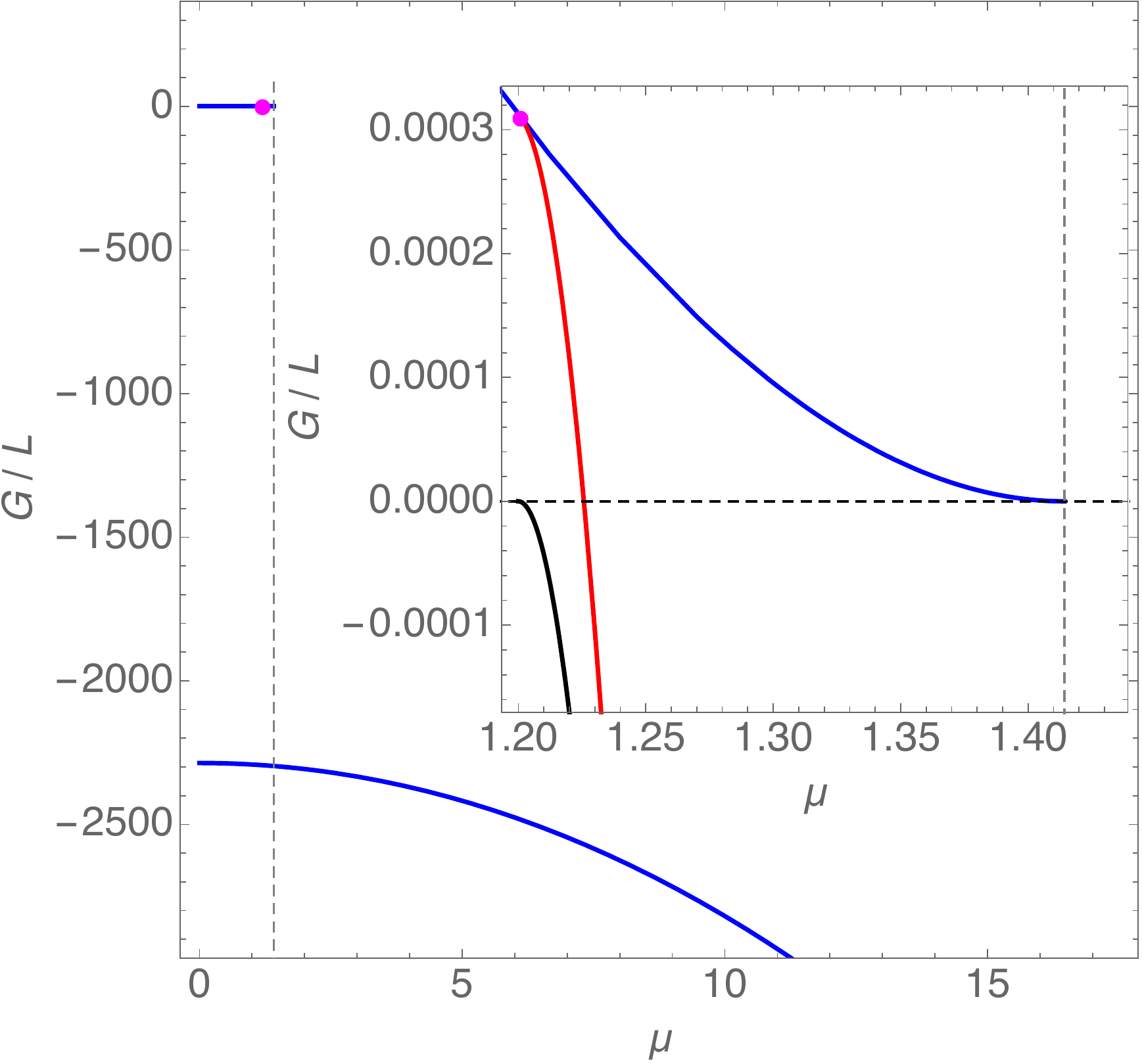}
\hspace{0.3cm}
\includegraphics[width=.45\textwidth]{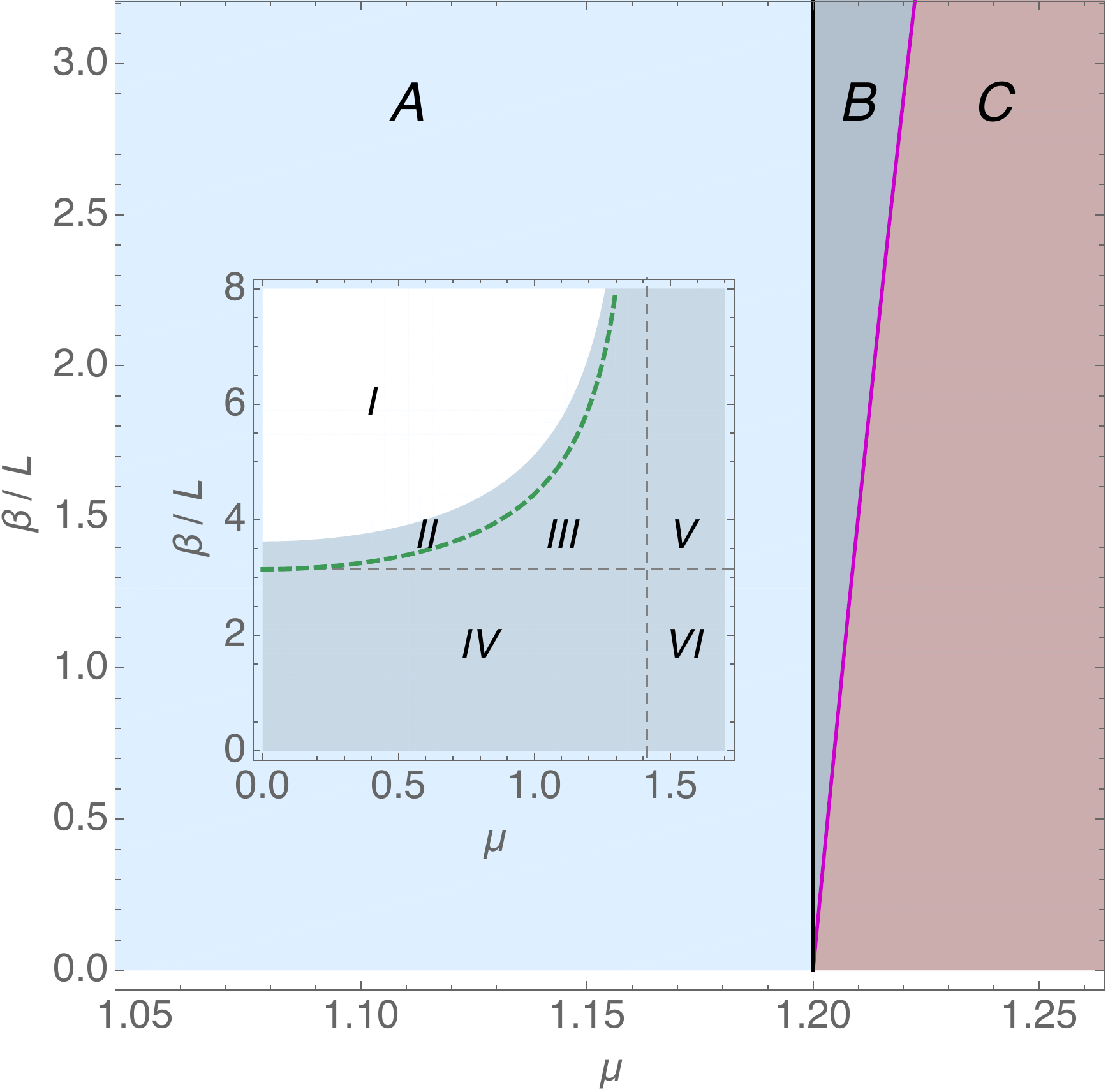}
}
\caption{{\bf Left Panel:}  Grand-canonical ensemble phase diagram $G$ vs $\mu$ for $T L=5$ and $e=2.5$. In the main plot, the upper (lower) curve describes small (large) RN-AdS$_4$ black holes. The vertical dashed line is at $\mu=\sqrt{2}$. The magenta point signals the onset of superradiance. The inset plot, is a zoom around the superradiant onset. It now also includes the hairy black hole (red curve) that branches-off from the small  RN-AdS$_4$ black hole at the onset of superradiance, as well as the thermal soliton (black curve) that starts at $\mu=0,G=0$ and extends to negative values of $G$. The dashed horizontal line describes the Giggs free energy of thermal AdS$_4$.   
The dominant thermal phase has the lowest Gibbs free energy. {\bf Right Panel:}   Phase diagram  $\beta=\frac 1 T$ vs $\mu$ of the grand-canonical ensemble  (shown for $e=2.5$). This is discussed in the text.}
\label{fig:GCphaseDiag}
\end{figure}

In the grand-canonical ensemble, the appropriate phase diagram  to describe the thermal phases of the theory is $G(\mu,T)$. To make the presentation clear, in  Fig. \ref{fig:Grand_RN_energy} we plot the $G$ of the RN-AdS$_4$ black hole as a function of $\mu$ for several fixed temperatures. At extremality ({\it left panel}), the RN-AdS$_4$ black hole dominates over thermal AdS$_4$. Above extremality, large black holes always have lower $G$ than small  RN-AdS$_4$ black holes whenever they co-exist ($G$ of small black holes is always non-negative). For  $0<T<T_{c_2}=\frac{1}{\pi}\sim 0.32$,  large black holes have $G>0$ for $\mu<\mu_{HP}$ but $G<0$ for $\mu>\mu_{HP}$ (see middle and right panels). So thermal-AdS$_4$ is preferred over RN-AdS$_4$ for $\mu<\mu_{HP}$, and a Hawking-Page phase transition occurs at $\mu=\mu_{HP}$, with large black holes becoming the dominant phase. However, for  $T>T_{c_2}=\frac{1}{\pi}$ (not shown in this figure; case represented in the left panel of Fig. \ref{fig:GCphaseDiag}),  there is no Hawking-Page phase transition since large RN-AdS$_4$ black holes always have $G\leq 0$ and dominate the ensemble. 

We can now add the hairy thermal phases to the grand-canonical phase diagram. Our perturbative analysis describes hairy black holes that bifurcate, at the onset of superradiance, from small $R_+$ RN-AdS$_4$ black holes. It follows from Figs. \ref{fig:Grand_radius_T}  that our perturbative computation is valid for large temperatures $T\gg T_{c}=\frac{\sqrt{3}}{2\pi}$ and chemical potential $\mu<\sqrt{2}\sim 1.41$. Therefore, as an illustrative example, in the left panel of Fig. \ref{fig:GCphaseDiag} we fix the temperature to be $T=5>T_{c_2}>T_c$ and the scalar field charge to be $e=2.5$, and we show all the thermal phases (small and large RN-AdS$_4$, thermal AdS$_4$, thermal soliton and hairy black hole) in the phase diagram $G$ vs $\mu$. In the main plot we find the large RN-AdS$_4$ black hole branch which exists for any $\mu$ and $-$ since we are at $T>T_{c_2}$ $-$ always has $G<0$. We also  display the small RN-AdS$_4$ that has $G\geq 0$ and exists for $\mu\leq \sqrt{2}$. The magenta point in this small RN-AdS$_4$ branch signals the zero-mode of superradiance, with small RN-AdS$_4$ to the right of this point being unstable. The inset plot then zooms the phase diagram to capture the region where the small RN-AdS$_4$ black holes are unstable. It also shows the hairy black hole branch (red curve) that bifurcates from the small RN-AdS$_4$ at the onset of superradiance. These black holes have lower $G$, and are thus preferred, than small  RN-AdS$_4$ black holes, whenever the two phases coexist. For sufficiently large $\mu$, the hairy black holes also have lower $G$ than thermal AdS$_4$ (which, recall, has $G=0$). However, hairy black holes always have higher $G$ than  large RN-AdS$_4$ black holes, when they co-exist. Thermal solitons (black curve) have $G<0$\footnote{For $T<T_{c_2}\sim 0.32$, solitons can be the overall preferred phase for small $\mu$ over $\frac 3 e$.}.

These qualitative properties extend to all other values of $T$ (and $e$) where our perturbative analysis is valid. Therefore, we conclude that in the perturbative regime where our hairy results hold and the solutions co-exist, large RN-AdS$_4$ black holes are always the preferred thermal phase over the hairy solutions in the grand-canonical ensemble.

Given that in the grand-canonical ensemble we keep the temperature $T$ and chemical potential $\mu$ fixed, it is instructive to summarize the conclusions above  for the thermal phases of the theory in a phase diagram $T$ vs $\mu$, or $\beta=1/T$ vs $\mu$. This effectively describes a projection of the 3-dimensional plot $G(T,\mu)$. This phase diagram is given in the  right panel of Fig. \ref{fig:GCphaseDiag} (for $e=2.5$). Recall that the perturbative results are valid for large values of the temperature, i.e. small $\beta=1/T$ (say $\beta \lesssim 3$) and $\mu<\sqrt{2}$. Consider first the main plot in the  right panel of Fig. \ref{fig:GCphaseDiag}. This is for $\mu<\sqrt{2}\sim 1.4$ and regions $A$, $B$ and $C$ are populated both by small and large RN-AdS$_4$ black holes. The onset curve of superradiance is the magenta boundary of regions $B$ and $C$. Small RN-AdS$_4$ black holes to the right of this curve are unstable to superradiance. Hairy black holes with lower free energy than the small  RN-AdS$_4$ solutions exist in region $C$.  Thermal solitons exist in regions $B$ and $C$ and have lower $G$ than thermal AdS$_4$, small RN-AdS$_4$ and hairy black holes. However, the solution with lowest Gibbs free energy in all regions $A$, $B$ and $C$ is the large RN-AdS$_4$ that always dominates the grand-canonical ensemble for $\beta<\beta_{c_2}=\pi$. Consider now the inset plot. This extends the main plot to values of $\beta$ higher than $\beta_{c_2}=\pi$ (the horizontal dashed line)  and $\mu>\sqrt{2}$, but does {\it not} include information about the hairy solutions. It allows to identify clearly the Hawking-Page phase transition (the main plot is a subset of region $IV$). Regions $I - VI$ describe the domain of existence of thermal AdS$_4$. Small RN-AdS$_4$ black holes exist on regions $II$, $III$ and $IV$ i.e for $\mu<\sqrt{2}$ (vertical dashed line), while large RN-AdS$_4$ black holes exist in  regions $II-VI$. The boundary between regions $I$ and $II$ corresponds to the cusp in the middle panel of Fig. \ref{fig:Grand_RN_energy}. The green dashed curve is the Hawking-Page phase transition curve. Below it, large RN-AdS$_4$ black holes dominate the grand-canonical ensemble. 

So far we have  found that small hairy black holes are never the preferred global thermodynamic phase of the grand-canonical phase diagram. However, they still dominate over small RN-AdS$_4$ black holes so, for completeness, we can discuss their local thermodynamic stability. In the grand-canonical we keep the intensive variables $T$ and $\mu$ fixed. Therefore, the local thermodynamic stability condition in this ensemble requires that the inverse of the Weinhold Hessian matrix, namely \cite{Chamblin:1999tk,Chamblin:1999hg,Monteiro:2008wr,Monteiro:2009tc}
\begin{equation}\label{HessianGC}
g_W^{\mu \nu} = -\frac{\partial^2 G}{\partial y_\mu \,\partial y_\nu} = 
\left[
\begin{array}{cc}
\beta C_\mu & \eta \\
\eta & \epsilon_T
\end{array}
\right],
\qquad y_\mu=(T,\mu),
\end{equation}
is positive definite. Here, $G$ is the Gibbs free energy,  and $C_\mu$ and $\epsilon_T$ are, respectively, the specific heat at constant chemical potential and the isothermal permittivity (i.e. capacitance) and given by   
\begin{equation}
C_\mu  = - T \left(\frac{\partial^2 G}{\partial T^2}\right)_\mu= T \left( \frac{\partial S}{\partial T} \right)_\mu, \qquad \epsilon_T=  - \left(\frac{\partial^2 G}{\partial \mu^2}\right)_T= \left( \frac{\partial Q}{\partial \mu} \right)_T,
\end{equation}
where we used the first law $dG=-S dT-Q d\mu$ to rewrite  $C_\mu$ and $\epsilon_T$ in terms of first derivative quantities. Finally, the off-diagonal term is $\eta=\left(  \frac{\partial Q}{\partial T} \right)_\mu = \left(  \frac{\partial S}{\partial \mu} \right)_T$. The equality of these two expressions is a Maxwell relation that follows from the symmetry of the Hessian matrix.
The Hessian matrix \eqref{HessianGC} can be written in the new basis $(dT,d\mu) \to (dT,d\mu + \epsilon_T^{-1} \eta d T)$, where it diagonalizes as $\hbox{diag}\{\beta C_\mu, \epsilon_T\}$.
We find that small hairy black holes have $C_\mu>0$ and $\epsilon_T<0$. That is,  $C_\mu$ and $\epsilon_T$ have opposite signs, and thus the Hessian has a negative eigenvalue. Small hairy black holes are thus locally unstable in the grand-canonical ensemble. Essentially, this is because they have $\epsilon_T<0$ and are thus electrically unstable: our theory has charged scalar field quanta and a small electric fluctuation that increases the charge of the black hole reduces its chemical potential.


\vskip .5cm
\centerline{\bf Acknowledgements}
\vskip .2cm
We wish to thank Jorge Santos and Marika Taylor for discussions. O.J.C.D. and R.M. acknowledge financial support from the STFC Ernest Rutherford grants ST/K005391/1 and ST/M004147/1.

\begin{appendix}

\section{Coefficients of the superradiant expansion}\label{App:Superradiance}

\subsection{Far region\label{AppendixSuperradiance:Far}}
Here, we give the coefficients of the far region expansion \eqref{superradiantExpansion}. The integration constants and  frequency coefficients $\omega_1$ and $\omega_2$ have already been determined by the boundary conditions and matching conditions as described in Section \ref{sec:superradiance}.
\begin{align}\label{superradiantExp:far}
\phi^{out}_0(R)=&\frac{e^{-i (\mu e-3) \arctan R}}{\left(R^2+1\right)^{3/2}},\nonumber\\
\phi^{out}_1(R)=&\frac{e^{-i (\mu {e}-3) \arctan R}}{96 R (R^2+1)^{5/2}} \Big[(R^2+1) (-24 i (\mu^2+2) R (\mu {e}-3) \log (\frac{R^2}{R^2+1})\nonumber\\
&-48 R \arctan R (3 \mu^2 -2 \mu {e}-2 i \omega_1+6)-4 R (3 R^2+12 i \pi +17) \omega_1+3 (R^4+6 R^2\nonumber\\
&-3) \omega_1 (2 i \log (R-i)-i \log (4+(R-i)^2)+2 \tan ^{-1}(\frac{2}{R-i})))+24 (-i \mu^3 {e} R\nonumber\\
&+3 \mu^2 (-1+R (\pi  R^2-2 R+i +\pi ))-2 \mu {e} (-2+R (\pi  R^2-2 R+i+\pi ))\nonumber\\
&+6 (-1+R (\pi  R^2-2 R+i+\pi )))\Big],\\
\phi^{out}_2(R)=&\frac{1}{R}\Big(-\frac{3}{16} \pi  \omega_2+\frac{1}{4608 \pi ^2}\Big(-256 i (3 \mu^2-4 \mu e+6)^2 (2 \tanh ^{-1}(2)-\log (3))\nonumber\\
&+32 \pi  (3 \mu^2-4 \mu e+6) (9 \mu^3 e (2 \tanh ^{-1}(2)-\log (3))-3 \mu^2 (52-9 \log (3)\nonumber\\
&+18 \tanh ^{-1}(2))+2 \mu e (88-9 \log (3)+18 \tanh ^{-1}(2))-6 (52-9 \log (3)\nonumber\\
&+18 \tanh ^{-1}(2)))+9 \pi ^3 (27 \mu^4-264 \mu^3 e+4 \mu^2 (56 e^2+117)-528 \mu e+108)\nonumber\\
&+288 i \pi ^2 (7 \mu^5 e-\mu^4 (4 e^2+21)+32 \mu^3 e-4 \mu^2 (2 e^2+15)+52 \mu e-84)\Big)\Big)+\mathcal O(R^0) \nonumber
\end{align}
Due to the length of the expression, we only show the leading term of the small $R$ expansion of $\phi^{out}_2(R)$ that is needed to determine $\omega_2$.

\subsection{Near region\label{AppendixSuperradiance:Near}}
Here we present the coefficients of the near region expansion \eqref{superradiantExpansion}. The integration constants have already been determined by the boundary conditions and by the matching conditions, as described in Section \ref{sec:superradiance}.
\begin{align}\label{superradiantExp:near}
\phi^{in}_0(y)=&1,\nonumber\\
\phi^{in}_1 (y)=&\frac{1}{36 \pi }\Big(9 \pi ^2 (-2 e \mu +3 \mu ^2+6)-64 e \mu -18 i \pi  (\mu ^2+2) (e \mu -3) (\log (y-\frac{\mu ^2}{2})+\log (R_p))\nonumber\\
&-3 i \pi  (e \mu  (3 \mu ^2+38)+12 y (e \mu -3)-33 (\mu ^2+2))+48 (\mu ^2+2)\Big),\\
\phi^{in}_2 (y)=&y^2 (-\frac{1}{2} e^2 \mu ^2+3 e \mu -6)+\frac{y}{36 \pi }\Big( (i (9 \pi ^2 (2 e^2 \mu ^2-3 e (\mu ^2+4) \mu +9 (\mu ^2+2)\nonumber)\\
&+3 i \pi  (e^2 (3 \mu ^2+38) \mu ^2-6 e (7 \mu ^2+36) \mu +144 (\mu ^2+2))+16 e \mu  (4 e \mu -3 (\mu ^2+2)))\nonumber\\
&-18 \pi  (\mu ^2+2) (e \mu -3)^2 \log (R_p)-18 \pi  (\mu ^2+2) (e \mu -3)^2 \log (y))\Big)+\mathcal O\left(y^0\right)\nonumber
.
\end{align}

Due to the length of the expression, we only show the large $y$ expansion of $\phi^{in}_2(y)$ needed to determine $\omega_2$.
However, note that the next order contributions to $\phi^{in}_2(y)$ and $\phi^{out}_2(R)$ are required to fix a remaining integration constant. 

\section{Coefficients of the soliton expansion}\label{App:Soliton}
In this Appendix we give the final coefficients of the soliton expansion \eqref{solitonexp}. The integration constants have already been fixed by the boundary conditions as described in Section \ref{sec:Soliton}.
\begin{align}
f_0(R)=&R^2+1,\nonumber\\
f_2(R)=&-\frac{3 \left(3 R^2+5\right)}{8 \left(R^2+1\right)^2}-\frac{9 \arctan R}{8 R},\nonumber\\
f_4(R)=& -\frac{315 \pi ^2 (2 e^2-9) (R^2+1)}{1024}\\
&+\frac{1}{12800 R^2 (R^2+1)^5} \Bigg(15 (R^2+1)^2 \arctan R \Big(R (90 \pi ^2 (e^2-9) (R^2+1)^3\nonumber\\
&+e^2 (4200 R^6+10899 R^4+8398 R^2+1339) (R^2+1)-18 (1050 R^8 +3689 R^6\nonumber\\
&+4447 R^4+1887 R^2+159))+30 (R^2+1) \arctan R (-4 (e^2 -9) R (R^2+1)^2 \arctan R\nonumber\\
&+e^2 (70 R^8+210 R^6+200 R^4+54 R^2+2)-9 R^2 ((35 (R^2 +3) R^2+93) R^2+15))\Big)\nonumber\\
&+R^2 \Big(450 \pi ^2 (e^2-9) (3 R^2+5) (R^2+1)^3+e^2 (31500 R^{10}+163485 R^8+340630 R^6\nonumber\\
&+361612 R^4+198846 R^2+45751)-18 (7875 R^{10}+39585 R^8+78480 R^6\nonumber\\
&+76232 R^4+35821 R^2+5751)\Big)\Bigg),\nonumber
\end{align}
\begin{align}
A_0(R)=&\frac{3}{e},\nonumber\\
A_2(R)=&\frac{21 (e^2-3)}{32 e}-\frac{e (3 R^2+5)}{8 (R^2+1)^2}-\frac{3 e \arctan R}{8 R},\\
A_4(R)=&\frac{1}{614400 e} \Bigg(\frac{1}{(R^2+1)^5}\Big(16 e^2 (450 \pi ^2 (e^2-9) (3 R^2+5) (R^2+1)^3+4 e^2 (6015 R^6\nonumber\\
&+19355 R^4+21443 R^2+8217) (R^2+1)\nonumber\\
&+\frac{1}{R}\Big(15 (R^2+1)^2 \arctan R (90 \pi ^2 (e^2-9) (R^2+1)^3+4 e^2 (851 R^4+1312 R^2\nonumber\\
&+311) (R^2+1)+60 \arctan R ((R^3+R) (e^2 (30 R^4+54 R^2+20)-9 (15 R^4+24 R^2\nonumber\\
&+5))-2 (e^2-9) (R^2+1)^3 \arctan R)-9 (1513 R^6+3579 R^4+2259 R^2+353))\Big)\nonumber\\
&-9 (9195 R^8+36610 R^6+53524 R^4+33830 R^2+6761))\Big)-3 \Big(339814 e^4\nonumber\\
&-3159903 e^2+2100 \pi ^2 (4 e^4+42 e^2-297)+7714818\Big)\Bigg).\nonumber
\end{align}
\newline
\begin{align}
\phi_1(R)=& \frac{1}{(R^2+1)^{3/2}},\nonumber\\
\phi_3(R)=&-\frac{3}{320 R (R^2+1)^{9/2}}\Bigg(20 (R^2+1)^2 \arctan R \Big(-2 e^2 (R^2+1)-(e^2-9) R (R^2\nonumber\\
&+1) \arctan R+18 R^2+9\Big)+R \Big(5 \pi ^2 (e^2-9) (R^2+1)^3+e^2 (22 R^4+48 R^2+26)\nonumber\\
&+54 R^4+36 R^2+22\Big)\Bigg),\\
\phi_5(R)=&\Big(\dots\Big).\nonumber
\end{align}
The expression for $\phi_5(R)$ is too long and not illuminating. We note that the integration constants arising from its equation of motion are determined by imposing that $\phi_5(R)\Big|_{R\to \infty}=\mathcal O(R^{-4})$. Also, imposing regularity at $R=0$ directly gives the expression for $\mu_4$ in \eqref{ThermoSoliton}.

\section{Coefficients of the hairy black hole expansion}\label{App:hairyBH}

\subsection{Far region expansion\label{AppendixBH:Far}}
Below, we give the coefficients of the hairy black hole far region expansion \eqref{RNexp}. The integration constants have already been fixed by the boundary conditions and by the matching with the near region as discussed in Section \ref{sec:Hairy BH}.
\begin{align}
f_{00}^{out}(R)=& R^2+1,\nonumber\\
f_{01}^{out}(R)=&-\frac{\frac{9}{e^2}+2}{2 R},\nonumber\\
f_{02}^{out}(R)=&-\frac{3 (16 e^2 R-3 \pi  e^2-72 R)}{2 \pi  e^4 R^2},\nonumber\\
f_{20}^{out}(R)=&-\frac{3 (R (3 R^2+5)+3 (R^2+1)^2 \arctan R)}{8 R (R^2+1)^2},\nonumber\\
f_{21}^{out}(R)=&\frac{1}{32 \pi  e^2 R^2 (R^2+1)^3}\Bigg(3 (R^2+1) \arctan R \Big(6 R (R^2+1)^2 (16 (2 e^2-9) R (R^2+1)\nonumber\\
&+27 \pi ) \arctan R+3 \pi  (2 e^2 (R^2+1)^2 (3 (R^4+R^2)+2)+153 R^2-27 (R^4+3 R^2\nonumber\\
&-1) R^4)+8 (2 e^2-9) R (24 R^6+73 R^4+66 R^2+9)-162 \pi ^2 R (R^2+1)^2\Big)\nonumber\\
&+\frac{1}{2} R \Big(16 (2 e^2-9) R (R^2+1) (3 R^2+5) (12 R^2+17)\nonumber\\
&-9 \pi ^2 R (R^2+1) (19 (2 e^2 (R^2+1)^3-9 R^4 (R^2+3))-405 R^2+9)+6 \pi  (e^2 (18 R^8\nonumber\\
&-3 R^6-69 R^4-R^2+31)+9 (-9 R^8+6 R^6+72 R^4+72 R^2+7))\Big)\Bigg),\end{align}
\begin{align}
f_{40}^{out}(R)=&\frac{1}{12800 R^2 (R^2+1)^5}\Bigg(-1800 (e^2-9) R (R^2+1)^5 \arctan R^3\nonumber\\
&+450 (R^2+1)^3 (e^2 (70 R^8+210 R^6+200 R^4+54 R^2+2)-9 R^2 (35 R^6+105 R^4\nonumber\\
&+93 R^2+15)) \arctan R^2+15 R (R^2+1)^2 \Big(90 \pi ^2 (e^2-9) (R^2+1)^3+e^2 (4200 R^8\nonumber\\
&+15099 R^6+19297 R^4+9737 R^2+1339)-18 (1050 R^8+3689 R^6+4447 R^4\nonumber\\
&+1887 R^2+159)\Big) \arctan R-\frac{1}{2} R^2 \Big(225 \pi ^2 (R^2+1)^3 (2 e^2 (35 R^6+105 R^4+99 R^2\nonumber\\
&+25)-9 (35 R^6+105 R^4+93 R^2+15))-2 e^2 (31500 R^{10}+163485 R^8+340630 R^6\nonumber\\
&+361612 R^4+198846 R^2+45751)+36 (7875 R^{10}+39585 R^8+78480 R^6\nonumber\\
&+76232 R^4+35821 R^2+5751)\Big)\Bigg).\nonumber
\end{align}

\begin{align}
A_{00}^{out}(R)=&\frac{3}{e},\nonumber\\
A_{01}^{out}(R)=&\frac{4 (2 e^2-9)}{\pi  e^3}-\frac{3}{e R},\nonumber\\
A_{02}^{out}(R)=&\frac{(2 e^2-9) (9 \pi ^2 (10 e^2-9) R+128 (21-2 e^2) R-128 \pi  e^2)}{32 \pi ^2 e^5 R},\nonumber\\
A_{20}^{out}(R)=&\frac{21 (e^2-3)}{32 e}-\frac{e (3 R^2+5)}{8 (R^2+1)^2}-\frac{3 e \arctan R}{8 R},\nonumber\\
A_{21}^{out}(R)=&\frac{1}{9600 \pi  e^3}
\Bigg(\frac{1}{R (R^2+1)^3}\Big(100 e^2 (8 (2 e^2-9) R (R^2+1) (57 R^2+77)+3 (R^2\nonumber \\
&+1) \arctan R (6 (R^2+1)^2 (8 (2 e^2-9) R+27 \pi ) \arctan R+8 (2 e^2-9) (25 R^4\nonumber\\
&+42 R^2 +9)+27 \pi  R (2 e^2 (R^2+1)^2-9 R^4-6 R^2+11)-162 \pi ^2 (R^2+1)^2)\nonumber\\
&-3 \pi  (e^2 (31 R^6 +79 R^4+21 R^2-11)-9 (20 R^6+71 R^4+66 R^2+7))\\
&-162 \pi ^2 R (3 R^4+8 R^2+5))\Big) -7650 \pi ^2 (2 e^2-9) e^2+75 \pi ^2 (236 e^4-1809 e^2+4212)\nonumber\\
&-8 (25330 e^4-264861 e^2 +685503)\Bigg)\nonumber\\
A_{40}^{out}(R)=&\frac{1}{614400}\Bigg(-\frac{3 (339814 e^4-3159903 e^2+2100 \pi ^2 (4 e^4+42 e^2-297)+7714818)}{e}\nonumber\\
&+\frac{138240 e}{(R^2+1)^5}+\frac{384 e (19 e^2+63)}{(R^2+1)^4}+\frac{128 e (389 e^2+1278)}{(R^2+1)^3}+\frac{160 e}{(R^2+1)^2}\Big( (524 e^2 \nonumber\\
&+90 \pi ^2 (e^2-9)+153)\Big)+\frac{240 e (1604 e^2+90 \pi ^2 (e^2-9)-5517)}{R^2+1}\nonumber \\
&+\frac{240 e \arctan R }{R (R^2+1)^3}\Big(90 \pi ^2 (e^2-9) (R^2+1)^3+4 e^2 (851 R^4+1312 R^2+311) (R^2+1)\nonumber
\end{align}
\begin{align}
&+60 \arctan R ((R^3+R) (e^2 (30 R^4+54 R^2+20)-9 (15 R^4+24 R^2+5))\nonumber\\
&-2 (e^2-9) (R^2+1)^3 \arctan R)-9 (1513 R^6+3579 R^4+2259 R^2+353)\Big)\Bigg)\nonumber
\end{align}

\begin{align}
\phi_{10}^{out}(R)=&\frac{1}{\left(R^2+1\right)^{3/2}} ,\nonumber\\
\phi_{11}^{out}(R)=&\frac{1}{12 \pi  e^2 R \left(R^2+1\right)^{5/2}}\Big(R (-2 (2 e^2-9) (3 R^2+17) (R^2+1)+3 \pi  R (2 e^2 (R^4+7 R^2+3)\nonumber\\
&-9 (R^4+7 R^2+9))+81 \pi ^2 (R^2+1))-6 (R^2+1) (2 e^2 (R^4+6 R^2-3)-9 (R^4 \nonumber\\
&+6 R^2-3 \pi  R-3)) \arctan R\Big) ,\nonumber\\
\phi_{12}^{out}(R)=&\frac{1}{288 \pi ^2 e^4}\Big(108 \pi ^2 \left(4 e^4+72 e^2-891\right) \log (R)+128 \left(28 e^4-324 e^2+891\right)\nonumber\\
&-36 \pi ^2 (28 e^4+90 e^2+432 \left(9-2 e^2\right) \log (2)+243)+6561 \pi ^4\Big)+\mathcal O(R),\nonumber\\
\phi_{30}^{out}(R)=&-\frac{3}{320 R \left(R^2+1\right)^{9/2}}\Big(20 (R^2+1)^2 \arctan R (-2 e^2 (R^2+1)\nonumber\\
&-(e^2-9) R (R^2+1) \arctan R+18 R^2+9)+R (5 \pi ^2 (e^2-9) (R^2+1)^3 \nonumber\\
&+e^2 (22 R^4+48 R^2+26)+54 R^4+36 R^2+22)\Big),\\
\phi_{31}^{out}(R)=&-\frac{1}{57600 \left(\pi  e^2\right)}\Big(-37800 \left(2 e^2-9\right) \left(5 e^2-36\right) \zeta (3)+42525 \pi ^4 \left(e^2-9\right)\nonumber\\
&+16 \left(14092 e^4-182394 e^2+528849\right)+30 \pi ^2 (4 e^4 (1260 \log (2)-773)\nonumber\\
&-16200 e^2 (\log (8)-2)+81 (900 \log (2)-307))\Big)+\mathcal O(R),\nonumber\\
\phi_{50}^{out}(R)=&\frac{1}{64512000}\Big(-39749042 e^4+533807289 e^2+307125 \pi ^4 \left(e^2-9\right)^2+3969000 (10 e^4\nonumber\\
&-84 e^2+135) \zeta (3)-18900 \pi ^2 (-292 e^4+2513 e^2+60 \left(10 e^4-84 e^2+135\right) \log (2)\nonumber\\
&-1341 )-1456792614\Big)+\mathcal O(R).\nonumber
\end{align}
$\phi_{12}^{out}(R)$, $\phi_{31}^{out}(R)$ and $\phi_{50}^{out}(R)$ have large expressions. We only display the small $R$ expansion that is needed for the matching with the near region. 

\subsection{Near region expansion\label{AppendixBH:Near}}
Next, we present the coefficients of the hairy black hole near region expansion \eqref{RNnear}. The integration constants have already been determined by the boundary conditions and by the matching with the far region as described in Section  \ref{sec:Hairy BH}.
\begin{align}
f_{00}^{in}(y)=&\frac{(y-1) \left(2 e^2 y-9\right)}{2 e^2 y^2},\nonumber\\
f_{01}^{in}(y)=&-\frac{12 \left(2 e^2-9\right) (y-1)}{\pi  e^4 y^2},\nonumber\\
f_{02}^{in}(y)=&\frac{(y-1) }{32 \pi ^2 e^6 y^2} \Big(128 (4 (e^2-27) e^2+405)+\pi ^2 (32 e^6 y (y^2+y+1)-540 e^4+2916 e^2\nonumber\\
&-2187)\Big), \nonumber
\end{align}
\begin{align}
f_{20}^{in}(y)=& -\frac{3 (y-1) (e^2 (32 y-11)-63)}{32 e^2 y^2},\\
f_{21}^{in}(y)=& -\frac{(y-1) }{3200 \pi  e^4 y^2}\Big(75 \pi ^2 (4 e^4 (57 y+8)+27 e^2 (26 y-97)+4212)+8 (-10 e^4 (2240 y\nonumber\\
&+563)+63 e^2 (1600 y+2697)-657153)\Big) ,\nonumber\\
f_{40}^{in}(y)=& -\frac{(y-1) }{204800 e^2 y^2}\Big(-2 e^4 (533888 y+91447)+27 e^2 (86784 y+291935)\nonumber\\
&+900 \pi ^2 (4 e^4 (19 y+9)-6 e^2 (9 y+145)+2079)-22747554\Big) .\nonumber
\end{align}

\begin{align}
A_{00}^{in}(y)=&\frac{3 (y-1)}{e y} ,\nonumber\\
A_{01}^{in}(y)=&\frac{4 \left(2 e^2-9\right) (y-1)}{\pi  e^3 y} ,\nonumber\\
A_{02}^{in}(y)=& \frac{\left(2 e^2-9\right) \left(-256 e^2+9 \pi ^2 \left(10 e^2-9\right)+2688\right) (y-1)}{32 \pi ^2 e^5 y},\nonumber\\
A_{20}^{in}(y)=&-\frac{\left(11 e^2+63\right) (y-1)}{32 e y} ,\\
A_{21}^{in}(y)=&\frac{\left(80 \left(10 \pi ^2-151\right) e^4+9 \left(50744-7275 \pi ^2\right) e^2+4212 \left(25 \pi ^2-434\right)\right) (y-1)}{3200 \pi  e^3 y} ,\nonumber\\
A_{40}^{in}(y)=&\frac{3 \left(-21666 e^4+860405 e^2+300 \pi ^2 \left(12 e^4-290 e^2+693\right)-2571606\right) (y-1)}{204800 e y} .\nonumber
\end{align}

\begin{align}
\phi_{10}^{in}(y)=&1 ,\nonumber\\
\phi_{11}^{in}(y)=&\frac{-32 e^2+81 \pi ^2+144}{12 \pi  e^2} ,\nonumber\\
\phi_{12}^{in}(y)=& \frac{1}{8 e^4}\Bigg(1458 \text{Li}_2\left(\frac{9-2 e^2 y}{9-2 e^2}\right)-6 e^2 y (2 e^2 (y-1)+45)+3 i \pi  (4 e^4+72 e^2-891)-\nonumber
\end{align}
\begin{align}
&1458 \log (2 e^2-9) \log \left(\frac{y-1}{2 e^2 y-9}\right)+\frac{1}{36 \pi ^2}\Big(108 \pi ^2 (243 \log ^2\left(\frac{2 e^2}{9 R_p-2 e^2 R_p}\right)\nonumber\\
&+\log \left(\frac{2 e^2}{R_p}\right) (243 \log \left(\frac{2 e^2}{R_p}\right)-486 \log \left(\frac{2 e^2}{(2 e^2-9) R_p}\right)-4 (e^2+18) e^2-486 i \pi \nonumber\\
&+891)+486 \log (\frac{1}{2 e^2}) \log (2 e^2-9))-108 i \pi ^3 (4 e^4+72 e^2-891)+128 (28 e^4\nonumber\\
&-324 e^2+891)-36 \pi ^2 (28 e^4+90 e^2+432 (9-2 e^2) \log (2)+243)+15309 \pi ^4\Big)\nonumber\\
&+3 \log (2 e^2 y-9) (4 e^4+486 \log (\frac{2 e^2 (y-1)}{2 e^2-9})-243 \log (2 e^2 y-9)+72 e^2+486 i \pi\nonumber\\
& -891)\Bigg),
\end{align}
\begin{align}
\phi_{30}^{in}(y)=&\frac{1}{320} (-3) \left(-14 e^2+5 \pi ^2 \left(e^2-9\right)+202\right) ,\nonumber\\
\phi_{31}^{in}(y)=&\frac{3 \pi  \left(28 e^4-270 e^2+405\right) \log (2)}{32 e^2}-\frac{21 \left(2 e^2-9\right) \left(5 e^2-36\right) \zeta (3)}{32 \pi  e^2}\nonumber\\
&+\frac{1}{57600 \pi  e^2}
\Big(225472 e^4-2918304 e^2+42525 \pi ^4 (e^2-9)-30 \pi ^2 (3092 e^4-32400 e^2\nonumber\\
&+24867)+8461584\Big),\nonumber\\
\phi_{50}^{in}(y)=&\frac{1}{64512000}\Big(-39749042 e^4+533807289 e^2+307125 \pi ^4 (e^2-9)^2+3969000 (10 e^4\nonumber\\
&-84 e^2+135) \zeta (3)-18900 \pi ^2 (-292 e^4+2513 e^2+60 (10 e^4-84 e^2+135) \log (2)\nonumber\\
&-1341)-1456792614\Big) .\nonumber
\end{align}
\end{appendix}

\bibliography{refs}{}
\bibliographystyle{JHEP}

\end{document}